\documentclass[]{jfm} 

\usepackage{graphicx,footmisc}
\usepackage[caption=false]{subfig}
\usepackage{color}
\usepackage{natbib,url}
\usepackage{amsmath}
\usepackage{float}
\usepackage[normalem]{ulem}
\usepackage{longtable}
\usepackage{placeins}
\usepackage{epstopdf}
\makeatletter
\newcommand{\raisemath}[1]{\mathpalette{\raisem@th{#1}}}
\newcommand{\raisem@th}[3]{\raisebox{#1}{$#2#3$}}
\makeatother
\newcommand{\tavg}[1]{\overline{\raisebox{0pt}[1.1\height]{$\mkern+4mu #1 \mkern+4mu$}}^{\raisemath{-0.5pt}{\, t}}}
\newcommand{\ztavg}[1]{\overline{\raisebox{0pt}[1.1\height]{$\mkern+2mu #1 \mkern+2mu$}}^{\raisemath{-0.5pt}{\,\phi,z,t}}}
\newcommand{\shellavg}[1]{\overline{\raisebox{0pt}[1.1\height]{$\mkern+2mu #1 \mkern+2mu$}}^{\raisemath{-0.5pt}{\,\phi,\theta, t}}}

\def\be{\begin{equation}}
\def\ee{\end{equation}}

\def\begineqn{\begin{equation*}}
\def\endeqn{\end{equation*}}

\def\beginar{\begin{eqnarray}}
\def\endar{\end{eqnarray}}
\def\beginarn{\begin{eqnarray*}}
\def\endarn{\end{eqnarray*}}

\def\lb{\left ( }
\def\rb{\right ) }
\def\lsq{\left [ }
\def\rsq{\right ] }

\def\ub{\mathbf{u}}

\def\ubp{\mathbf{u}^{\prime}}

\def\mub{\overline{\bf u}}

\def\dst{{\partial_t}}

\def\hz{{\bf\widehat z}}

\def\hr{{\bf \widehat r}}

\def\mT{\overline{T}}

\def\mur{\overline{u}_r}

\def\muth{\overline{u}_\theta}

\def\muph{\overline{u}_\phi}

\def\Rat{\widetilde{Ra}}

\def\Ret{\widetilde{Re}}

\title{Asymptotic scaling relations for rotating spherical convection with strong zonal flows}

\author{Justin A. Nicoski\aff{1}, Anne R. O'Connor\aff{1} \and Michael A. Calkins\aff{1}}
\affiliation{\aff{1}Department of Physics, University of Colorado, Boulder, Colorado 80309, USA}

\begin{document}
\maketitle
\begin{abstract}

We analyse the results of direct numerical simulations of rotating convection in spherical shell geometries with stress-free boundary conditions, which develop strong zonal flows. Both the Ekman number and the Rayleigh number are varied.
We find that the asymptotic theory for rapidly rotating convection can be used to predict the Ekman number dependence of each term in the governing equations, along with the convective flow speeds and the dominant length scales.
Using a balance between the Reynolds stress and the viscous stress, together with the asymptotic scaling for the convective velocity, we derive an asymptotic prediction for the scaling behaviour of the zonal flow with respect to the Ekman number, which is supported by the numerical simulations. 
We do not find evidence of distinct asymptotic scalings for the buoyancy and viscous forces and, in agreement with previous results from asymptotic plane layer models, we find that the ratio of the viscous force to the buoyancy force increases with Rayleigh number. 
Thus, viscosity remains non-negligible and we do not observe a trend towards a diffusion-free scaling behaviour within the rapidly rotating regime.

\end{abstract}


\section{Introduction}
\label{S:intro}

Rotating convection plays an important dynamical role in stars and planets, where it is believed to be one of the primary drivers of global scale magnetic fields \citep{sS10,cJ11b,jmA15}, and possibly gives rise to coherent large-scale flows such as zonal jets and vortices, as observed on the giant planets \citep{mH22,lS22}. Understanding the physics of turbulence driven by rotating convection remains challenging due to the vast range of spatiotemporal scales. As a result, the parameter space accessible to direct numerical simulation (DNS) and laboratory experiments remains distant from that which characterises natural convective systems. An approach that is often taken to overcome this limitation is to identify asymptotic scaling behaviour in model output so that extrapolation to the conditions of natural systems is possible \citep[e.g.][]{uC02,jmA07}. The development of asymptotically reduced equation sets is another strategy that has been particularly helpful for improving understanding of rotating convective turbulence and dynamos in simplified planar geometries since more extreme parameter regimes can be accessed in comparison to DNS \citep{mS06,mC13, mC15b,mY22b}. In general, excellent agreement has been found between the results of plane layer asymptotic models and DNS when an overlap in parameter space is possible \citep{mP16,mY22b}. However, asymptotic models have not yet been developed for global spherical geometries. Towards this end, the present investigation utilises DNS in spherical shell geometries to determine the asymptotic scaling behaviour of various key quantities, including force balances, length scales, and convective and zonal flow speeds. 

We consider Boussinesq convection in a rotating spherical shell with angular velocity $\Omega$. The geometry is specified by the aspect ratio $\eta = r_i/r_o$, where $r_i$ is the radius of the inner sphere, and $r_o$ is the radius of the outer sphere. The fluid is forced via a temperature contrast $\Delta T$ between the inner and outer boundaries, and gravity varies linearly with radius. For this system, the convection dynamics are determined by the sizes of several non-dimensional parameters, including the Rayleigh number and Ekman number, defined by, respectively,
\be
Ra = \frac{g_o \alpha \Delta T H^3}{\kappa \nu}, \qquad Ek = \frac{\nu}{\Omega H^2},
\ee
where $g_o$ is the gravitational acceleration at the outer boundary, $\alpha$ is the coefficient of thermal expansion, $H = r_o - r_i$ is the depth of the fluid layer, $\kappa$ is the thermal diffusivity and $\nu$ is the kinematic viscosity. The most unstable state consists of convective Rossby waves that align with the rotation axis and drift in the prograde direction \citep{fB70}. Asymptotic theory, valid in the limit $Ek \rightarrow 0$, has shown that the critical Rayleigh number scales as $Ra_c = O(Ek^{-4/3})$, and the critical azimuthal (zonal) wavenumber scales as $m_c = O(Ek^{-1/3})$ \citep{pR68,fB70,cJ00,eD04}. Thus, large Rayleigh numbers are needed to drive rotating convection and the subsequent motions become increasingly smaller scale as the Ekman number is reduced.

As the Rayleigh number is increased beyond critical, strong zonal flows develop in rotating spherical convection provided stress-free mechanical boundary conditions are applied on the inner and outer spherical surfaces \citep[e.g.][]{jmA01b,uC02}. These zonal flows are characterised by alternating regions of prograde and retrograde motion that are approximately invariant in the direction of the rotation axis. For a fixed value of $Ek$, the number of jets is controlled both by the Rayleigh number and the shell aspect ratio, $\eta$. In full sphere and small aspect ratio geometries ($\eta \lesssim 0.6$), simulations typically find a single prograde jet in the equatorial region and retrograde jets at higher latitudes \citep{jmA01b,uC02,yL21}. As $\eta$ and $Ra$ are increased there is a tendency for multiple jets to form at higher latitudes, leading to a banded structure that is reminiscent of the flows observed on the gas giant planets \citep{uC01,mH05,cJ09b,tG14,mH22}. The number of zonal jets that appear can be related to the Rhines length scale \citep{pR75,mH05,tG14}. For sufficiently large Rayleigh numbers there is an eventual transition to a retrograde equatorial jet and prograde high latitude jets \citep{jmA07b,kS19}.

%
%

Steady zonal flows are driven by Reynolds stresses and damped by global scale viscous stresses. Thus, the scaling behaviour of the zonal flow is intrinsically linked to the scaling of the correlations of the convective flows \citep{uC02}. Such correlations are not known a priori, though various scaling theories have been presented in the literature. For perfectly correlated small-scale velocity components, as relevant near the onset of convection, the zonal flow amplitude exhibits a quadratic dependence on the small-scale velocity \citep{jA01}. However, \cite{uC02} found that correlations decrease with increasing supercriticality, which causes the quadratic scaling of the zonal flow to eventually break down. Later investigations have found that this loss of correlation in the small-scale velocity is a monotonically decreasing function of $Ra$ \citep{aS11,tG12}. For sufficiently large Rayleigh numbers, \cite{uC02} and \cite{yL21} find evidence that the zonal flow scaling approaches a `diffusion-free' regime, so-called because of the lack of dependence on $\nu$ and $\kappa$, though within this regime the dynamics are no longer geostrophic since inertia and the Coriolis force are of the same order of magnitude.

Aside from their correlations, it is also important to understand the scaling of the convective flow speeds themselves. One scaling theory that is often invoked to explain convective flow speed scaling behaviour in rotating convection is the so-called Coriolis-Inertia-Archimedean (CIA) balance \citep[e.g.][]{cJ15,jmA20}. This balance predicts that the dominant convective length scale behaves like $\ell \sim \Rat^{1/2} Ek^{1/3}$, and the global scale Reynolds number should scale like $Re = U H/\nu \sim Ek Ra/Pr$, where $U$ is a characteristic flow speed, the reduced Rayleigh number is defined as $\Rat = Ra Ek^{4/3}$ and the thermal Prandtl number is $Pr = \nu/\kappa$. \cite{eK13b} analyzed length scales and flow speeds in a broad suite of numerical dynamo simulations in spherical geometries and concluded that viscous effects remained important in controlling these quantities. CIA theory is often contrasted with the scaling of the linear instability scale, $\ell \sim Ek^{1/3}$. However, it is important to emphasise that, from the point of view of asymptotics, both the linear `viscous' length scale and the length scale predicted by CIA theory are of the size $O(Ek^{1/3})$ given that $\Rat$ is an order unity asymptotic parameter.

Several previous investigations have tested these CIA scaling predictions in spherical geometries with no-slip boundary conditions \citep{tG16,cG19,rL20}, laboratory experiments in rotating cylindrical geometries \citep{mM21}, as well as asymptotic models of plane layer convection with stress-free boundary conditions \citep{sM21,tO23}. \cite{tG16} carried out a comprehensive survey of rotating spherical convection and found that the length scale for their smallest Ekman number cases ($Ek=3\times 10^{-7}$) approached the Rhines scaling predicted by the CIA balance, and that the interior dissipation could also be approximated through a CIA balance. However, as we discuss in the present study, \cite{tG16} did not investigate the convection and zonal flow separately. A similar approach was taken by \citet{rL20} in which constant heat flux thermal boundary conditions were used. 
\citet{cG19} simulated rotating spherical convection with no-slip boundary conditions and $Pr=0.01$ using both a two-dimensional quasi-geostrophic model and three-dimensional DNS, and found that the length scale increases with Rayleigh number for all parameter ranges used; they find evidence of length scales and flow speeds approaching the CIA scaling predictions as the Ekman number is reduced. 

The rotating convection experiments of \cite{mM21} exhibited an increase of the integral length scale with Rayleigh number, but at a slower rate than that predicted by CIA theory. In asymptotic models, \cite{sM21} found flow speed scaling behaviour consistent with CIA theory over a finite range of Rayleigh numbers; the deviation from CIA theory at large Rayleigh numbers was attributed to the effects of the large scale vortex (LSV) that is generated in this system. \cite{tO23} also found that certain measures of the convective length scales show an increase with $\Rat$, but at a rate that is slower than the exponent of $1/2$. Importantly, however, the asymptotic models do not find a CIA force balance in the fluid interior. Instead, the buoyancy force is only comparable to the Coriolis and inertial forces within the thermal boundary layers, though viscous effects are equally important in these regions of the flow domain. Moreover, the ratio of the viscous force to the buoyancy force was found to be an increasing function of $\Rat$, indicating that the CIA balance is never achieved in plane layer convection \citep{sM21,tO23}. The conclusion from these asymptotic studies is that convective length scales remain viscously controlled. This same conclusion was reached by \cite{mY22b} who found similar behaviour in DNS of rapidly rotating convection driven dynamos in the plane layer geometry. An additional important finding in the studies of \cite{mM21,mY22b,tO23} is that the viscous dissipation length scale remains approximately constant with increasing $\Rat$ -- this indicates that length scale evolution in rotating convective turbulence is fundamentally different than non-rotating convective turbulence where the dissipation length decreases strongly with increasing Rayleigh number \citep[e.g.][]{mY21}. We observe similar behaviour for the length scales and force balances in the spherical simulations reported in the present investigation.

The scaling behaviour of key quantities such as flow speeds and length scales is linked to the force balances in the governing equations. To our knowledge, the asymptotic scaling behaviour of the force balances (and terms in the heat equation) in spherical convection simulations have not been reported to date, though several previous dynamo studies have computed these forces over a range of parameters. For an Ekman number of $Ek=10^{-4}$ in a spherical dynamo, \citet{kS12} noted that the Lorentz force was smaller than the inertial term and that the Lorentz force did not significantly alter the convective flow as compared to the purely hydrodynamic model. However, \citet{kS12} also noted that at smaller Ekman numbers, the Lorentz force seemed to make larger changes to the convection. In general, simulations find that the force balance in the mean equations is thermal wind to leading order, with the Lorentz force entering at higher order in the zonal component of the mean momentum equation \citep{jA05,mC21,rO21}. For the small scale convective dynamics, dynamo studies find that the zeroth order force balance is geostrophic, the first order force balance is between the ageostrophic Coriolis force, the buoyancy force and the Lorentz force, and inertial and viscous forces enter at the next order \citep{rY16} (note, however, these authors did not separate the mean and fluctuating dynamics). Here, we find a similar sequence of balances in the fluctuating momentum equation, though there does not appear to be an asymptotic difference between the buoyancy force, the viscous force and the inertial force, which is similar to plane layer rotating convection. Also in agreement with previous asymptotic studies, we find that the ratio of the viscous force to the buoyancy force is an \textit{increasing} function of $\Rat$. For the large-scale dynamics, we find that the flows are geostrophically balanced to leading order.

In this paper, we investigate the asymptotic behaviour of rapidly rotating convection and the associated zonal flows in a spherical shell with stress-free boundary conditions. Knowledge of such scaling behaviour is crucial for the development of asymptotic models. Overall, we find excellent agreement between the asymptotic predictions and the results of the nonlinear simulations. We develop a prediction for the asymptotic scaling of the amplitude of the zonal flow and conclude that the zonal flow must remain dependent on viscosity given that viscous stresses are the sole saturation mechanism for this component of the flow. The paper is organized as follows. In section \ref{S:model}, we describe the model and governing equations. We give a brief overview of the asymptotic theory in section \ref{S:theory} and numerical results are analysed in section \ref{S:results}. A discussion is provided in section \ref{S:discuss}.

\section{Model}
\label{S:model}
The governing equations consist of the conservation laws for momentum, thermal energy and mass. We non-dimensionalise these equations using the length $H$, the large-scale viscous diffusion time $H^2/\nu$ and the temperature scale $\Delta T$. With this non-dimensionalisation, the governing equations are given by
\be
\dst \ub + \ub \cdot \nabla \ub = -\frac{2}{Ek} \hz \times \ub - \frac{1}{Ek}\nabla P + \frac{Ra}{Pr} \lb \frac{r}{r_o} \rb T \, \widehat{\mathbf{r}}+ \nabla^2 \ub,
\ee
\be
\dst T + \ub \cdot \nabla T = \frac{1}{Pr} \nabla^2 T,
\ee
\be
\nabla \cdot \ub = 0,
\ee
where $\ub=\langle u_r, u_\theta, u_\phi \rangle$ is the fluid velocity, $T$ is the temperature, $P$ is the pressure and $r$ is radius. The `axial' direction points in the direction of the rotation axis and the axial and radial unit vectors are denoted by $\hz$ and $\hr$, respectively. In all of the simulations presented here we fix $Pr = 1$.

The boundary conditions are impenetrable ($u_r=0$), stress-free and fixed temperature.
We use the code Rayleigh to numerically solve the governing equations \citep{featherstone_et_al_2022}. Rayleigh is a pseudo-spectral code which uses spherical harmonics to represent data on spherical shells and Chebyshev polynomials to represent data in the radial direction. A $2/3$ de-aliasing is used for both the spherical harmonics and the Chebyshev polynomials. Time-stepping is carried out using a second-order semi-implicit Crank-Nicolson method for the linear terms and a second-order Adams-Bashforth method for the nonlinear terms. The time step is chosen adaptively to maintain numerical stability.

\subsection{Notation and outputs}

Due to the symmetry of the model setup around the rotation axis, it is convenient to define mean and fluctuating components of some scalar quantity $X$ relative to an azimuthal or zonal average, i.e.
\be
\overline{X} = \int_0^{2\pi}  X d\phi,
\ee
with $X'=X-\overline{X}$. 

We define the Reynolds number as 
\be
Re = \tavg{\sqrt{\langle \ub \cdot \ub \rangle}},
\ee
where $\langle \cdot \rangle$ denotes a volume average and $\tavg{(\cdot)}$ denotes a time average. We further define the mean and fluctuating (convective) Reynolds numbers as, respectively,
\be
Re_z =  \tavg{\sqrt{\langle \mub \cdot \mub \rangle}}, \qquad  Re_c =  \tavg{\sqrt{\langle \ub' \cdot \ub' \rangle}}.
\ee
In all of the simulations, we find that the zonal component of the mean flow dominates and we therefore refer to the mean Reynolds number as the `zonal' Reynolds number.

We analyse several different length scales in this paper. \cite{uC06} \citep[see also][]{tG16,rL20} define the spherical harmonic length scale as
\be
\left(\ell_{sh}\right)^{\, -1} = \tavg{\left(\dfrac{\sum_{l=0}^{l_{max}} \sum_{m=0}^{l} l\mathcal{E}_l^m}{\pi \sum_{l=0}^{l_{max}} \sum_{m=0}^{l} \mathcal{E}_l^m}\right)},
\label{E:length}
\ee
where $l_{max}$ is the maximum spherical harmonic degree in the simulation and $\mathcal{E}_l^m $ is the radially averaged kinetic energy density of spherical harmonic degree $l$ and order $m$. Thus, the volume averaged kinetic energy density is given by
\be
\frac{1}{2} \langle \ub \cdot \ub \rangle = \sum_{l=0}^{l_{max}} \sum_{m=0}^{l} \mathcal{E}_l^m.
\ee
Due to the strong influence of the zonal flow in our simulations, we separate the spherical harmonic length scale into a zonal and a non-zonal component. We define these length scales as
\be
 \quad \left(\ell_{sh}'\right)^{\, -1} = \tavg{\left(\dfrac{\sum_{l=1}^{l_{max}} \sum_{m=1}^{l} l\mathcal{E}_l^m}{\pi \sum_{l=1}^{l_{max}} \sum_{m=1}^{l} \mathcal{E}_l^m}\right)}, \\   \left(\overline{\ell}_{sh}\right)^{\, -1} = \tavg{\left(\dfrac{\sum_{l=0}^{l_{max}}l\mathcal{E}_l^{m=0}}{\pi \sum_{l=0}^{l_{max}} \mathcal{E}_l^{m=0}}\right)}.
 \label{E:length_fluct_mean}
\ee
We define the fluctuating and mean Taylor microscales as
\be
\qquad \ell'_{tm} = \tavg{\sqrt{\dfrac{\langle \ub' \cdot \ub' \rangle }{\langle (\nabla \times \ub') \cdot (\nabla \times \ub') \rangle}}}, \qquad \overline{\ell}_{tm} = \tavg{\sqrt{\dfrac{\langle \mub \cdot \mub \rangle }{\langle (\nabla \times \mub) \cdot (\nabla \times \mub) \rangle}}},
\label{E:tm}
\ee
respectively. The Taylor microscale can be considered a viscous dissipation length scale since it characterises the length scale at which viscous effects become important.

The Nusselt number is calculated according to 
\be
Nu = \frac{ \partial_r \shellavg{T} \bigr\rvert_{r=r_o}}{\partial_r T_c\bigr\rvert_{r=r_o}},
\ee
where $\shellavg{\left(\cdot\right)}$ is a shell and time average, and $T_c$ is the conductive temperature profile, which satisfies
\be
\nabla^2 T_c = 0, \quad T_c(r_i) = 1, \quad T_c(r_o) = 0.
\ee
We also define the viscous dissipation rates of the mean and fluctuating velocity fields according to
\be
\overline{\varepsilon} = \tavg{\langle \left(\nabla \times \overline{\mathbf{u}} \right)^2 \rangle} \qquad \textnormal{and} \qquad \varepsilon' = \tavg{\langle \left( \nabla \times \mathbf{u}' \right)^2 \rangle},
\ee
respectively.

\section{Theory}
\label{S:theory}
Here we provide arguments for the scaling behaviour of various quantities. We use the term asymptotic to mean rotationally constrained motions in which the Ekman number and Rossby number, $Ro = U/ \lb 2 \Omega H \rb$, are both small relative to unity, i.e.~$(Ro, Ek) \ll 1$. The dimensional flow speed $U$ characterises the magnitude of fluctuating velocity field. We expect many aspects of the asymptotic theory for rotating convection in spherical geometries presented in \cite{cJ00} and \cite{eD04} to hold here, though some of these scalings must be modified to account for nonlinear terms in the governing equations. In particular, the scaling of the convective flow speeds and temperature perturbation need to be reduced by a factor of $Ek^{1/3}$ relative to \cite{cJ00} and \cite{eD04}, though this difference does not influence the leading order force balance in the fluctuating momentum equation when zonal flows are weak. The scaling of the large-scale zonal flow depends on the small-scale convective velocity, so we first consider theoretical scaling laws for the convective velocity and the corresponding convective length scale. Such scaling laws can be found by examining the fluctuating momentum equation and the fluctuating heat equation, which are given by, respectively,
\begin{align}
\begin{split}
\left[\dst \ubp + \mub \cdot \nabla \ubp\right] &+ \ubp \cdot \nabla \mub  + \ubp \cdot \nabla \ubp - \overline{\ubp \cdot \nabla \ubp} = \\ &-\frac{2}{Ek} \hz \times \ubp - \frac{1}{Ek}\nabla P' + \frac{Ra}{Pr} \lb \frac{r}{r_o} \rb T' \, \widehat{\mathbf{r}}+ \nabla^2 \ubp,
\end{split}
\end{align}
\be
\left[\dst T' + \mub \cdot \nabla T'\right] + \ubp \cdot \nabla \overline{T} + \ubp \cdot \nabla T' - \overline{ \ubp \cdot \nabla T' } = \frac{1}{Pr} \nabla^2 T'.
\ee
We find that the terms in brackets can be large compared to some terms due to the large amplitude of the zonal flow. However, summing the terms in brackets leads to results smaller than the individual terms in the brackets, which physically means that the Lagrangian time derivative is smaller than the Eulerian time derivative. We will therefore consider the sum of the terms in brackets rather than each individually. It is well known from linear theory that the Coriolis force and pressure gradient force are dominant terms in the limit $\lb Ro, Ek \rb \rightarrow 0$. We will see below that the zonal flow can also modify the leading order force balance. In order to study the first-order effects, we eliminate the pressure gradient by taking the curl of the fluctuating momentum equation. This operation yields
\begin{align}
\begin{split}
\nabla \times\left[\dst \ubp + \mub \cdot \nabla \ubp\right] &+\nabla \times (\ubp \cdot \nabla \mub) + \nabla\times (\ubp \cdot \nabla \ubp - \overline{\ubp \cdot \nabla \ubp}) = \\ &-\frac{2}{Ek} \frac{\partial \ubp}{\partial z} +\frac{Ra}{Pr} \nabla \times \lb \frac{r}{r_o} T' \, \widehat{\mathbf{r}} \rb + \nabla^2 \lb \nabla \times \ubp \rb.
\end{split}
\end{align}
Assuming length scales of order one in the $z$-direction, length scales of order one for azimuthally averaged terms, and length scales of order $\ell$ otherwise, the vorticity equation and heat equation can be approximately written as
\be
\nabla \times\left[\dst \ubp + \mub \cdot \nabla \ubp\right] + \frac{\mub \ubp}{\ell} + \frac{\left(\ubp\right)^2}{\ell^2} - \nabla\times \overline{\ubp \cdot \nabla \ubp} \sim -\frac{\ubp}{Ek} + \frac{Ra}{Pr}\frac{T'}{\ell} +\frac{\ubp}{\ell^3},
\label{E:fluct_momentum_approx}
\ee
\be
\left[\dst T' + \mub \cdot \nabla T'\right] + \ubp \overline{T} + \frac{\ubp T'}{\ell} - \overline{ \ubp \cdot \nabla T' } \sim \frac{1}{Pr}\frac{T'}{\ell^2}.
\label{E:fluct_thermal_eqn_approx}
\ee
where factors of order one have been dropped. We now follow \citet{jmA20} and assume a balance between $\ubp \overline{T}$ and $\ubp T'\ell^{-1}$ in the temperature equation. Using $\overline{T}=O(1)$ yields $T' \sim \ell$. Plugging this relation in for $T'$ in the momentum equation and assuming a CIA balance where the fluctuating-fluctuating advection term is used yields
\be
\frac{\left(\ubp\right)^2}{\ell^2} \sim -\frac{\ubp}{Ek} \sim \frac{Ra}{Pr},
\ee
which can be solved for the convective velocity and length scale to give
\be
\ubp \sim \frac{Ra Ek}{Pr}, \qquad \ell \sim \sqrt{\frac{Ra Ek^2}{Pr}}.
\ee
We can rewrite these expressions in terms of the reduced Rayleigh number as
\be
\ubp \sim Ek^{-1/3}\frac{\Rat}{Pr}, \qquad \ell \sim Ek^{1/3} \sqrt{\frac{\Rat}{Pr}}.
\label{E:fluct_scaling}
\ee
Here again we note that both the CIA and viscous length scale have the same $Ek^{1/3}$ dependence.
One of the points that we stress in the present study is that all length scales in this system are viscously selected to leading order, with order one variations away from this viscous scale since $\Rat = O(1)$  \citep[e.g.][]{mY22b,tO23}. 

We can understand why these different balances produce the same Ekman dependence by making a different set of assumptions than the assumptions used for a CIA balance. The first assumption we make is that the Ekman dependence for any term can be written in terms of a power law. We will assume that the Ekman dependence of the convective flow speeds, the length scale, and the fluctuating temperature can be written as $\ubp = O\left(Ek^{x_u}\right)$, $\ell = O\left(Ek^{x_\ell}\right)$, $T' = O\left(Ek^{x_T}\right)$. The second assumption is that ratios of certain terms in the momentum and heat equations do not change when the Ekman number is reduced. We will assume that the advection of fluctuating velocity by fluctuating velocity, viscosity, and buoyancy follow the same Ekman number scaling in the fluctuating momentum equation. In the heat equation we will assume that conduction and advection of the mean temperature by the fluctuating velocity follow the same Ekman number scaling. These assumptions produce the system of equations given by
\be
2x_u-x_\ell = x_u-2x_\ell = -4/3 + x_T, \qquad x_T-2x_\ell = x_u.
\ee
Note that we have assumed the mean temperature and the length scale of the mean temperature do not depend on the Ekman number. Solving this system of equations yields $x_u=-1/3$, $x_\ell = 1/3$, and $x_T = 1/3$, which implies that $\ubp = O\left(Ek^{-1/3}\right)$, $\ell = O\left(Ek^{1/3}\right)$, and $T' = O\left(Ek^{1/3}\right)$. Note that the Ekman number dependence derived here is the same as the Ekman number dependence derived using the CIA balance written in terms of the reduced Rayleigh number. Therefore, we can get the $Ek^{1/3}$ scaling for the length scale by assuming that various terms follow the same Ekman number scaling without actually assuming any balances a priori. These scalings are equivalent to the scalings used to derive the asymptotic model of rotating convection in a Cartesian geometry (e.g. \citep{mS06}). 

An asymptotic constraint on the amplitude of the zonal flow can now be obtained upon examination of the mean momentum equation. The mean momentum equation is given by
\be
\dst \mub + \mub \cdot \nabla \mub + \overline{ \ubp \cdot \nabla \ubp } + \frac{2}{Ek} \, \hz \times \mub = - \nabla \overline{P} + \frac{Ra}{Pr} \lb \frac{r}{r_o} \rb   \mT \, \widehat{\mathbf{r}} + \nabla^2 \mub .
\ee
The zonal component is then 
\be
\dst \muph+ \lsq \mub \cdot \nabla \mub \rsq_{\phi} + \lsq \overline{ \ubp \cdot \nabla \ubp } \rsq_{\phi} + \frac{2}{Ek} \, \lsq \hz \times \mub \rsq_{\phi} =  \lsq \nabla^2 \mub \rsq_{\phi} ,
\label{E:NS_mean_phi}
\ee
where the square brackets and corresponding subscript are used for brevity. The above equation allows for a straightforward interpretation of the zonal flow dynamics. Time dependence and advection of the zonal flow by mean meridional flows are captured by the first two terms on the left-hand side. The term 
$\lsq \overline{ \ubp \cdot \nabla \ubp } \rsq_{\phi}$ includes the divergence of the Reynolds stresses and acts as the primary source of the zonal flow. The zonal component of the mean Coriolis force can be written as $\lsq \hz \times \mub \rsq_{\phi} = \cos \theta \, \muth + \sin \theta \, \mur$ such that only meridional circulation appears in this term. Averaging equation \eqref{E:NS_mean_phi} along the z-direction eliminates the Coriolis term due to continuity, and averaging equation \eqref{E:NS_mean_phi} in time eliminates the time derivative term. Therefore, averaging equation \eqref{E:NS_mean_phi} along the z-direction and in time leaves only the two advective terms and the viscous term. For a careful derivation of the balance between the averaged advection and viscosity in spherical coordinates, see \cite{wD17}. Other studies have found that the mean-mean advection term is small \citep[e.g.][]{wD17}, so there must be a balance between the fluctuating advection term and the viscous term. Letting $\ztavg{\left(\cdot\right)}$ denote an average over $\phi$, $z$, and time, the preceding argument implies that
\be
\lsq \ztavg{ \ubp \cdot \nabla \ubp } \rsq_{\phi}  \approx  \lsq \nabla^2 \ztavg{\mathbf{u}} \rsq_{\phi} .
\label{E:bal}
\ee
The above balance is expected to hold for all values of $Ek$ and $Ra$; the zonal flow is therefore intrinsically dependent on viscosity and we should not expect its scaling behaviour to be `diffusion-free'. We note that a similar balance holds for mean flows in planar geometries \citep[e.g.][]{jN22}. Finally, since averaged quantities vary on order one length scales, this balance suggests
\be
\mub_\phi = O \lb C_R u' u' \rb ,
\label{E:zonal_scaling}
\ee
where $C_R$ represents the correlation of the fluctuating velocity components. We might expect that this correlation gets weaker as the reduced Rayleigh number is increased and the flow becomes less constrained by rotation. \citet{uC02} confirmed this weakening of the correlation as the Rayleigh number is increased by directly calculating the correlation and by noting that the zonal flow amplitude only scales as the square of the small-scale velocity for Rayleigh numbers near the onset of convection. However, the correlation cannot be greater than one and we would not expect the correlation to get closer to zero as $Ek$ is decreased with $\Rat$ fixed. We might therefore anticipate that $C_R$ does not depend on the Ekman number for small values of the Ekman number, in which case we can use equation \eqref{E:zonal_scaling} to find a scaling relationship for the zonal flow with Ekman number. The asymptotic analysis predicts that the convective velocity scales as $\mathbf{u}' = O\left(Ek^{-1/3}\right)$, which implies that
\be
\mub_\phi = O \lb Ek^{-2/3} \rb,
\label{E:scale}
\ee
thus indicating that the zonal flow is intrinsically dependent on viscosity, albeit in an asymptotic sense. 
For time-dependence in the zonal flow, we require that the time derivative is comparable to the viscous force so that
\be
\dst \mub_\phi = O \lb Ek^{-2/3} \rb, 
\ee
which, along with equation \eqref{E:scale}, indicates that the zonal flow time scale is $O(1)$ in our non-dimensional, large-scale viscous diffusion units. Thus, the zonal flow varies on a large-scale viscous diffusion time. This property is one of the reasons that computations of rotating spherical convection with stress free boundary conditions are so demanding -- very long integration is necessary to saturate the amplitude of the zonal flow and reach a statistically stationary state. 

The asymptotic constraint on the zonal flow amplitude allows for additional insight into the force balance by which it is constrained. In the limit $Ek \rightarrow 0$, the Rayleigh number must scale as $Ra = O \lb Ek^{-4/3} \rb$ to generate convection \citep{pR68}. Along with the fact that the magnitude of the mean temperature is independent of the Ekman number, this indicates that the mean buoyancy force scales as $O \lb Ek^{-4/3} \rb$. The radial and co-latitudinal components of the mean Coriolis force both contain $\muph$, thus indicating that these components scale as $O \lb Ek^{-5/3} \rb$, which is larger than the mean buoyancy force by a factor of $O \lb Ek^{-1/3} \rb$. Thus, the zonal flow is geostrophically balanced to leading order, i.e.
\be
-2 \sin \theta \, \muph \approx  - \partial_{r} \overline{P}, \qquad  
- 2  \cos \theta \, \muph \approx  - r^{-1} \partial_{\theta} \overline{P} .
\ee

It is informative to compare the scaling behaviour of zonal flows that are geostrophically balanced with a zonal flow that is in thermal wind balance \citep[e.g.][]{mC21}.  
An order of magnitude estimate for the scaling of the thermal wind component of the zonal flow can be obtained if we balance the mean Coriolis force with the mean buoyancy force,
\be
- \frac{2 \sin \theta }{Ek} \, \muph^{tw} \approx  \frac{Ra}{Pr} \lb \frac{r}{r_o} \rb   \mT \qquad \Rightarrow \qquad \muph^{tw} \sim \frac{Ra Ek}{Pr} .
\ee
Using the definition of the reduced Rayleigh number this becomes
\be
\muph^{tw} \sim \frac{\Rat}{Pr} Ek^{-1/3}.
\ee
And since $\Rat = O(1)$ this implies $\muph^{tw} = O(Ek^{-1/3})$, which is of the same order as the convective flow speeds. Thus, zonal flows that satisfy a thermal wind balance are substantially weaker than those that are geostrophic. Interestingly, the above scaling also provides an estimate for the scaling of the thermal wind with Rayleigh number and represents a `diffusion-free' scaling in the sense that it indicates that the thermal wind does not depend on either $\nu$ or $\kappa$. The thermal wind scaling seems to be loosely consistent with the zonal flows present in the dynamos of \citep[][]{mC21}; when a magnetic field is present the Lorentz force strongly damps the geostrophic component of the zonal flow. 




\section{Numerical Results}
\label{S:results}

\subsection{Overview}

To test the theoretical arguments given in the previous section, we perform a suite of direct numerical simulations of convection across a range of Ekman number and Rayleigh number. We consider the aspect ratios $\eta=0.35$ and $\eta=0.7$. 
 The smallest Ekman numbers that were simulated are $Ek=10^{-6}$ and $Ek=3\times 10^{-5}$ for the $\eta=0.35$ and $\eta=0.7$ aspect ratios, respectively.
Details of the simulations are contained in Tables \ref{T:sims_thick_v0}-\ref{T:sims_thin_v0} in the appendix.

\begin{figure}
 \begin{center}
 \subfloat[][]{\includegraphics[width=0.33\textwidth]{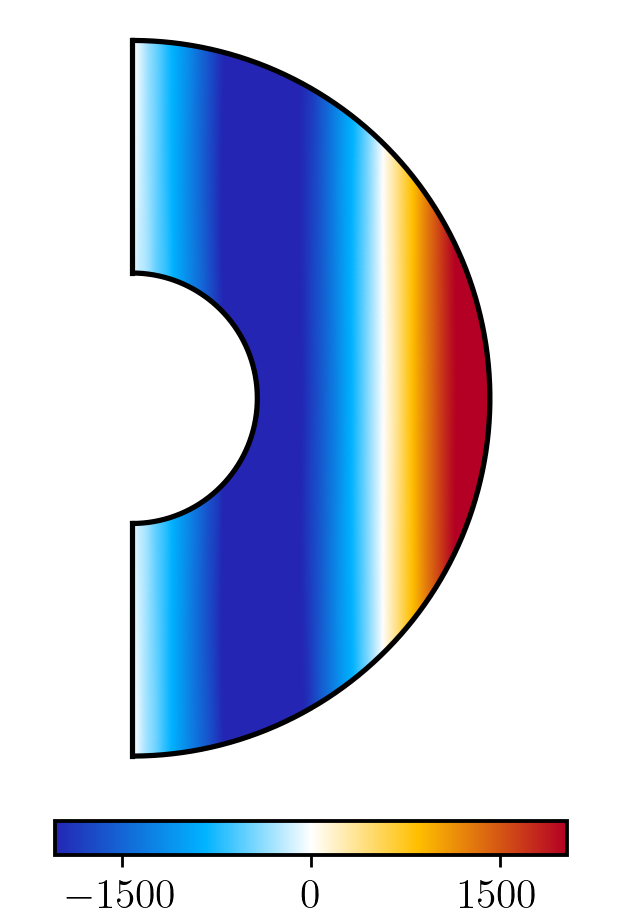}}
\subfloat[][]{\includegraphics[width=0.33\textwidth]{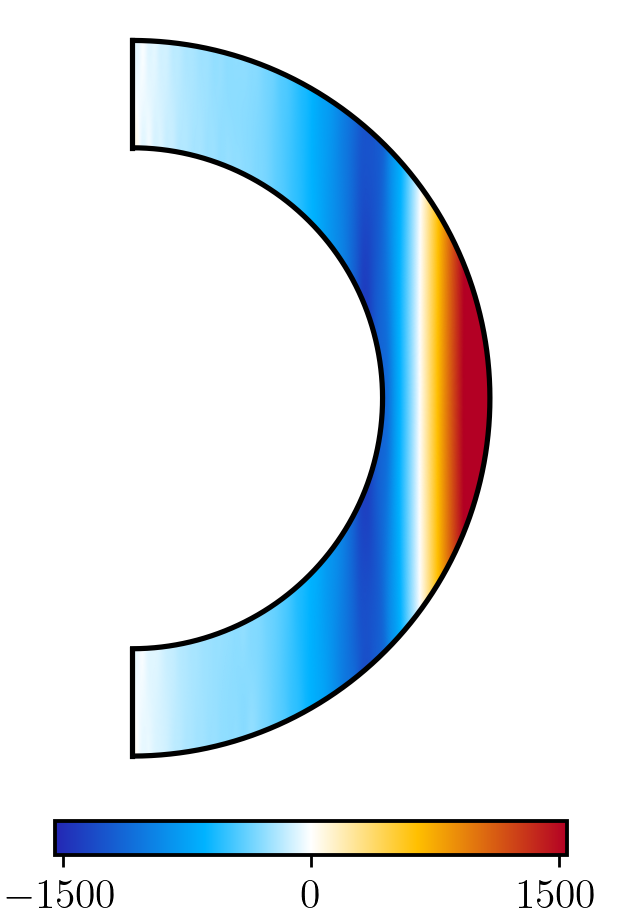}}
\subfloat[][]{\includegraphics[width=0.33\textwidth]{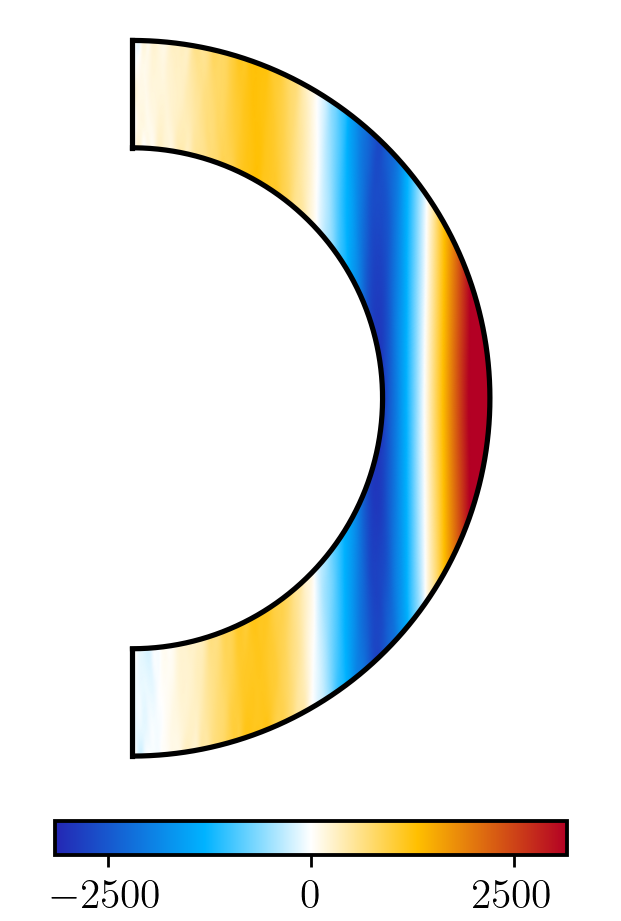}}
\caption{Instantaneous visualisations of the zonal flow where red indicates prograde motion and blue indicates retrograde motion: (a) $\eta=0.35$, $Ek=10^{-5}$, $Ra=2\times 10^8$ ($\Rat \approx 43$);  (b) $\eta=0.7$, $Ek=3\times 10^{-5}$, $Ra=3.2\times 10^7$ ($\Rat \approx 30$); (c) $\eta=0.7$, $Ek=3\times 10^{-5}$, $Ra=6.4\times 10^{7}$ ($\Rat \approx 60$).}
\label{F:azavg_zonal_flow}
\end{center}
\end{figure}

The qualitative nature of the zonal flows was similar for much of the parameter space covered, and representative cases for both shell thicknesses are shown in figure \ref{F:azavg_zonal_flow}. The zonal flows we observe in this study are similar to zonal flows observed in previous works. These zonal flows are characterised by a nearly invariant structure in the axial ($z$) direction, and consist of a single prograde jet at the equator with retrograde jets at higher latitudes. A small number of the thin shell ($\eta=0.7$) cases at low Ekman number and high Rayleigh number developed high latitude jets, as shown in figure \ref{F:azavg_zonal_flow}(c). A subset of our cases exhibit relaxation oscillations in which the convection mainly occurs during short bursts; this behaviour was also observed in previous work \citep[e.g.][]{uC02}. 
Movie 1 from the supplementary material shows the radial velocity in the equatorial plane over the course of one relaxation oscillation with the convective Reynolds number shown for reference. From this movie, we see that during times of weak convection, the convection is strongest near the inner boundary in the equatorial plane. However, during times of strong convection, the convection fills the whole region in the equatorial plane.

We find that using $\Rat$, as opposed to the supercriticality measure, $Ra/Ra_c$, results in improved collapse of our data when comparing with asymptotic predictions; \cite{uC02} also found that using $\Rat$ improved the collapse of some data.
This effect likely arises from the slow rate of convergence of the critical Rayleigh number to the predicted asymptotic scaling of $Ra_c \sim Ek^{-4/3}$ \citep[e.g.][]{eD04,aB23}. 



%

\subsection{Flow speeds}

\begin{figure}
 \begin{center}
 \subfloat[][]{\includegraphics[width=0.48\textwidth]{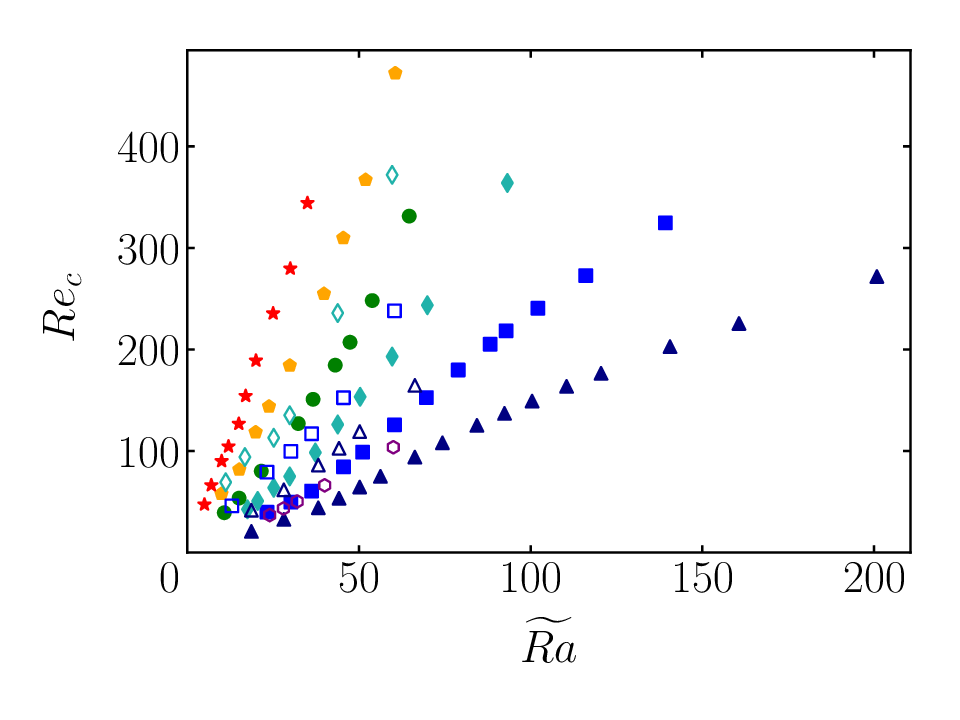}}
\subfloat[][]{\includegraphics[width=0.48\textwidth]{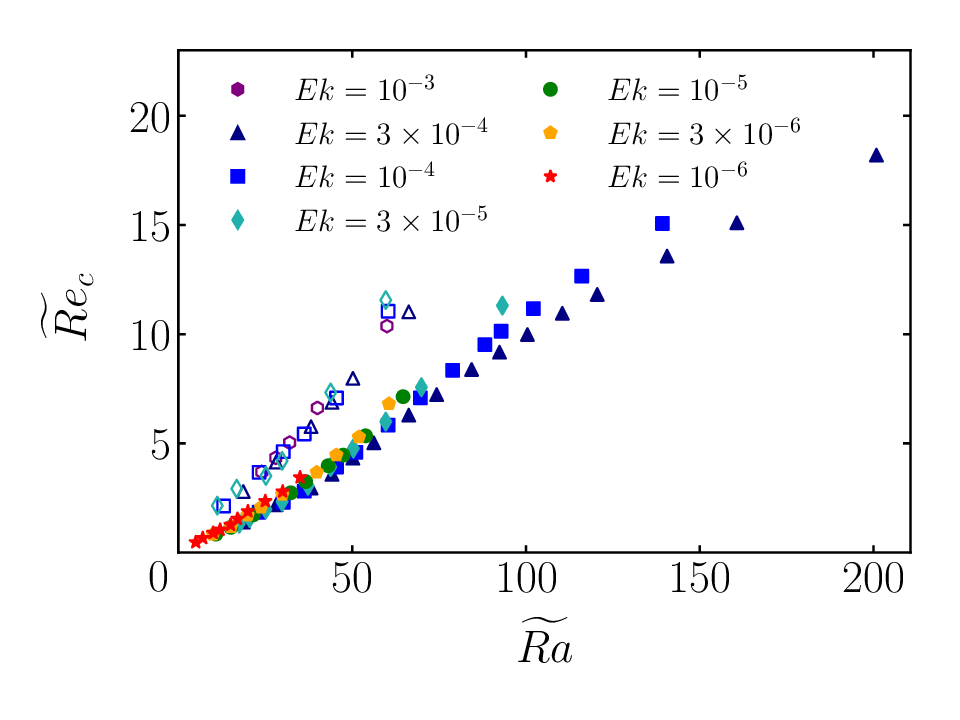}}
\hspace{0mm}
\subfloat[][]{\includegraphics[width=0.48\textwidth]{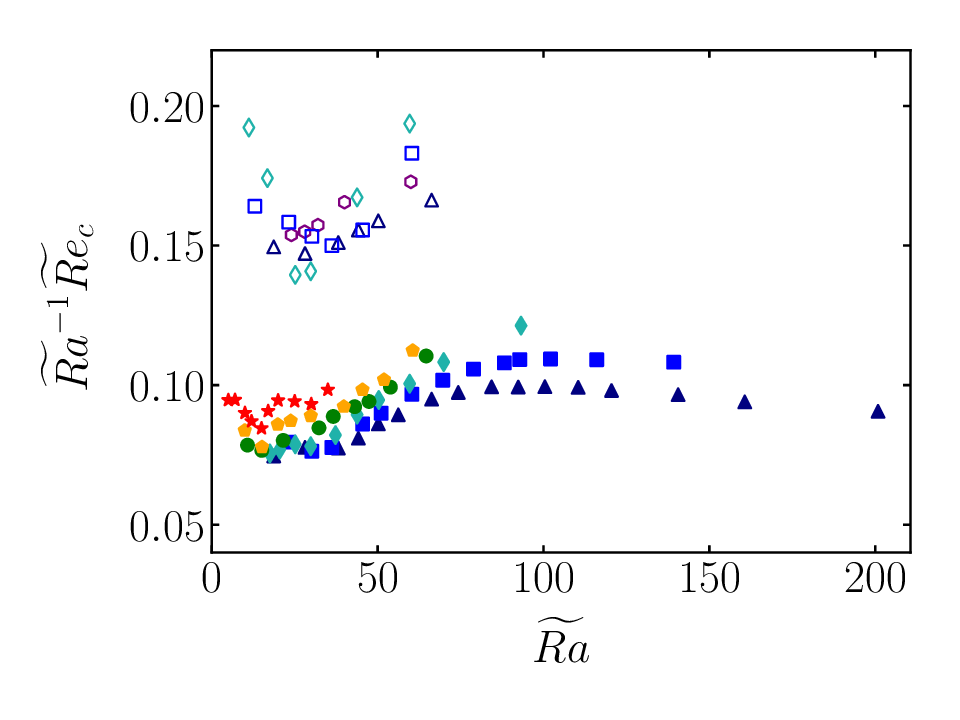}}
\subfloat[][]{\includegraphics[width=0.48\textwidth]{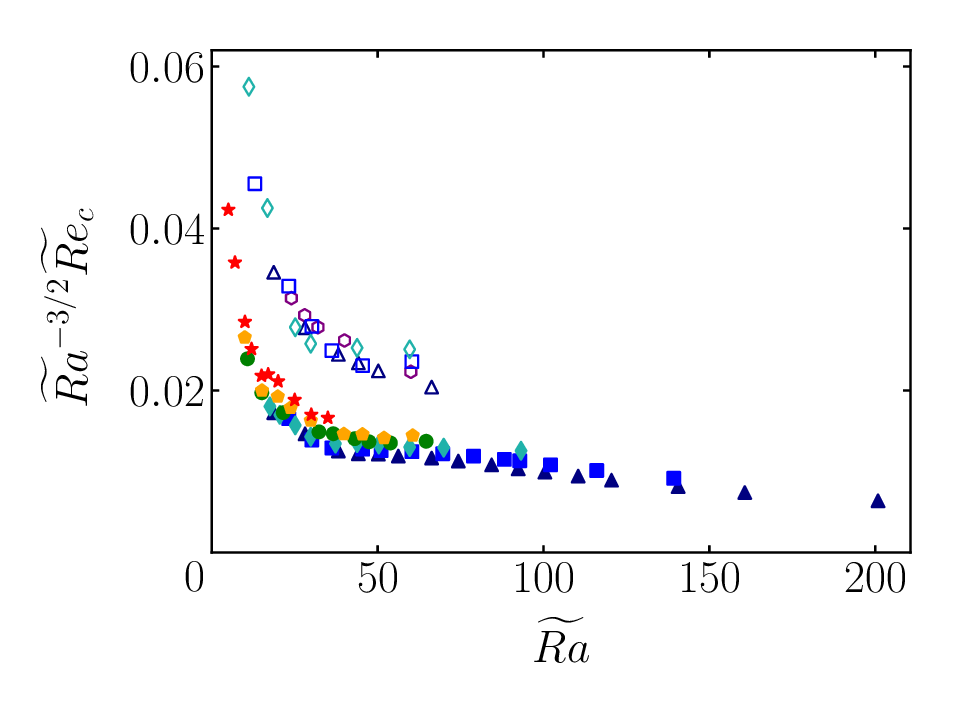}}
\caption{Reynolds number characterising the flow speeds of the fluctuating (convective) velocity versus the reduced Rayleigh number, $\Rat$: (a) the convective Reynolds number $Re_c$; (b) the rescaled convective Reynolds number $\Ret_c = Ek^{1/3} Re_c$; (c) the compensated convective Reynolds number $\Ret_c \Rat{\vphantom{Ra}}^{-1}$; (d) the compensated convective Reynolds number $\Ret_c \Rat{\vphantom{Ra}}^{-3/2}$. The filled symbols represent $\eta=0.35$ cases and the hollow symbols represent $\eta=0.7$ cases. }
\label{F:Re_fluct}
\end{center}
\end{figure}

Global rms values of both the fluctuating and mean velocity are computed to determine their scaling behaviour with respect to the Ekman number. Note that the mean velocity is dominated by the zonal flow (i.e.~the $\phi$-component of the mean velocity). Figure \ref{F:Re_fluct}(a) shows the fluctuating Reynolds number for both the $\eta=0.35$ cases and the $\eta=0.7$ cases. Figure \ref{F:Re_fluct}(b) shows the asymptotically rescaled fluctuating Reynolds number, i.e.~$\widetilde{Re}_c = Ek^{1/3} Re_c$, for the two different aspect ratios. We find that the rescaled data is order unity and collapses onto a single curve, which supports the $\ubp = O\left(Ek^{-1/3}\right)$ asymptotic scaling for the fluctuating velocity. However, we note that the Ekman number scaling of the convective Reynolds number might be time dependent. If we calculated the convective Reynolds number using data only during the convective peaks of the relaxation oscillations, we would obtain a steeper scaling closer to $Ek^{-1/2}$. This suggests that the time series for the convective velocity becomes more strongly peaked at lower Ekman number. Note that we expect deviation from this asymptotic scaling behaviour as the system loses rotational constraint; this deviation is particularly noticeable in figure \ref{F:Re_fluct}(b) for the high Rayleigh number regime ($\Rat \gtrsim 100$) for the two largest Ekman numbers used in the $\eta=0.35$ simulations, $Ek=3\times10^{-4}$ and $Ek=10^{-4}$. Figure \ref{F:Re_fluct}(c,d) shows two versions of the compensated convective Reynolds number: $\Ret_c \Rat{\vphantom{Ra}}^{-1}$ and $\Ret_c \Rat{\vphantom{Ra}}^{-3/2}$. We observe in figure \ref{F:Re_fluct}(c) that the compensated Reynolds number $\Ret_c \Rat{\vphantom{Ra}}^{-1}$ becomes nearly horizontal for our large Rayleigh number cases at large Ekman number and small aspect ratio, which suggests these cases may be scaling as $Re_c \sim \Rat$.  However, this scaling behaviour may be localised in $\Rat$ space. For sufficiently small Ekman number and large Rayleigh number, the compensated plot for $\Ret_c \Rat{\vphantom{Ra}}^{-3/2}$ collapses the data well, though the scaling appears slightly weaker than $\Rat^{3/2}$ which suggests that the convective Reynolds number scales approximately as $Re_c \sim \Rat{\vphantom{Ra}}^{3/2}$ in this regime.

\begin{figure}
 \begin{center}
 \subfloat[][]{\includegraphics[width=0.48\textwidth]{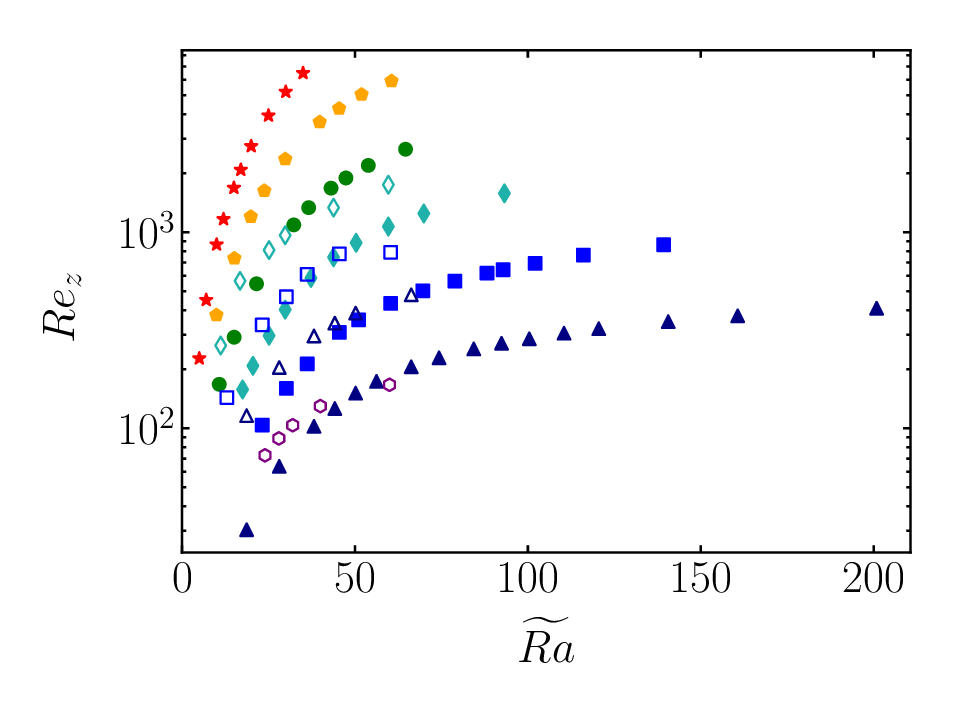}}
\subfloat[][]{\includegraphics[width=0.48\textwidth]{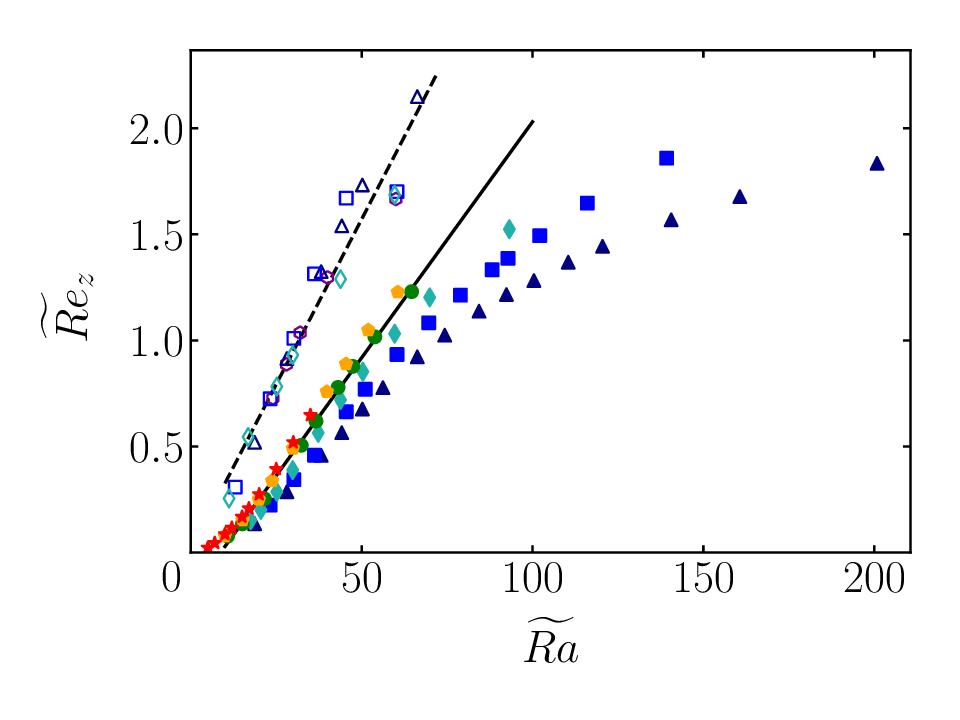}}
\caption{Reynolds number characterising the flow speeds of the mean (zonal) velocity field versus $\Rat$: (a) the zonal Reynolds number $Re_z$; (b) the rescaled zonal Reynolds number $\widetilde{Re}_z=Ek^{2/3} Re_z$. The solid line is a least squares fit of the data for the $\eta=0.35$, $Ek=10^{-5}$ cases to a line and is given by $Re_z =(0.022\Rat-0.19)Ek^{-2/3}$; the dashed line is a least squares fit of all the $\eta=0.7$ data to a line and is given by $Re_z=(0.031\Rat-0.016)Ek^{-2/3}$. The symbols are the same as defined in figure \ref{F:Re_fluct}. }
\label{F:Re_mean}
\end{center}
\end{figure}

Figure \ref{F:Re_mean}(a) shows the mean Reynolds number as a function of $\Rat$, and figure \ref{F:Re_mean}(b) shows the corresponding asymptotically rescaled mean Reynolds number, $\widetilde{Re}_z = Ek^{2/3} Re_z$. As mentioned previously, the mean flow is dominated by the zonal component in all of our simulations. While there is clearly some spread in the rescaled data for the thick shell cases, there is an indication that the data collapses to an asymptotic state as $Ek \rightarrow 0$. Moreover, the rescaled values are order unity. There appears to be better collapse for the thin shell cases, indicating that the fluid depth may play an important role. Also note the excellent collapse for the three thin shell data points near $\Rat \approx 60$ -- the two lower Ekman number cases of these three develop prograde high latitude jets as shown in figure 1(c). Taken together, the data seems to support that the zonal flow scales as $Ek^{-2/3}$ in the rapidly rotating regime. We note that \cite{tG14} found an empirical scaling for the zonal Rossby number of $Ro_{zon} \sim Ra^{0.6} Ek^{0.99}$, which, when converted to a Reynolds number and using $Ra \sim Ek^{-4/3}$ gives $Re_z \sim Ek^{-0.81}$. This result broadly agrees with the scaling of $Re_z \sim Ek^{-2/3}$ derived in this paper.

While the balance between viscosity and Reynolds stresses predicts a zonal flow scaling of $Ek^{-2/3}$, this balance is unable to predict how the zonal flow scales with the reduced Rayleigh number due to the fact that the correlation of the fluctuating velocity components is an unknown function of $\Rat$. Thus, we make an empirical fit of $Re_z$ with respect to the reduced Rayleigh number, which is shown in figure \ref{F:Re_mean}(b). We find that a line does a good job of fitting our small Ekman number cases. However, our larger Ekman cases at $\eta=0.35$ are run to larger values of the reduced Rayleigh number and are not well fit by a line at larger values of $\Rat$. One possibility for this effect is that these larger Ekman number cases at large Rayleigh number are no longer in the rapidly rotating regime and therefore follow a different trend. Another possibility is that the affine dependence we have used to fit the data only holds near the onset of convection and that the zonal Reynolds number follows a different trend for large enough values of the reduced Rayleigh number, even as the Ekman number is decreased. 

\begin{figure}
 \begin{center}
 \includegraphics[width=0.48\textwidth]{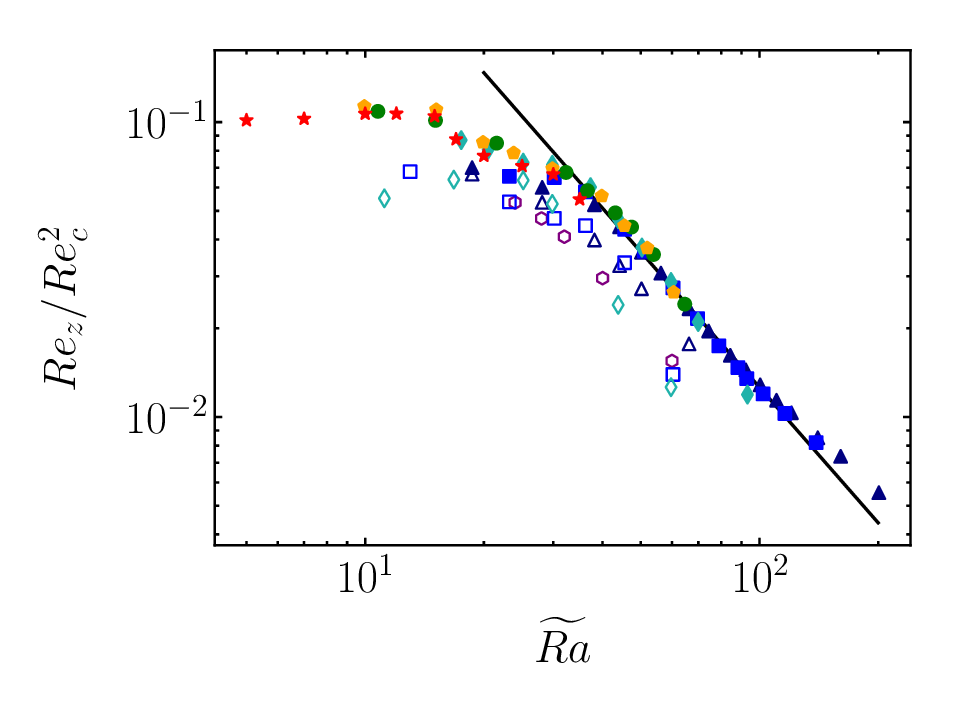}
\caption{Ratio of the zonal Reynolds number to the square of the convective Reynolds number. The least squares fit to a power law scaling calculated using the $\eta=0.35$, $Ek=10^{-4}$ cases is shown as a black line, and is given by $C_R = 14.21\Rat^{-1.526}$. The symbols are the same as defined in figure \ref{F:Re_fluct}. }
\label{F:Rez_Rec_ratio}
\end{center}
\end{figure}

It is also interesting to consider the relationship between the zonal and convective Reynolds numbers. Equation \eqref{E:zonal_scaling} suggests that the relevant quantity to consider is $Re_z/Re_c^2$, which we show in figure \ref{F:Rez_Rec_ratio}. We predicted that this ratio is independent of the Ekman number, which we see is a good approximation for our range of parameters. \citet{uC02} noted that the relation $Re_z \sim Re_c^2$ holds only when the correlation between the fluctuating velocity components is constant, which occurs near the onset of convection. Anelastic simulations of rotating spherical convection also find that $Re_z/Re_c^2$ is nearly constant near the onset of convection, and that $Re_z/Re_c^2$ decreases with increasing $Ra$ for sufficiently large Rayleigh number \citep{tG12}. Our simulations are also consistent with this behaviour, showing that $Re_z/Re_c^2$ is approximately constant up to $\Rat \approx 15$. For $\Rat \gtrsim 15$, we observe that $Re_z/Re_c^2$ decreases with increasing $\Rat$. This trend of $Re_z/Re_c^2$ decreasing as $\Rat$ is increased does not show a systematic dependence on the Ekman number, since all data points appear to collapse well. Because $Re_z/Re_c^2$ is related to the correlation of the convective velocity components, this suggests that the correlation of the convective velocity components decreases as $\Rat$ is increased, even at asymptotically small Ekman numbers.

\subsection{Length scales}

\begin{figure}
 \begin{center}
\subfloat[][]{\includegraphics[width=0.33\textwidth]{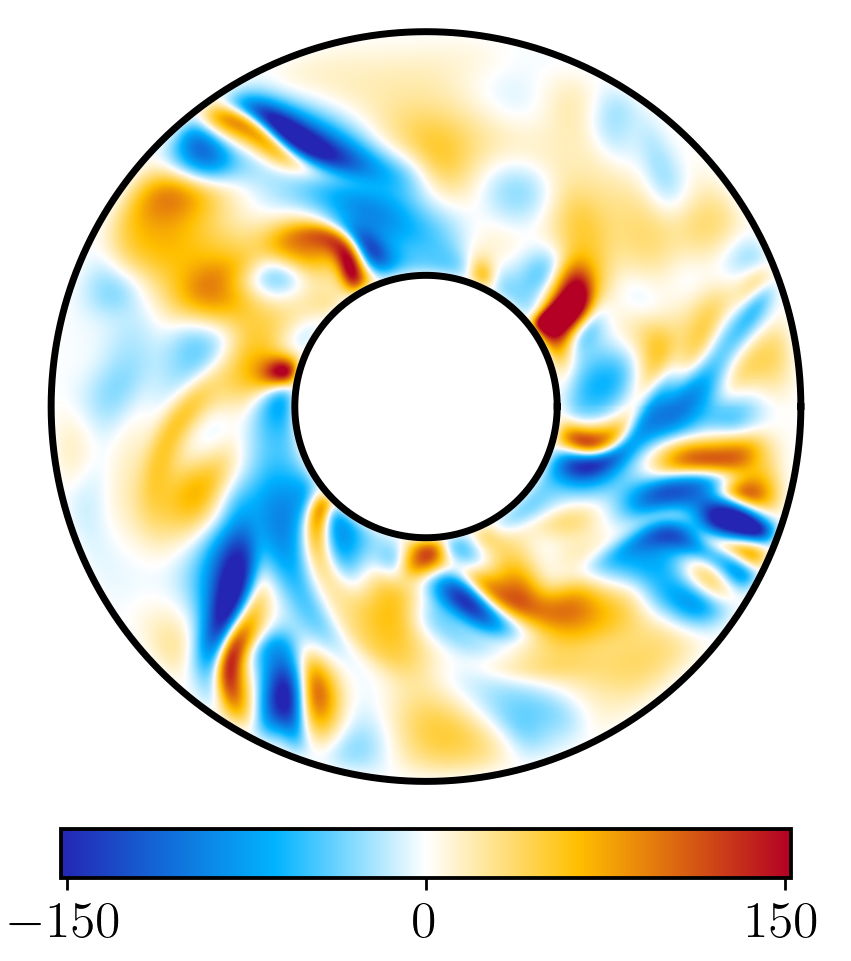}}
\subfloat[][]{\includegraphics[width=0.33\textwidth]{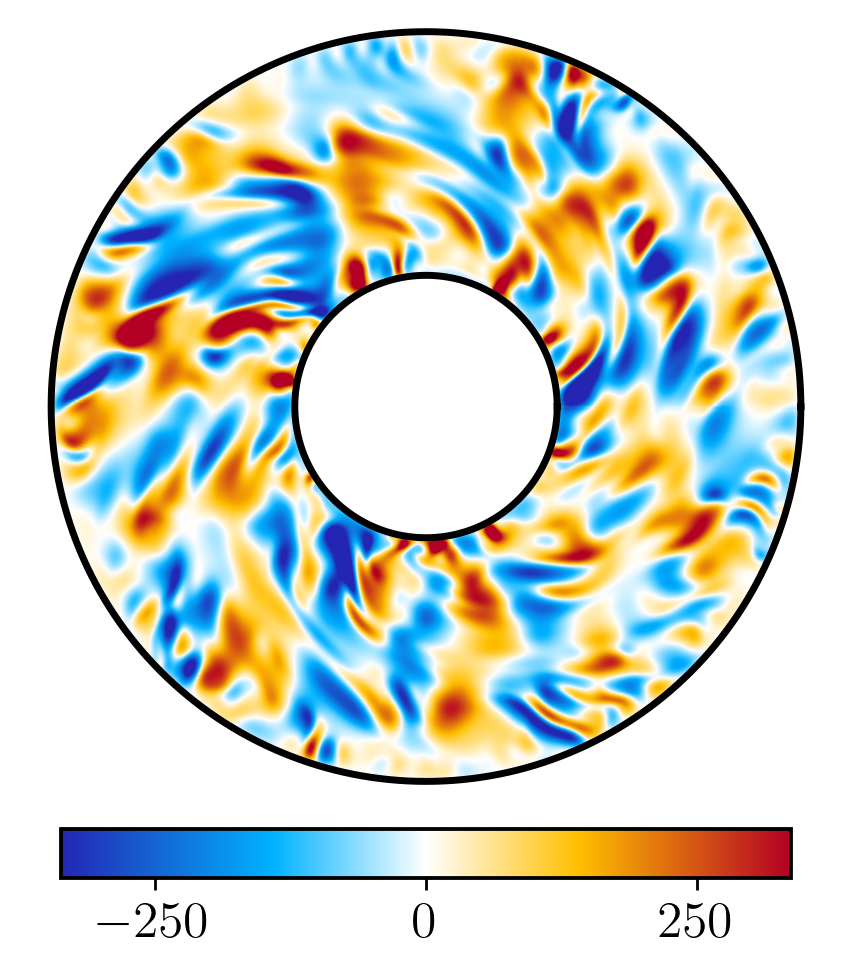}}
\subfloat[][]{\includegraphics[width=0.33\textwidth]{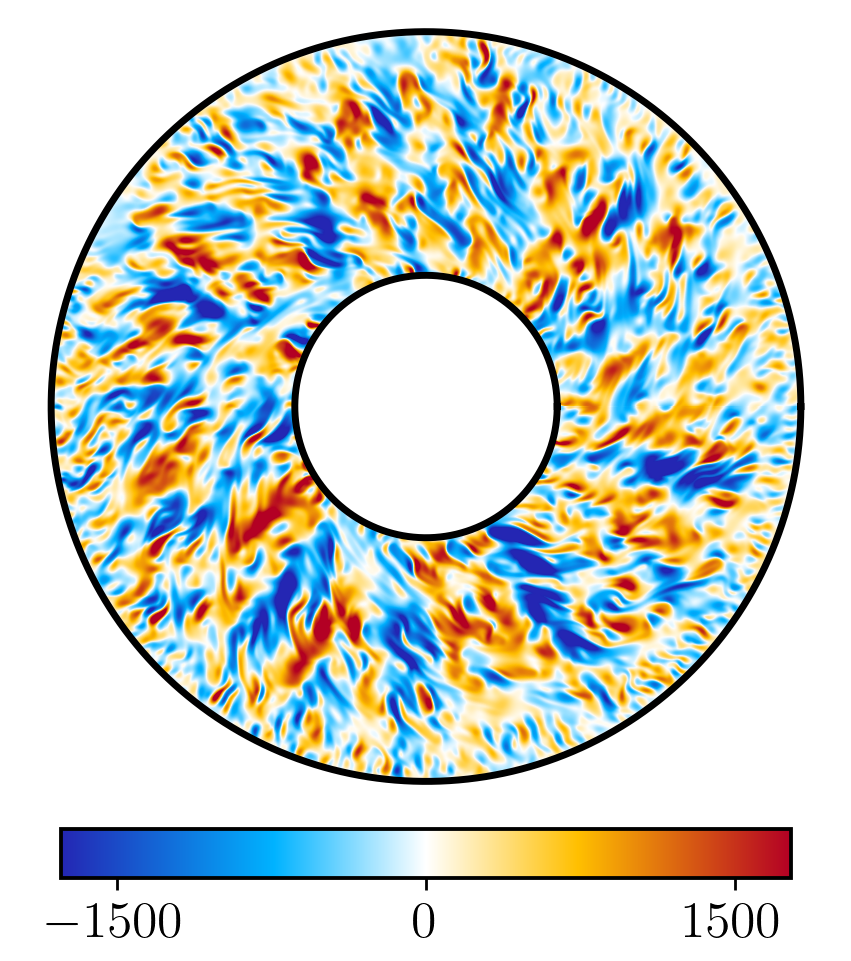}}
\caption{Visualisations of the radial velocity in the equatorial plane for three thick shell cases during a time interval of strong convection: (a) $Ek=3\times 10^{-4}$, $Ra=2.5\times 10^6$ ($\Rat \sim 50.2$);  (b) $Ek=3\times 10^{-5}$, $Ra=5.4\times 10^7$ ($\Rat \sim 50.3$); (c) $Ek=3\times 10^{-6}$, $Ra=1.2\times 10^{9}$ ($\Rat \sim 51.9$).}
\label{F:equatorial_slices}
\end{center}
\end{figure}


\begin{figure}
 \begin{center}
 \subfloat[][]{\includegraphics[width=0.48\textwidth]{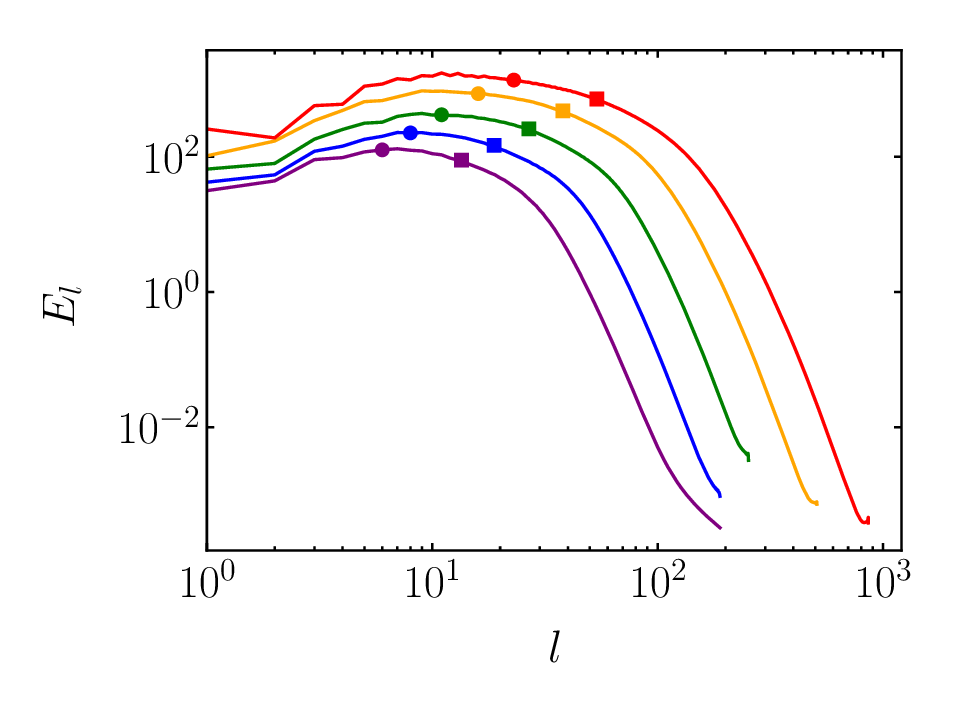}}
\subfloat[][]{\includegraphics[width=0.48\textwidth]{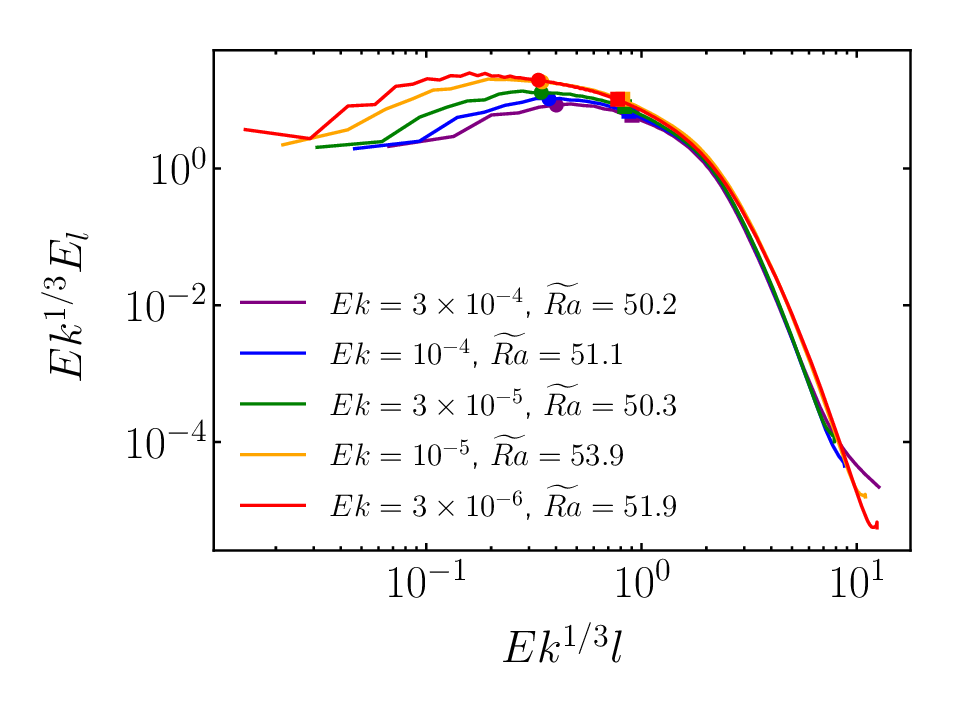}}
\hspace{0mm}
\subfloat[][]{\includegraphics[width=0.48\textwidth]{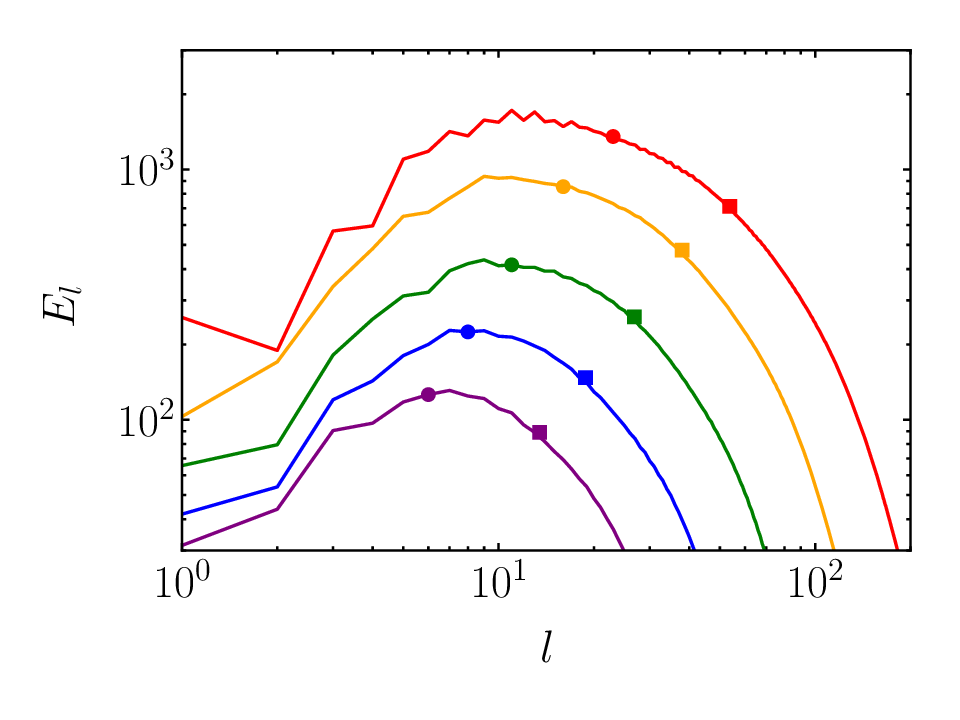}}
\subfloat[][]{\includegraphics[width=0.48\textwidth]{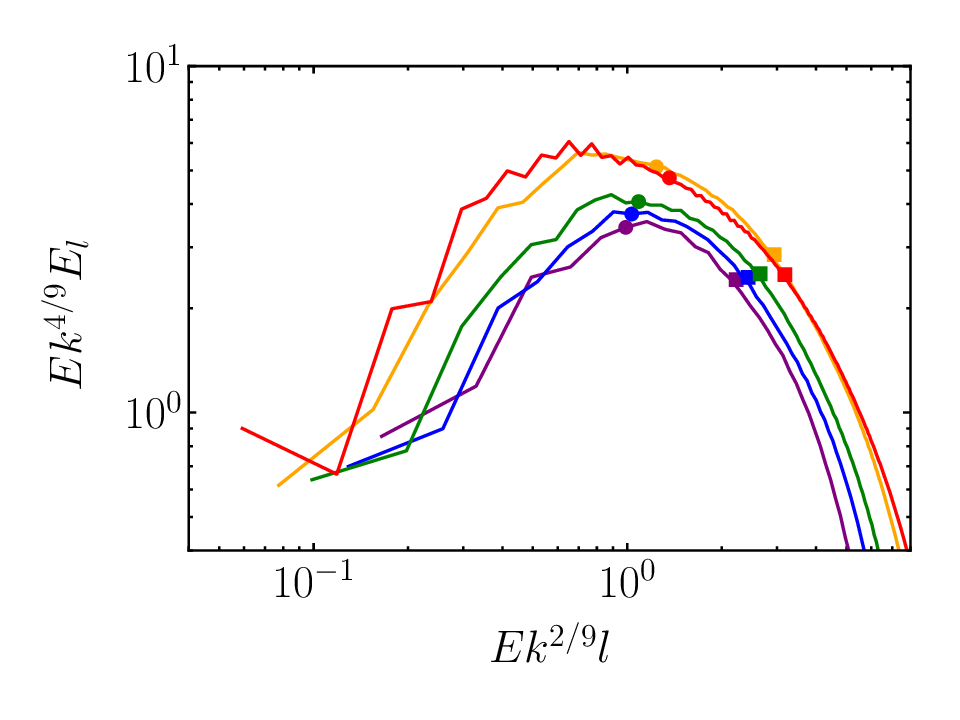}}
\caption{Time- and radially-averaged spherical harmonic kinetic energy spectra. Cases with $\eta=0.35$ and similar values of the reduced Rayleigh number are shown. Circles denote the critical azimuthal wavenumber ($m_c$) at the onset of convection for each parameter set and squares denote the degree $l$ derived from the spherical harmonic length scale $\ell_{sh}^{\prime} = \pi/l$. (a) Spherical harmonic spectra. (b) Rescaled spherical harmonic spectra. (c) Spherical harmonic spectra at small $l$. (d) Rescaled spherical harmonic spectra at small $l$.}
\label{F:spectra_Ek}
\end{center}
\end{figure}

\begin{figure}
 \begin{center}
 \subfloat[][]{\includegraphics[width=0.48\textwidth]{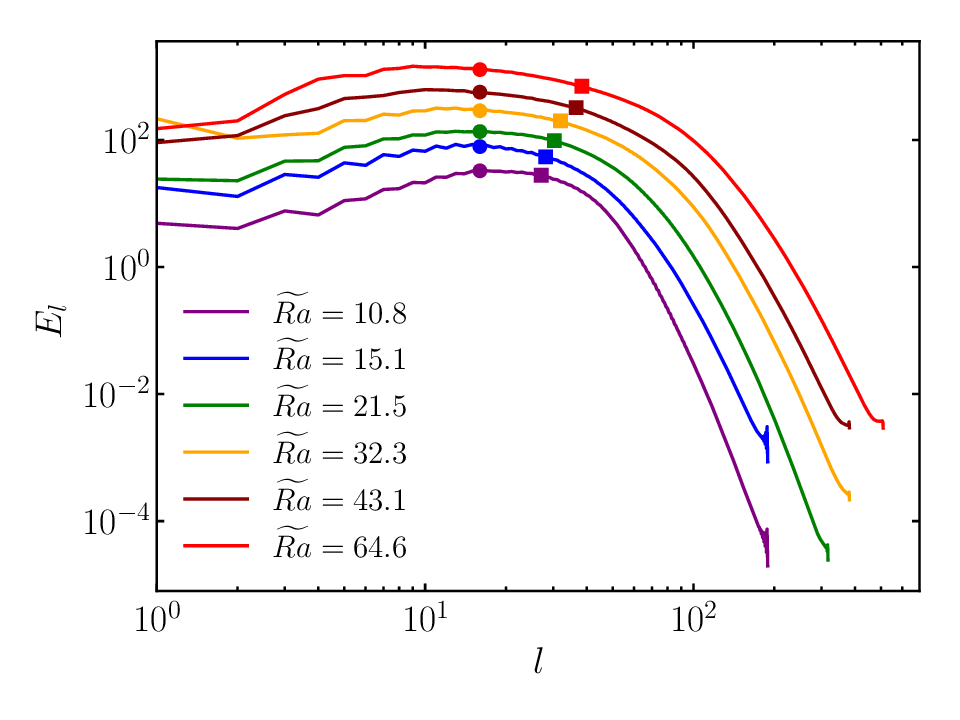}}
\subfloat[][]{\includegraphics[width=0.48\textwidth]{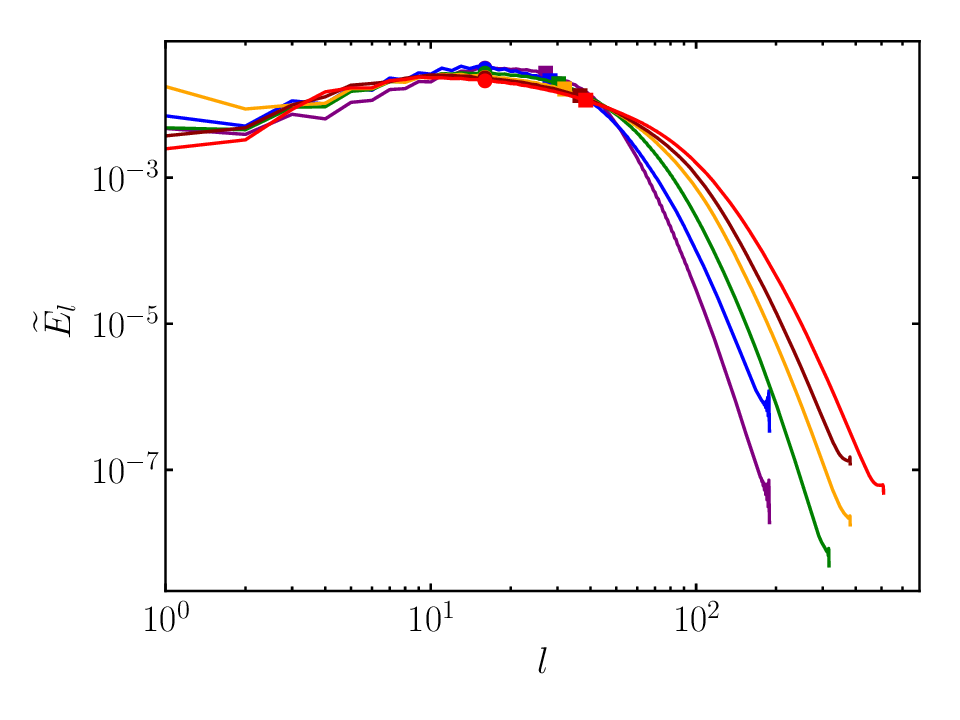}}
\caption{Time- and radially-averaged spherical harmonic spectra summed. Select cases from $\eta=0.35$, $Ek=10^{-5}$ are shown. Circles denote the critical azimuthal wavenumber at the onset of convection and squares denote the spherical harmonic degree calculated from $\ell_{sh}^{\prime}$, as in figure \ref{F:spectra_Ek}. (a) Spherical harmonic spectra. (b) Spherical harmonic spectra normalized to have an area of one.}
\label{F:spectra}
\end{center}
\end{figure}

In this section, we compute length scales for both the mean flow and the small-scale convection. To provide a physical picture of how these length scales vary with Ekman number, we show snapshots of the radial velocity in the equatorial plane for three different Ekman number cases with $\Rat \approx 50$ in figure \ref{F:equatorial_slices}. As expected, the typical length scale of the convection decreases with decreasing Ekman number.  

One way to quantitatively study how the length scale varies with Ekman number is to consider how the kinetic energy power spectrum varies with Ekman number. We define the sum of the kinetic energy power spectrum over the order $m$ and the normalization of the kinetic energy power spectrum summed over the order $m$ as
\be
E_l = \tavg{\sum_{m=1}^{l} \mathcal{E}_l^m}, \qquad \widetilde{E}_l = \tavg{\left(\frac{\sum_{m=1}^{l} \mathcal{E}_l^m}{\sum_{l=1}^{l_{max}} \sum_{m=1}^{l} \mathcal{E}_l^m}\right)}.
\ee
Due to the influence of the zonal flow, we again only sum over $m\geq 1$ modes to remove the influence of the zonal flow. Figure \ref{F:spectra_Ek} shows how these kinetic energy spectra vary with Ekman number. The critical azimuthal wavenumber $m_c$ is shown for reference, and $m_c$ takes the values $6$, $8$, $11$, $16$, and $23$ for the cases of $Ek=3\times10^{-4}$, $Ek=10^{-4}$, $Ek=3\times 10^{-5}$, $Ek=10^{-5}$, and $Ek=3\times 10^{-6}$, respectively. These values were calculated using the eigenvalue solver Kore \citep{aB23}. We see that the smaller Ekman number cases have more power in each mode and extend to higher values of $l$ in comparison to larger Ekman number cases. We attempt to collapse this data by multiplying the degree $l$ by $Ek^{1/3}$, the expected scaling for the length scale, and multiplying the amplitude of the data by $Ek^{1/3}$. We choose the amplitude scaling such that the area under the kinetic energy power spectrum, the volume averaged kinetic energy density, scales as $Ek^{-2/3}$, which is consistent with the scaling of our convective Reynolds number. This collapsed data for the kinetic energy power spectrum is shown in figure \ref{F:spectra_Ek}(b). We see that the collapse of the data is reasonable for values of $l$ greater than $m_c$, but not as good for values of $l$ less than $m_c$. This discrepancy might suggest that large length scales are following a different scaling than anticipated. We investigate the behaviour of the kinetic energy power spectrum for the small $l$ modes in figure \ref{F:spectra_Ek}(c,d). The power spectrum for these small $l$ modes appears to shift more slowly to the right with Ekman number than the large $l$ modes. Several papers on the linear onset of convection predict a longer length scale in the cylindrical radial direction than in the azimuthal direction. \citet{eD04} predicts a $Ek^{2/9}$ length scale for cases with differential heating, and a $Ek^{1/6}$ length scale for cases with certain boundary conditions and heating conditions. We show the collapse of our data for the predicted $Ek^{2/9}$ length scale in figure \ref{F:spectra_Ek}(d), which does a reasonable job of collapsing our data. However, we do not have a large enough range in Ekman number to discern between a $Ek^{1/6}$ and a $Ek^{2/9}$ scaling. 

 
Figure \ref{F:spectra} shows the kinetic energy spectra for the $\eta=0.35$, $Ek=10^{-5}$ cases averaged in time and radius. Figure \ref{F:spectra}(b) shows the spherical harmonic spectra normalized to have area one. We find that the kinetic energy contained in high $l$ values increases slightly with increasing $\Rat$, but that the overall shape of the spectra does not change significantly. This behaviour indicates that small-scale convection becomes  more prominent as the Rayleigh number is increased. 
 
\begin{figure}
 \begin{center}
 \subfloat[][]{\includegraphics[width=0.48\textwidth]{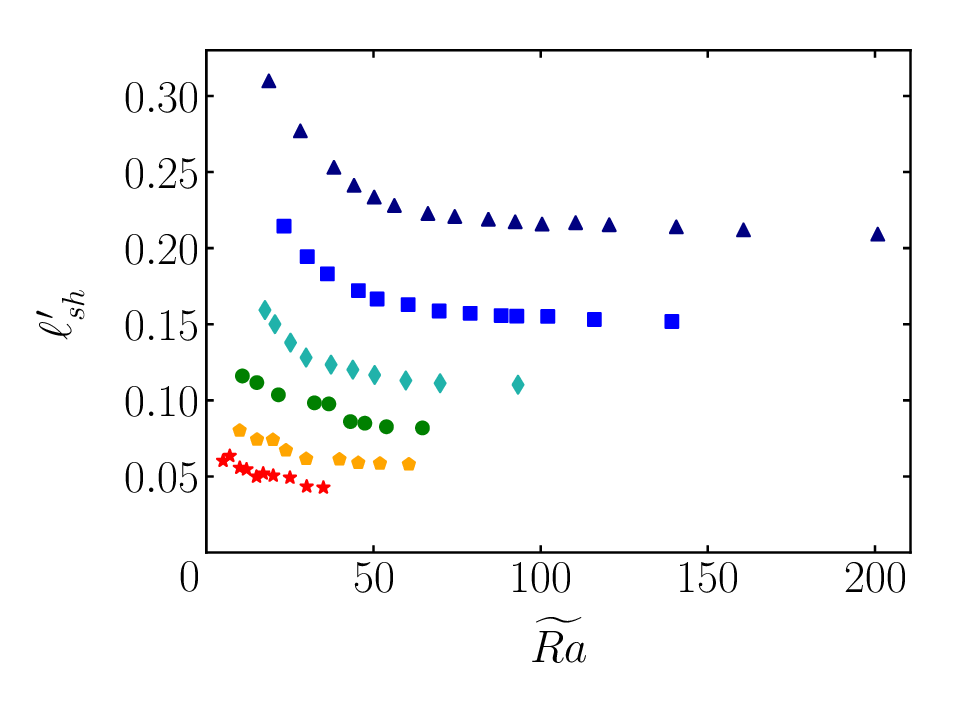}}
\subfloat[][]{\includegraphics[width=0.48\textwidth]{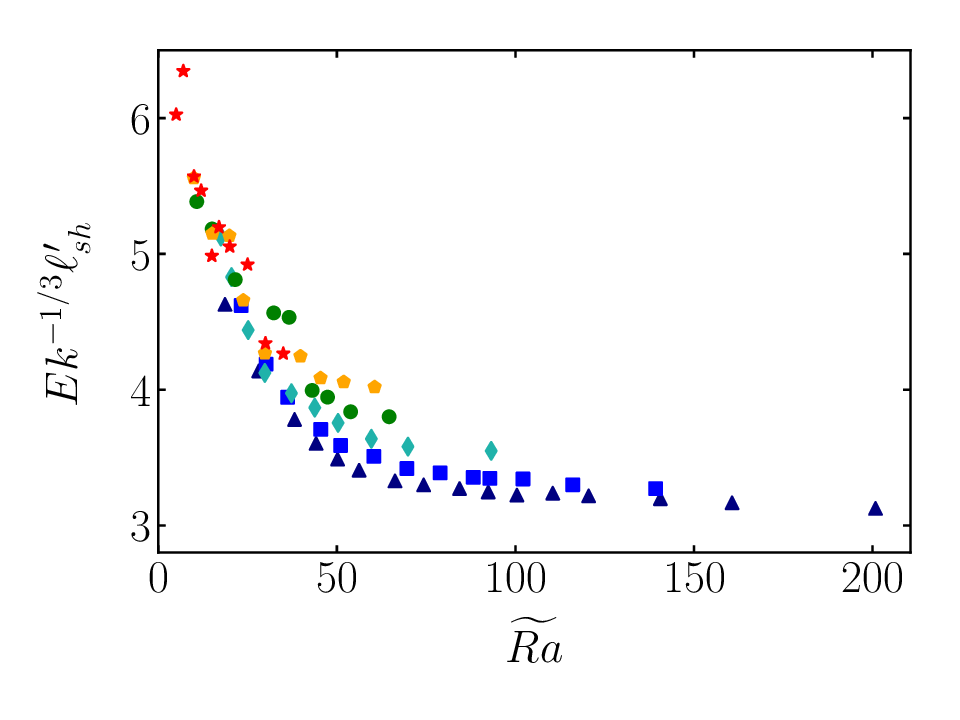}} \\
 \subfloat[][]{\includegraphics[width=0.48\textwidth]{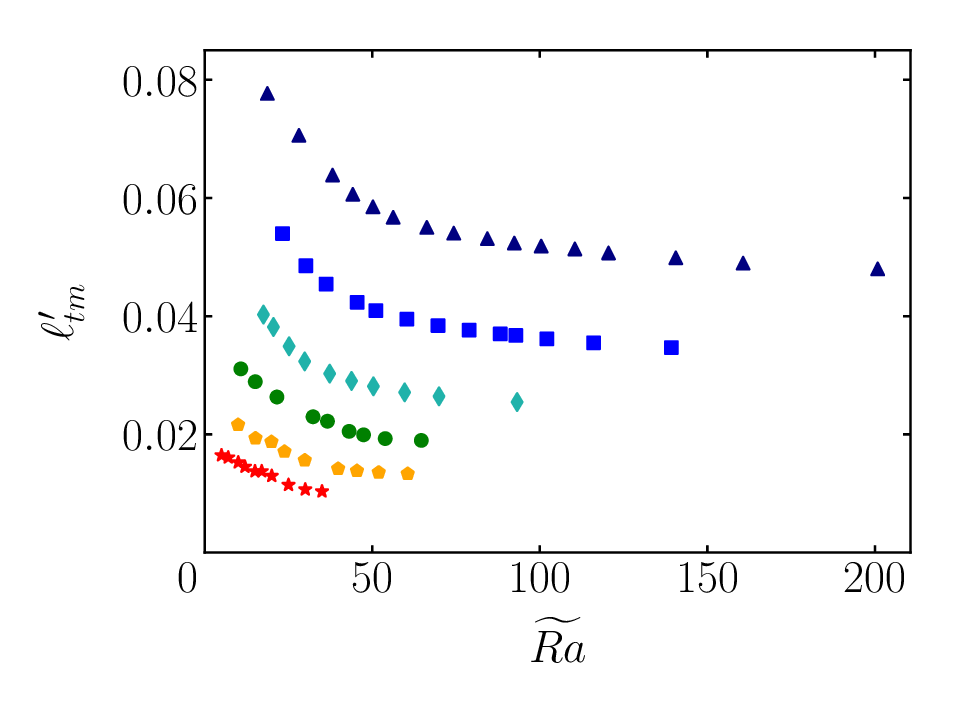}}
\subfloat[][]{\includegraphics[width=0.48\textwidth]{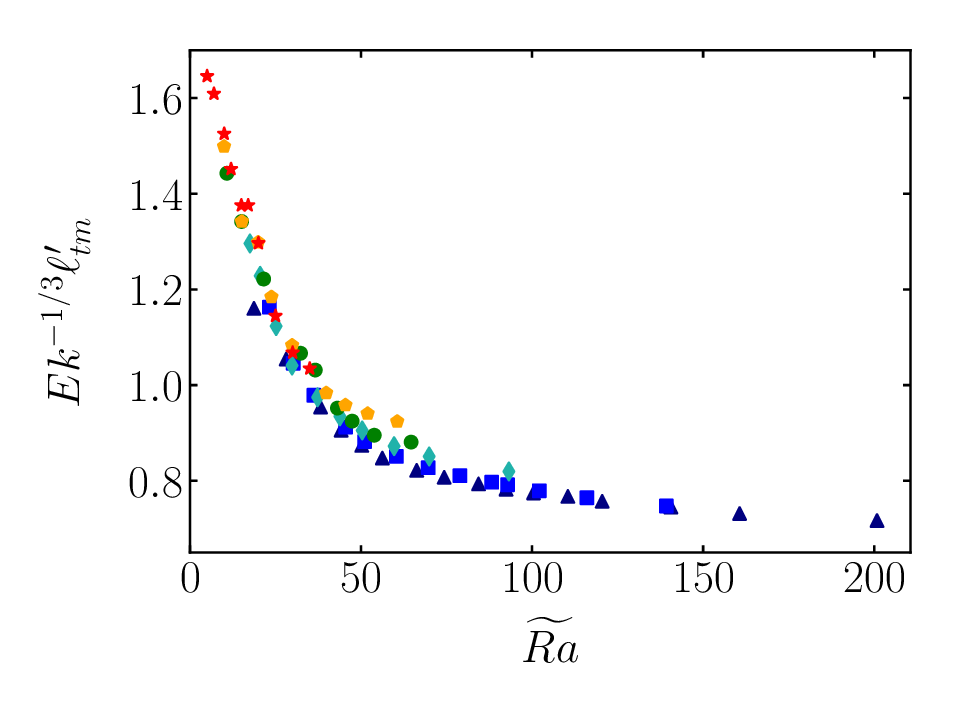}}
\caption{Convective length scales for the $\eta=0.35$ cases: (a), (b) spherical harmonic length scale;
(c), (d) Taylor microscale. Plots (b) and (d) show the asymptotically rescaled length scales. The symbols are the same as defined in figure \ref{F:Re_fluct}.}
\label{F:ls_sh}
\end{center}
\end{figure}

The computed length scales are all shown in figure \ref{F:ls_sh}. The spherical harmonic length scale is shown in panels (a) and (b), and the Taylor microscale is shown in panels (c) and (d); the asymptotically rescaled lengths are shown in panels (b) and (d).
The spherical harmonic length scale is observed to decrease rapidly for $\Rat \lesssim 50$, then levels off for larger values of $\Rat$; not surprisingly this behaviour is consistent with the trends observed in the kinetic energy spectra. There may be a trend suggesting that this length scale converges to a nearly constant value at the largest values of $\Rat$, though this behaviour may be occurring outside of the rapidly rotating regime. Both the decreasing and constant dependence of the length scale on Rayleigh number observed here is in contrast to the prediction made by the CIA balance where the length scale increases as the Rayleigh number increases. Figure \ref{F:ls_sh}(b) shows the asymptotically rescaled spherical harmonic length scale. As with the flow speeds, this rescaled quantity is order unity and we find significantly less scatter in the data when viewed in this rescaled coordinate, suggesting that this particular length scale does approximately scale as $Ek^{1/3}$. However, we also note that a subtle Ekman number dependence still exists in the rescaled length scale; larger Ekman number cases seem to level off at smaller rescaled length scales in comparison to the smaller Ekman number cases. 
This behaviour might indicate that the system is either converging slowly toward the $O(Ek^{1/3})$ length scale with decreasing Ekman number, or that the asymptotic dependence of this length scale is slightly weaker than $Ek^{1/3}$. Our data for the kinetic energy power spectrum indicates that a second length scale may be present at small $l$ values which follows a different Ekman number scaling. Thus, the spherical harmonic length scale may have a composite asymptotic dependence on $Ek$ since it may be capturing both the $Ek^{2/9}$ and $Ek^{1/3}$ asymptotic scalings.




Figure \ref{F:ls_sh}(c,d) shows the Taylor microscale, where we see that it follows a very similar trend in comparison to the spherical harmonic length scale. The collapse of the rescaled quantity indicates that the viscous dissipation length scale is consistent with a $Ek^{1/3}$ dependence across the entire range of parameters used. The Taylor microscale appears to decrease with $\Rat$ over a wider range of $\Rat$ than the spherical harmonic length scale, which becomes approximately constant with $\Rat$. If we interpret the spherical harmonic length scale as the energy containing length scale (similar to an integral length scale), then our computed length scales indicate that the scale separation present in these simulations is relatively small across the entire range of parameters since $\ell_{sh}^{\prime}$ and $\ell_{tm}^{\prime}$ are not too dissimilar in value. For instance, at $\Rat \approx 50$, $\ell_{sh}^{\prime} \approx 4$ and $\ell_{tm}^{\prime} \approx 0.9$. We note that the trends observed in both length scales contrast sharply with computed length scales in non-rotating Rayleigh-B\'enard convection, where both the energy containing length scale and the dissipation length scale decrease rapidly with increasing Rayleigh number \citep[e.g.][]{mY21}.

\begin{figure}
 \begin{center}
 \subfloat[][]{\includegraphics[width=0.48\textwidth]{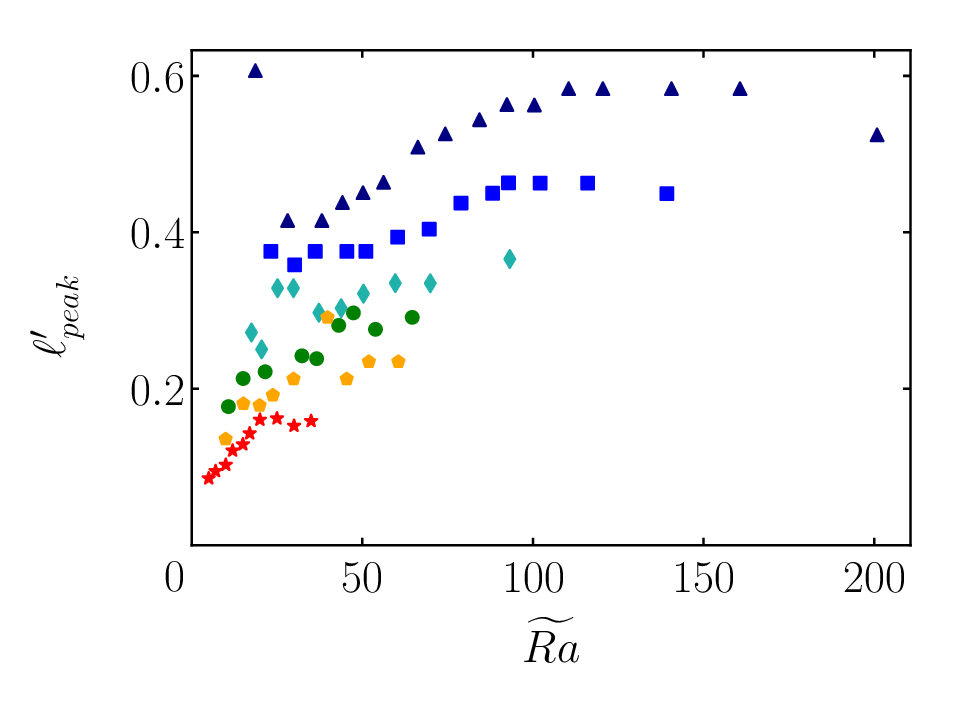}}
\subfloat[][]{\includegraphics[width=0.48\textwidth]{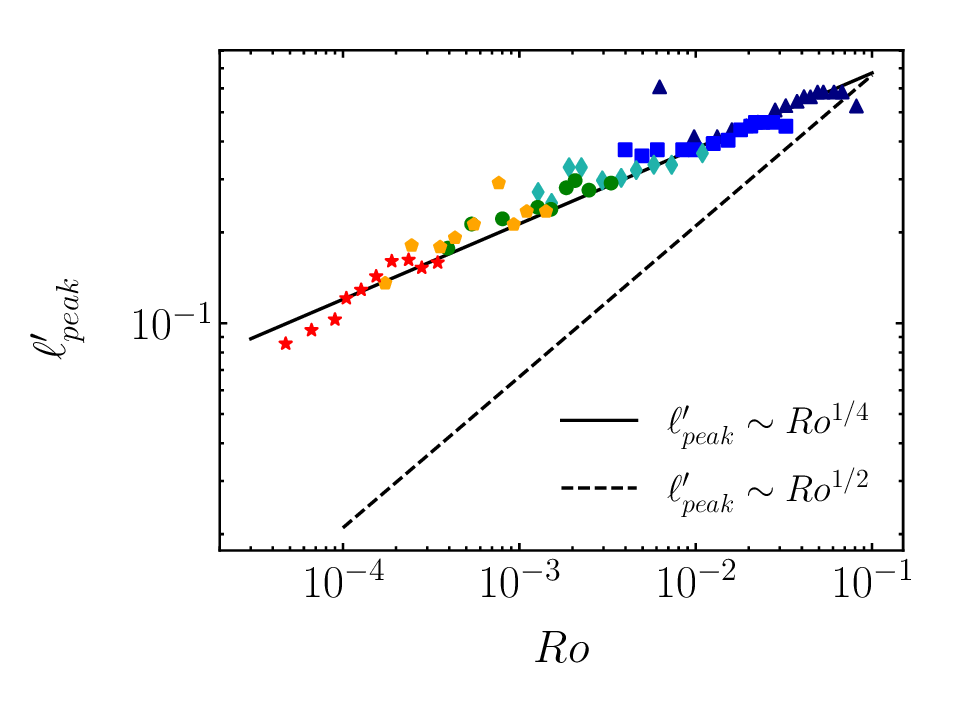}} \\
 \subfloat[][]{\includegraphics[width=0.48\textwidth]{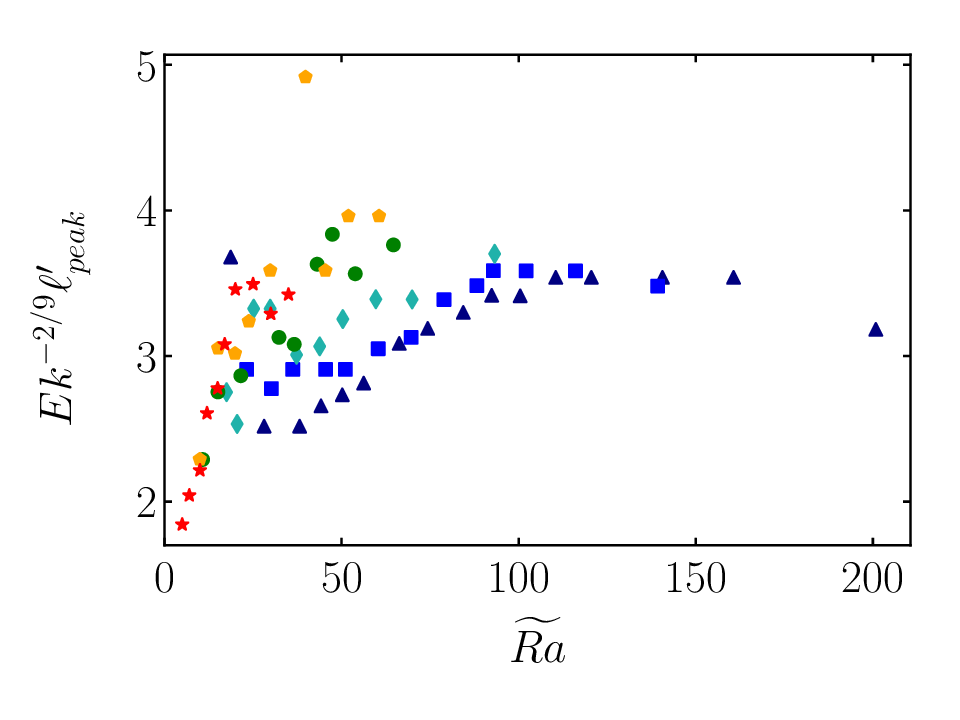}}
  \subfloat[][]{\includegraphics[width=0.48\textwidth]{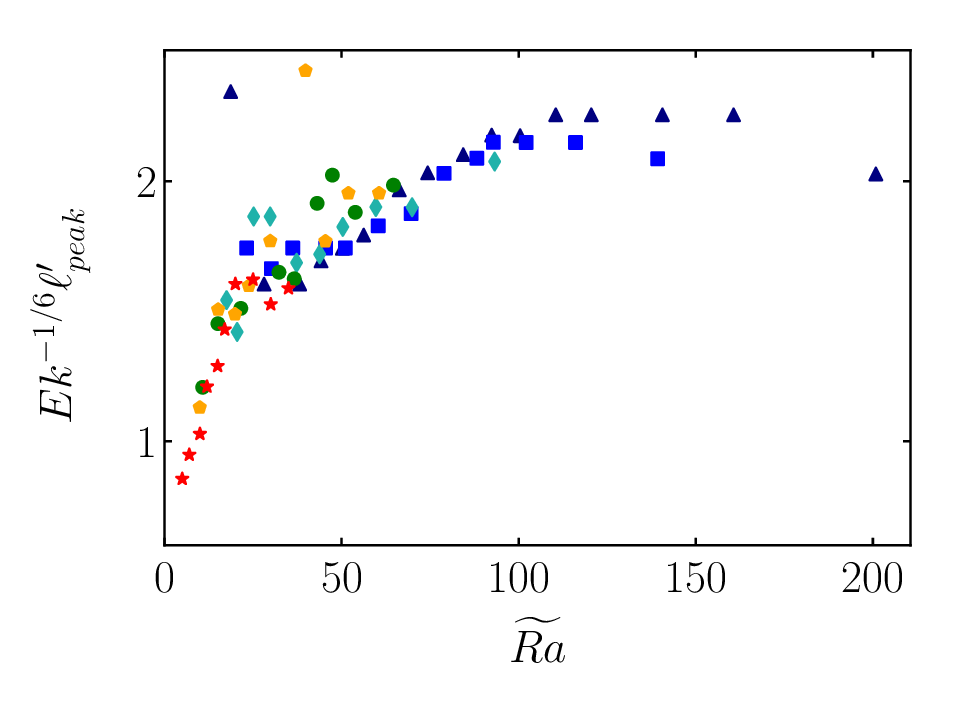}}
\caption{Length scale calculated from the degree $l_{peak}$ where the fluctuating kinetic energy power spectrum peaks: (a)  peak length scale; (b) peak length scale versus the Rossby number characterising the convective flow speeds (two Rossby number scalings are shown for reference); (c) peak length scale rescaled by $Ek^{-2/9}$; (d) peak length scale rescaled by $Ek^{-1/6}$. }
\label{F:ls_peak}
\end{center}
\end{figure}

In order to study the behaviour of the $Ek^{2/9}$ length scale found in the power spectrum, we define a length scale based on the degree $l_{peak}$ where the kinetic energy power spectrum (again with the $m=0$ mode removed) reaches a maximum. The peak length scale is then given by $\ell^{\prime}_{peak}=\pi/l_{peak}$. We fit a 30th degree polynomial to our kinetic energy power spectrum in order to smooth $l_{peak}$, although we obtained similar results when we fit our data with a 15th degree polynomial. Figure \ref{F:ls_peak}(a) shows the peak length scale and figure \ref{F:ls_peak}(c) shows the peak length scale rescaled by $Ek^{-2/9}$. We note that while the peak length scale approximately scales as $Ek^{2/9}$, we obtain a somewhat better fit for $Ek^{1/6}$ as shown in figure \ref{F:ls_peak}(d). Regardless, this peak length scale seems to scale differently than $\ell^{\prime}_{sh}$, which suggests that $\ell^{\prime}_{peak}$ does in fact correspond to a different length scale than the small-scale convective length scale. We also note that the length scale $\ell^{\prime}_{peak}$ increases with increasing Rayleigh number, in contrast to $\ell^{\prime}_{sh}$ and $\ell^{\prime}_{tm}$, which either decrease or remain constant as the Rayleigh number is increased. Figure \ref{F:ls_peak}(b) shows the peak length scale as a function of the Rossby number characterising the convective flow speeds, $Ro=EkRe_c$, where it can be seen that $\ell^{\prime}_{peak}$ approximately scales as $Ro^{1/4}$, which is in contrast to the $Ro^{1/2}$ scaling that is predicted from the CIA balance. To ensure the validity of our results with polynomial interpolation, we also estimated the peak length scale by replacing $\mathcal{E}_l^m$ in equation \ref{E:length_fluct_mean} with $\left(\mathcal{E}_l^m\right)^{10}$, which weights the length scale more towards the peak. Using this estimate of the peak length scale, we found qualitatively similar results to what is shown in figure \ref{F:ls_peak}.

\begin{figure}
 \begin{center}
 \subfloat[][]{\includegraphics[width=0.48\textwidth]{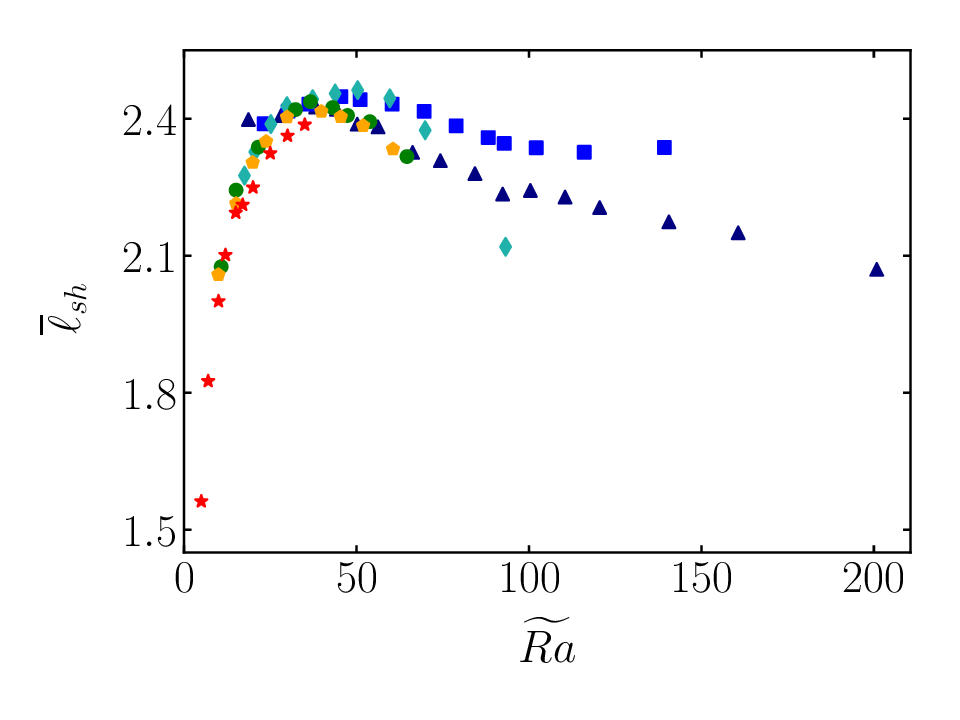}}
\subfloat[][]{\includegraphics[width=0.48\textwidth]{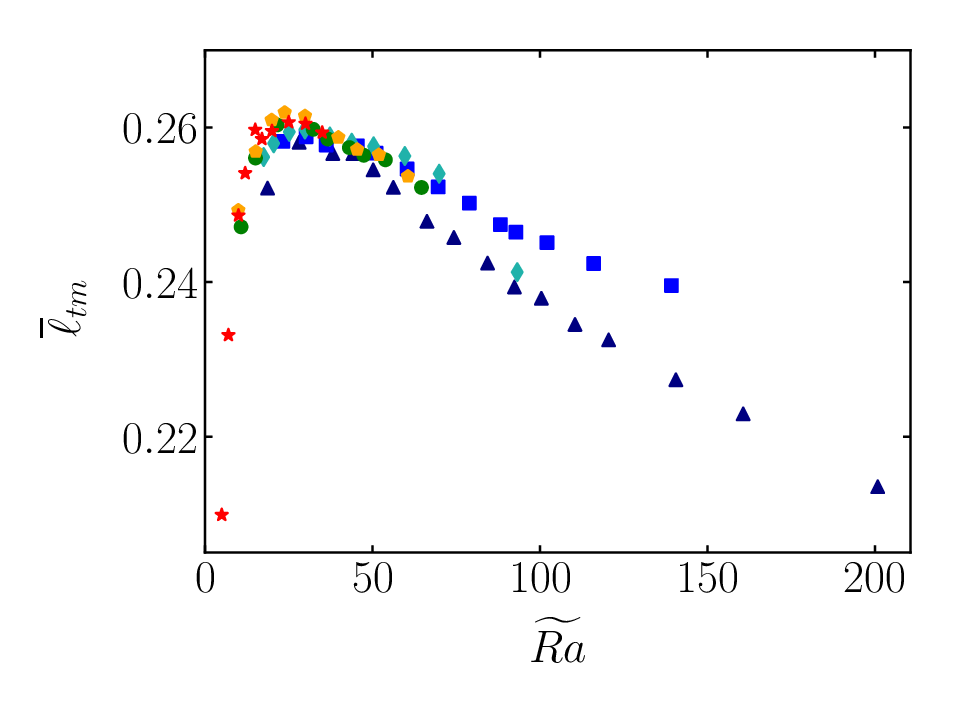}}
\caption{Length scales computed from the mean flow: (a) spherical harmonic length scale; (b) Taylor microscale. The symbols are the same as defined in figure \ref{F:Re_fluct}.}
\label{F:ls_m0}
\end{center}
\end{figure}

We now consider length scales of the zonal flow, which are shown in figure \ref{F:ls_m0}. Unlike the convective length scales, we expect these mean length scales to be order unity and approximately independent of the Ekman number. The mean length scale reaches a maximum at approximately $\Rat \sim 25$ and, for larger values of the reduced Rayleigh number, decays slowly over the range of Rayleigh numbers used in this study. We also computed a Taylor microscale for the mean flow, as shown in \ref{F:ls_m0}(b); although this length scale is slightly smaller than the spherical harmonic length scale, it shows a similar behaviour with $Ek$ and $\Rat$ in comparison to the spherical harmonic length scale.

\subsection{Heat equation balances}

\begin{figure}
 \begin{center}
 \subfloat[][]{\includegraphics[width=0.48\textwidth]{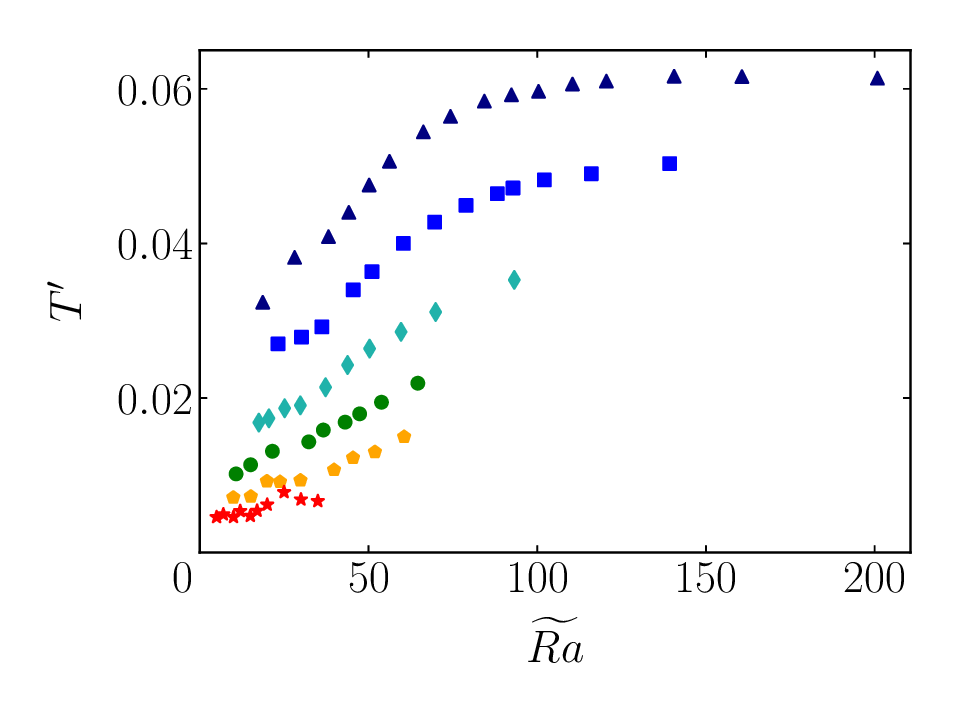}}
\subfloat[][]{\includegraphics[width=0.48\textwidth]{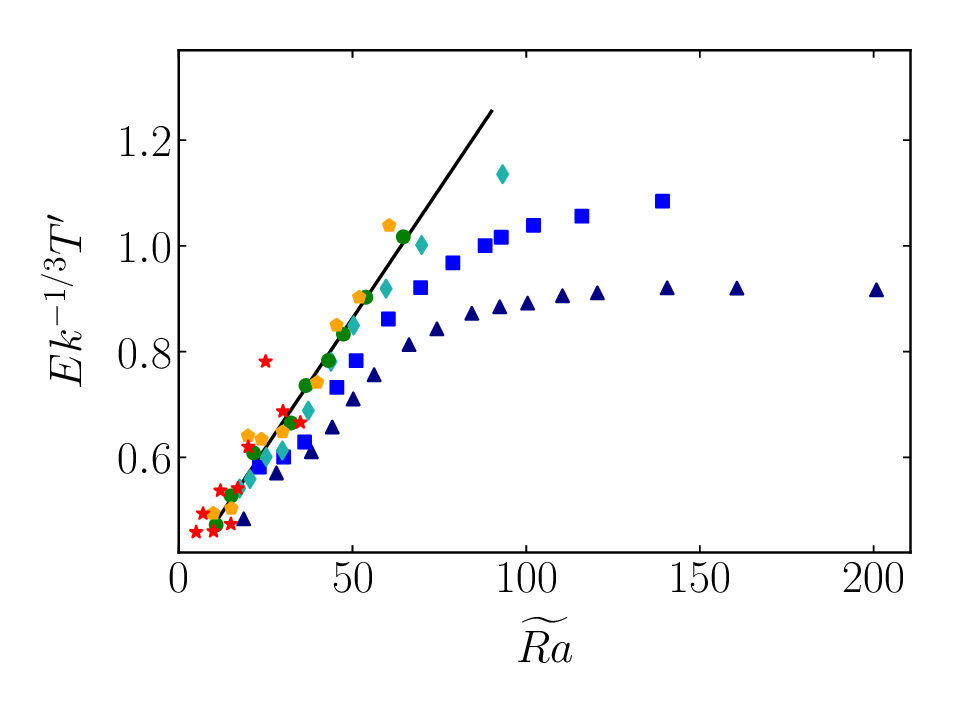}} \\
\subfloat[][]{\includegraphics[width=0.48\textwidth]{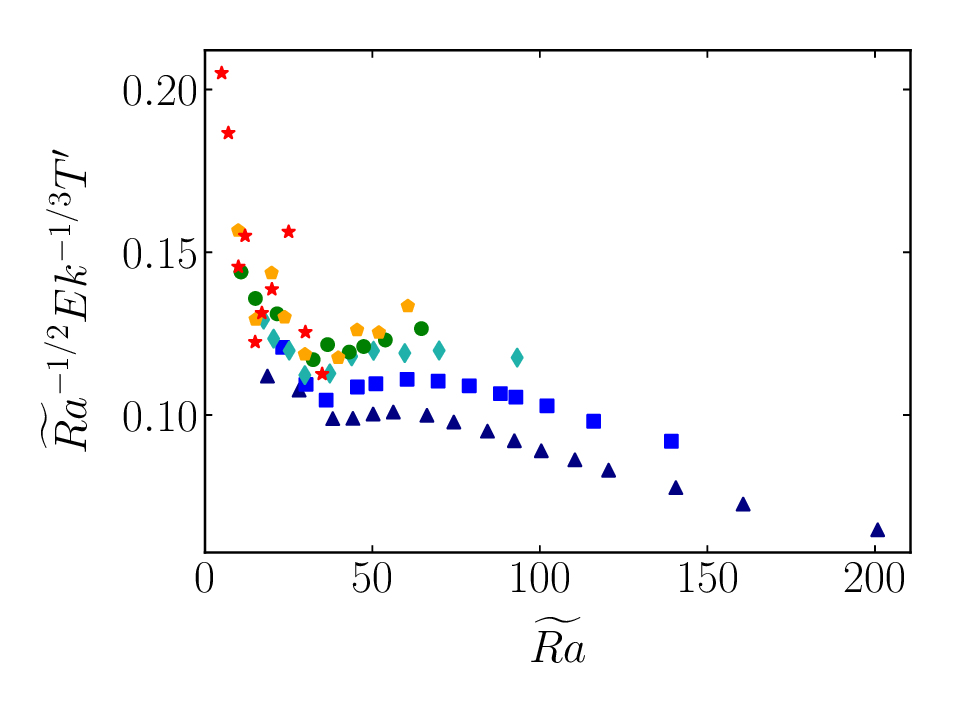}}
\caption{Time-averaged volume rms of the fluctuating temperature; (a) raw data; (b) asymptotically rescaled data; (c) compensated data. The line $Ek^{-1/3}T'=0.01\Rat+0.5$ is shown for reference in (b). The symbols are the same as defined in figure \ref{F:Re_fluct}.}
\label{F:temp_rms}
\end{center}
\end{figure}

\begin{figure}
 \begin{center}
 \subfloat[][]{\includegraphics[width=0.33\textwidth]{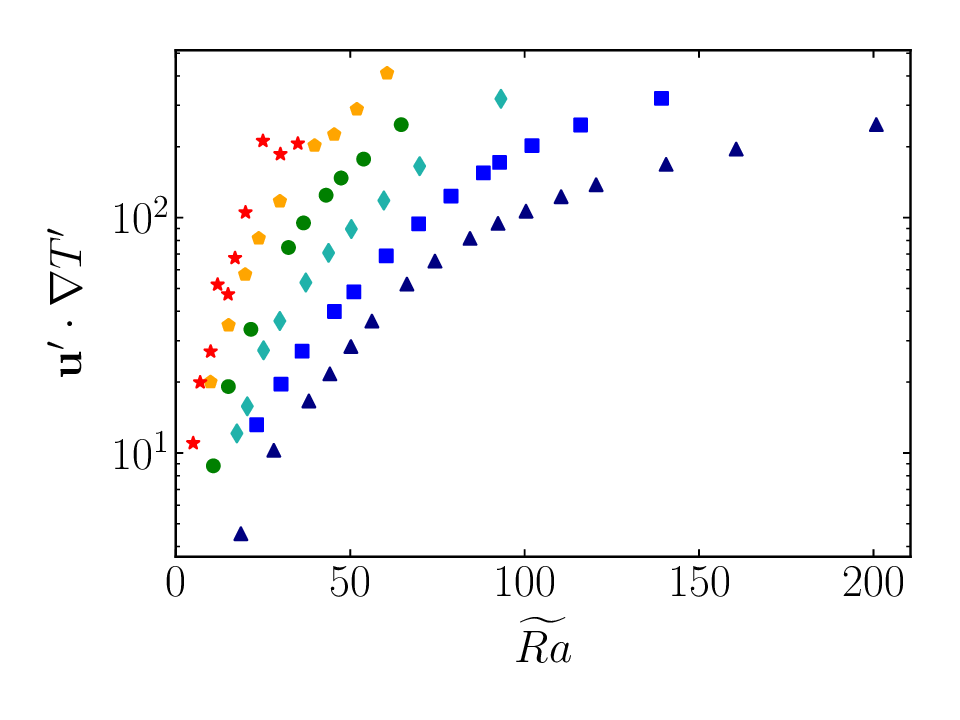}}
 \subfloat[][]{\includegraphics[width=0.33\textwidth]{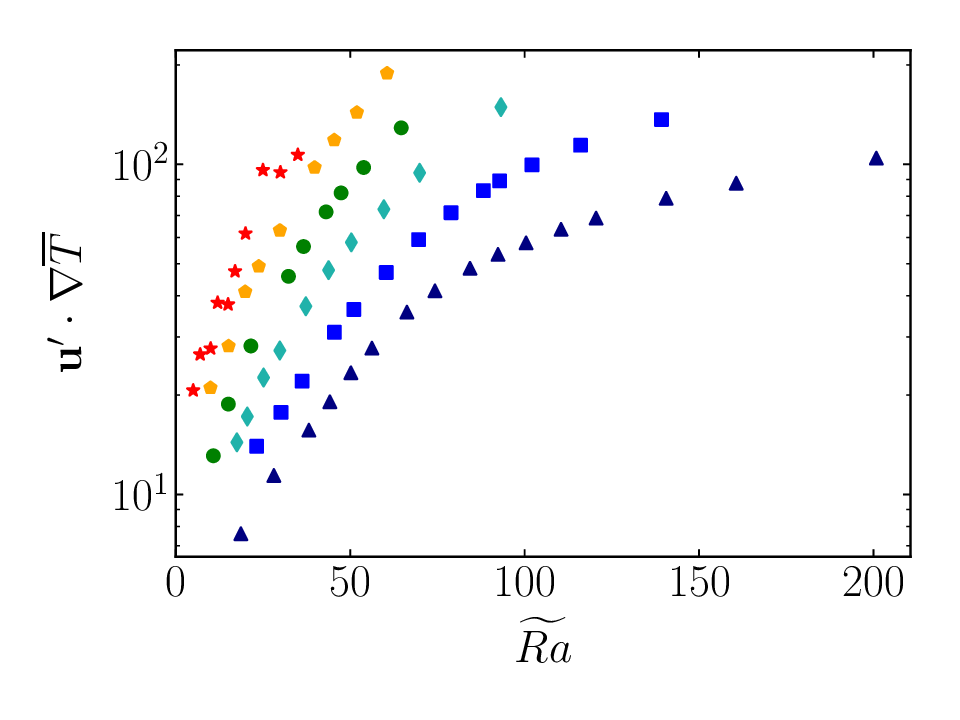}}
 \subfloat[][]{\includegraphics[width=0.33\textwidth]{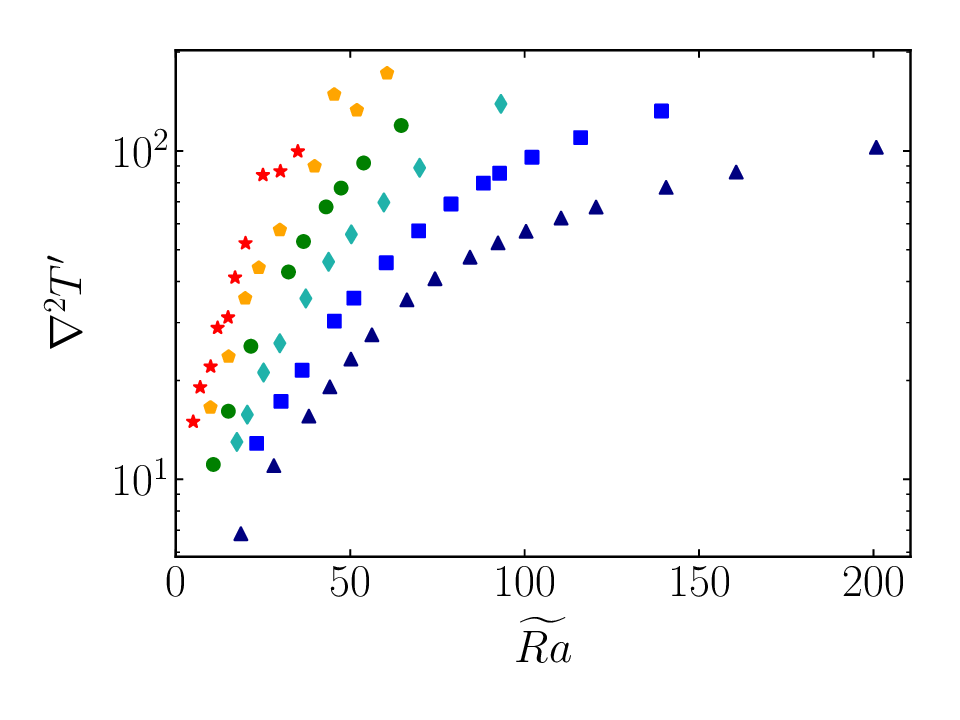}}
 \hspace{0mm}
 \subfloat[][]{\includegraphics[width=0.33\textwidth]{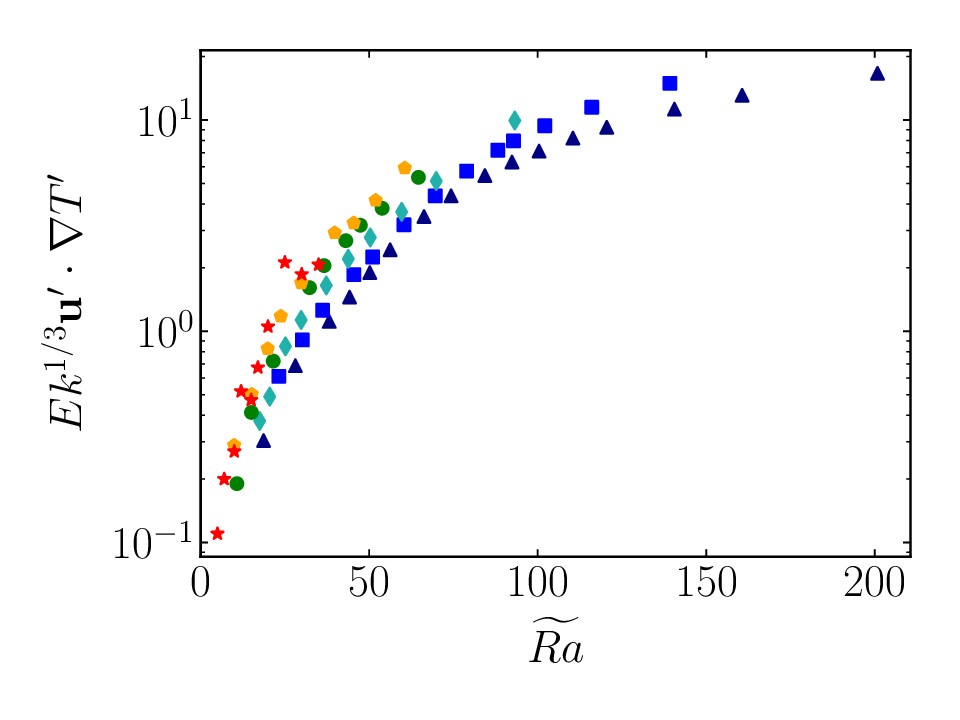}}
 \subfloat[][]{\includegraphics[width=0.33\textwidth]{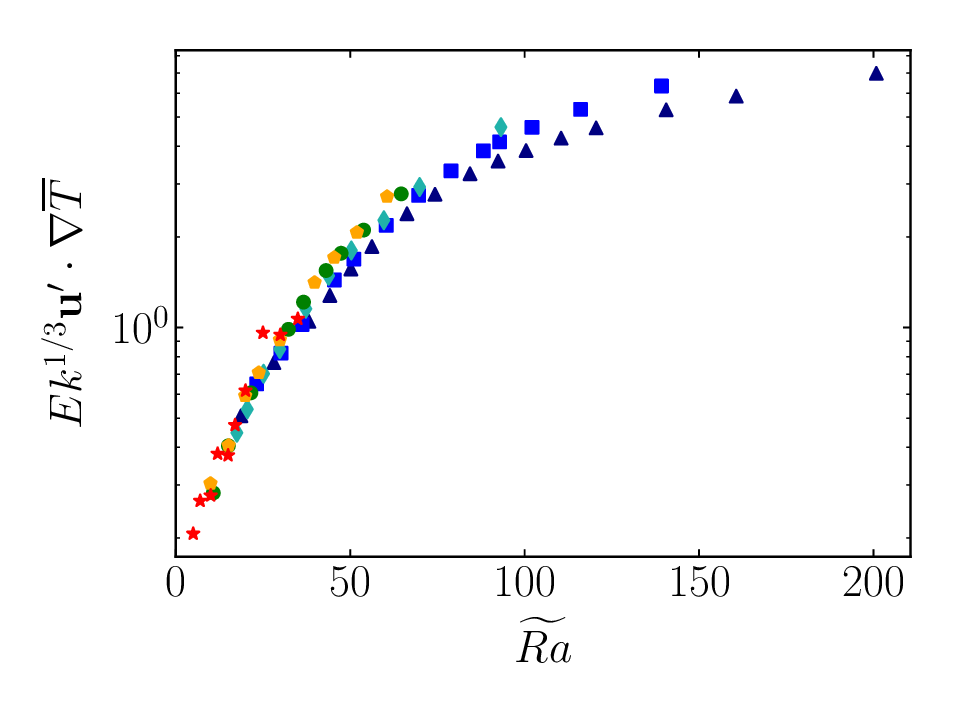}}
 \subfloat[][]{\includegraphics[width=0.33\textwidth]{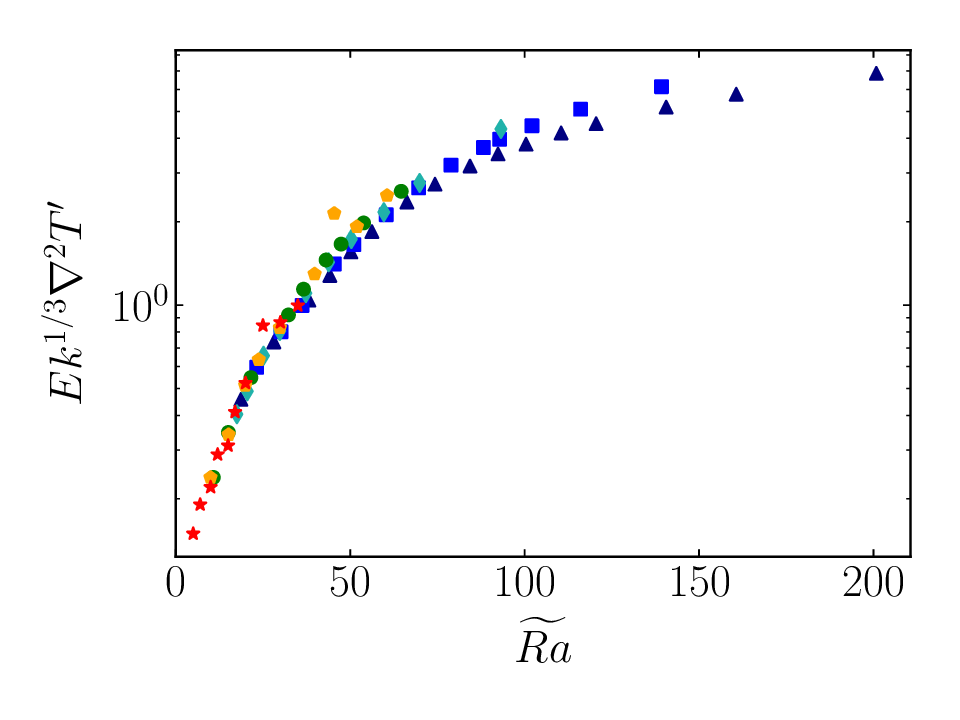}}
 \caption{Time-averaged volume rms values of various terms from the fluctuating heat equation. The terms shown are: (a) $\mathbf{u}' \cdot \nabla T'$; (b) $\mathbf{u}' \cdot \nabla \overline{T}$; (c) $\nabla^2 T'$. Asymptotically rescaled data: (d) $Ek^{1/3}\mathbf{u}' \cdot \nabla T'$; (e) $Ek^{1/3}\mathbf{u}' \cdot \nabla \overline{T}$; (f) $Ek^{1/3}\nabla^2 T'$. The symbols are the same as defined in figure \ref{F:Re_fluct}. }
\label{F:thermal_terms_scaling}
\end{center}
\end{figure}

 \begin{figure}
 \begin{center}
 \subfloat[][]{\includegraphics[width=0.48\textwidth]{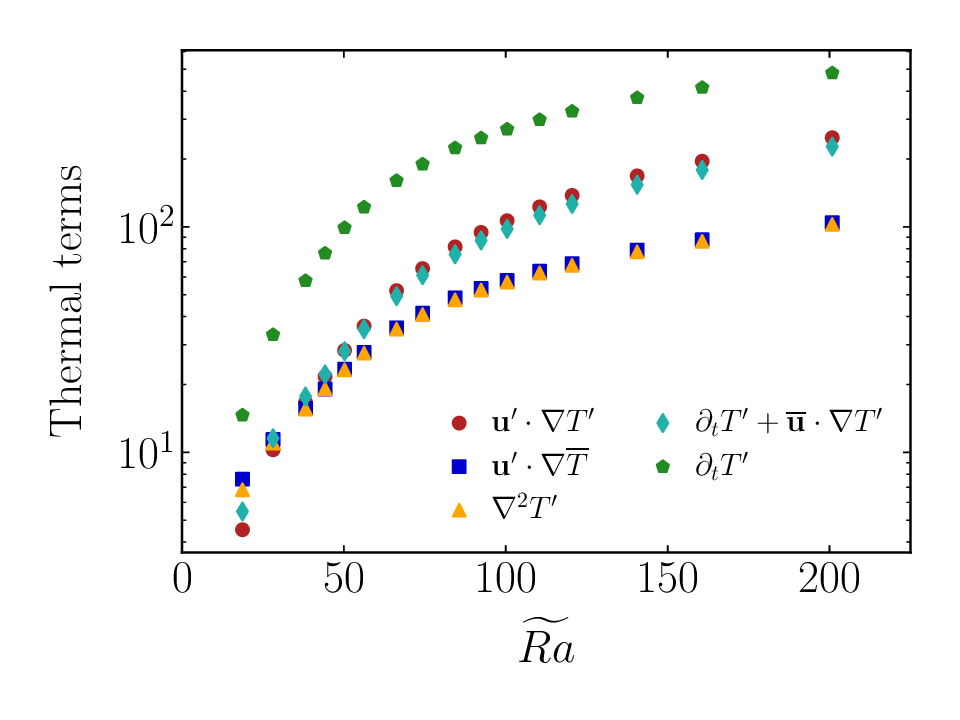}}
  \subfloat[][]{\includegraphics[width=0.48\textwidth]{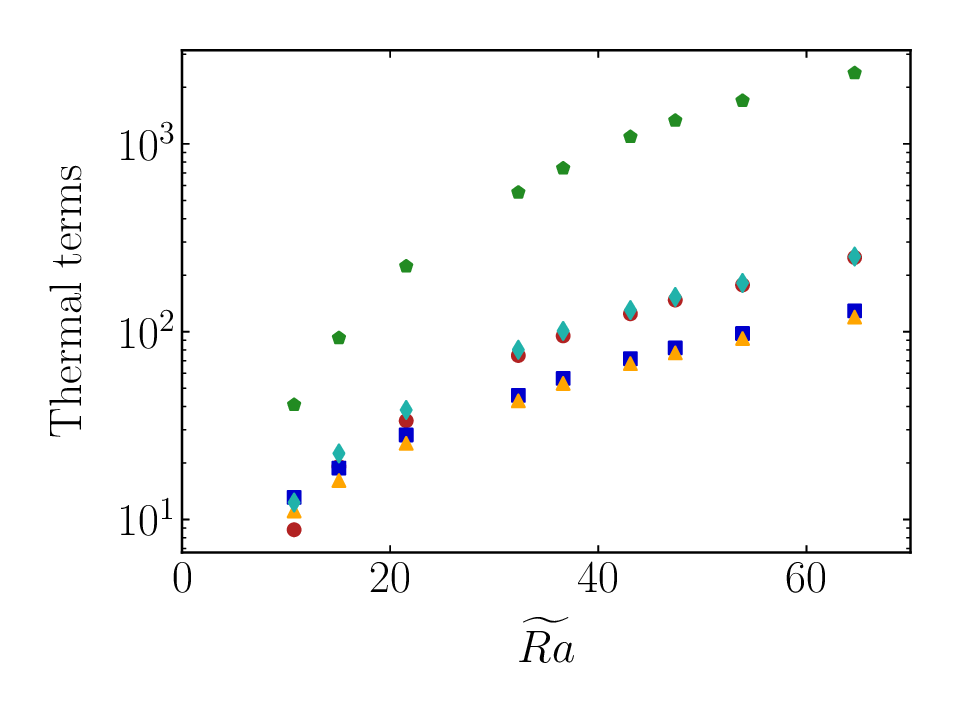}}
\caption{Time-averaged volume rms of several terms from the fluctuating heat equation averaged in time for thick shell simulations: (a) $Ek=3\times 10^{-4}$; (b) $Ek=10^{-5}$. }
\label{F:thermal_terms_Ra_comp}
\end{center}
\end{figure}

The asymptotic scaling behaviour of terms in both the momentum and heat equations are linked. For this purpose, we study the scaling of various terms in both equations beginning with the heat equation in the present section. Figure \ref{F:temp_rms} shows the rms of the fluctuating temperature for all of the thick shell cases. We observe a systematic decrease in the magnitude of the fluctuating temperature as the Ekman number is reduced. Figure \ref{F:temp_rms}(b) shows that the fluctuating temperature is well described by the relationship $T' = O(Ek^{1/3})$ for fixed $\Rat$, as predicted by asymptotic theory. A line gives a decent approximation for how the fluctuating temperature varies with $\Rat$, in contrast to the CIA balance, which predicts a reduced Rayleigh number dependence of $T'\sim \Rat{\vphantom{Ra}}^{1/2}$. Figure \ref{F:temp_rms}(c) shows the compensated fluctuating temperature using the reduced Rayleigh number dependence predicted by the CIA balance: $\Rat{\vphantom{Ra}}^{1/2}$. We see that the data is not very well described by this Rayleigh number dependence; for none of the Ekman numbers does there exist a large range of $\Rat$ where the compensated data is horizontal. We also note that for the $Ek=3\times 10^{-4}$ cases, the largest $\Rat$ cases exhibit a slight decrease in the rms value of $T'$ as compared to lower $\Rat$ cases, so $T'$ is not strictly increasing with $\Rat$ in our data.


Figure \ref{F:thermal_terms_scaling} shows how the various terms from the fluctuating heat equation depend on Ekman number and reduced Rayleigh number. Figure \ref{F:thermal_terms_scaling}(d,e,f) shows how well the asymptotic prediction of the scaling collapses the data. The diffusion term and the advection of the mean temperature by the fluctuating velocity term are well collapsed by the predicted $Ek^{-1/3}$ scaling, although the advection of the fluctuating temperature by the fluctuating velocity is not as well collapsed by the asymptotic scaling. The data suggests that a slightly stronger dependence on the Ekman number is present since the rescaled data from the smaller Ekman number cases lies above that of the larger Ekman number cases. However, we note that the magnitude of the rescaled terms are all comparable, thus providing support for the asymptotic theory.


Figure \ref{F:thermal_terms_Ra_comp} shows how the various terms from the fluctuating heat equation vary with the reduced Rayleigh number for two fixed values of the Ekman number. The large magnitude of the time derivative term is a consequence of the zonal flow; advection by the zonal flow is not balanced and causes large accelerations. Therefore, the sum of the advection by the zonal flow and the time derivative is smaller by comparison and this sum is approximately balanced by the term $\mathbf{u}' \cdot \nabla T'$. Another interesting observation from figure \ref{F:thermal_terms_Ra_comp} is that there is a near balance of the diffusion term and the advection of the mean temperature term for a wide range of $\Rat$. This balance was also noted by \cite{sM21}.

\subsection{Force balances}

\begin{figure}
 \begin{center}

\subfloat[][]{\includegraphics[width=0.33\textwidth]{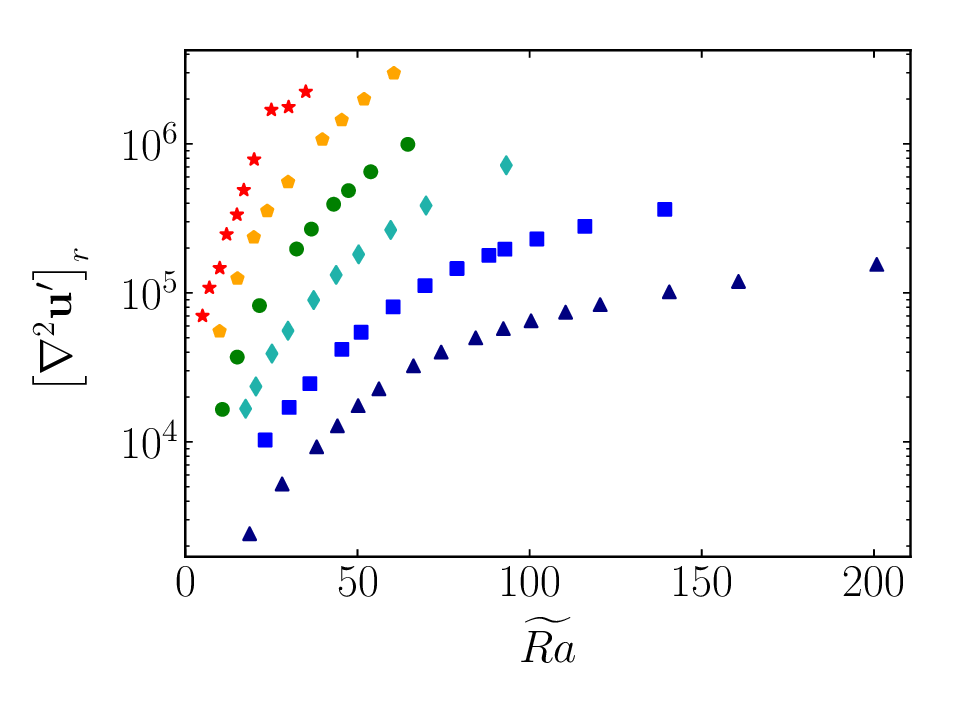}}
\subfloat[][]{\includegraphics[width=0.33\textwidth]{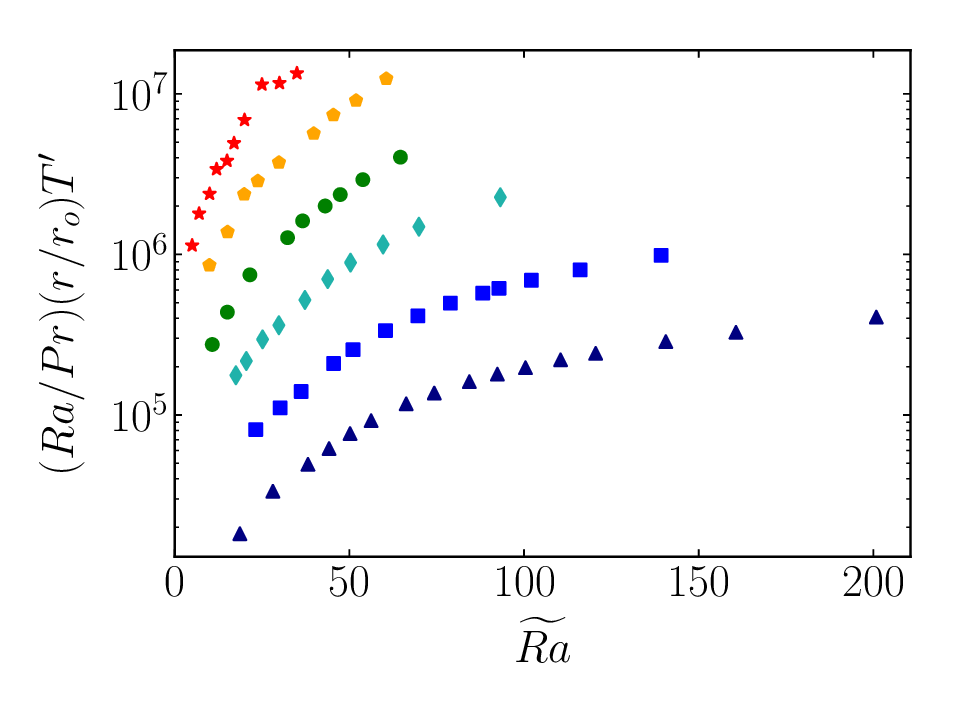}} 
\subfloat[][]{\includegraphics[width=0.33\textwidth]{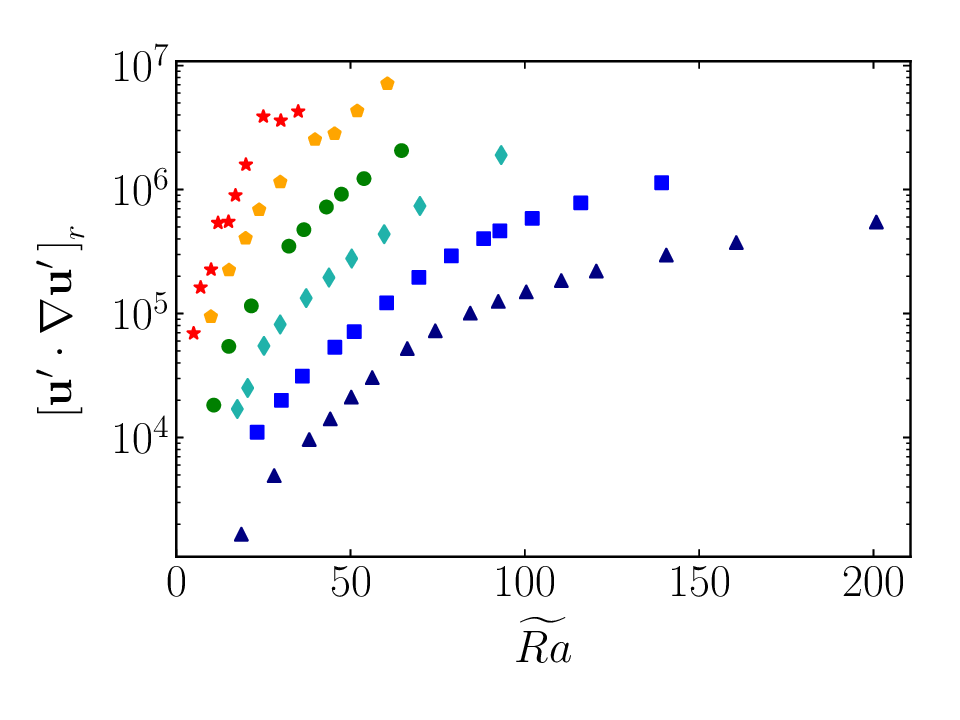}}
\hspace{0mm}
\subfloat[][]{\includegraphics[width=0.33\textwidth]{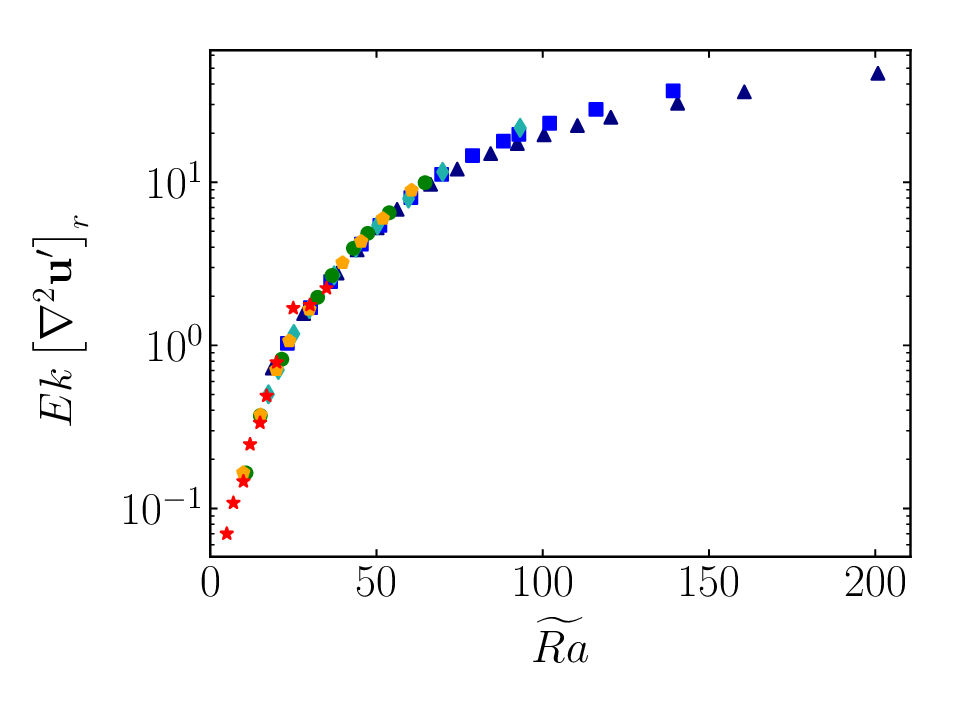}}
\subfloat[][]{\includegraphics[width=0.33\textwidth]{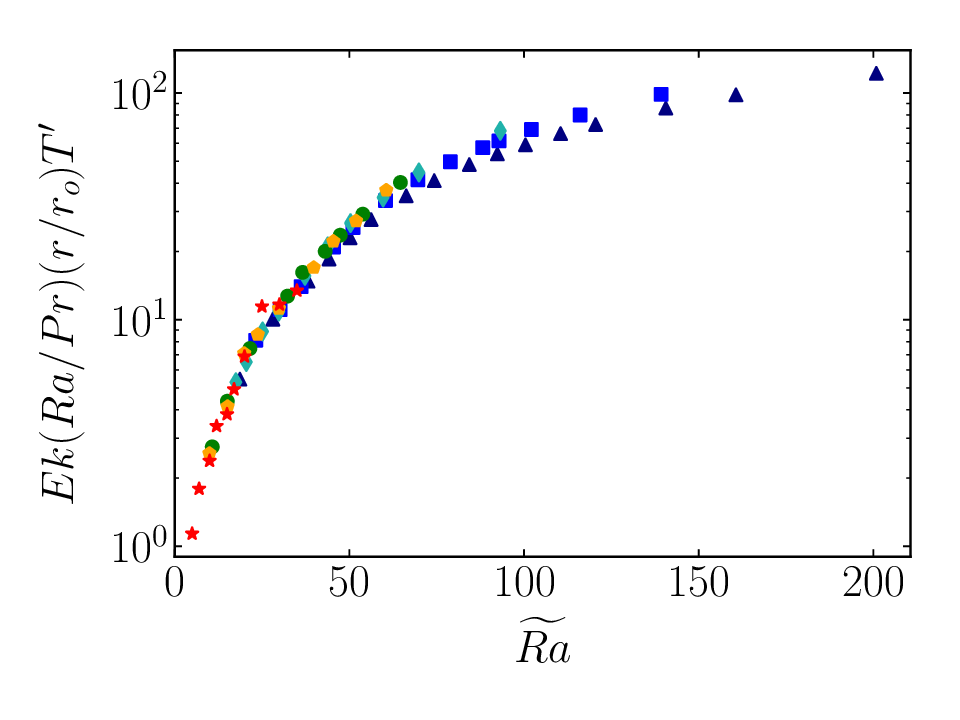}}
\subfloat[][]{\includegraphics[width=0.33\textwidth]{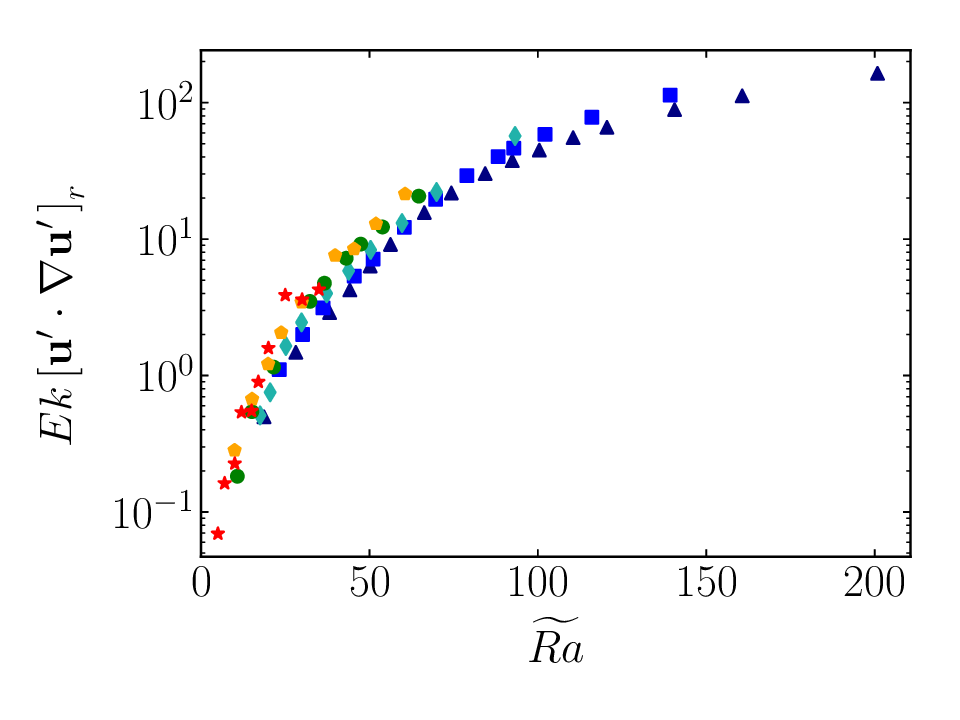}}

\caption{Time-averaged volume rms of various terms from the radial component of the fluctuating momentum equation: (a) viscous force;  (b) buoyancy force; (c) fluctuating advection term; (d) rescaled viscous force; (e) rescaled buoyancy force; (f) rescaled fluctuating advection term. The symbols are the same as defined in figure \ref{F:Re_fluct}. }
\label{F:fluct_forces_Ek_comp}
\end{center}
\end{figure}


\begin{figure}
 \begin{center}
\subfloat[][]{\includegraphics[width=0.33\textwidth]{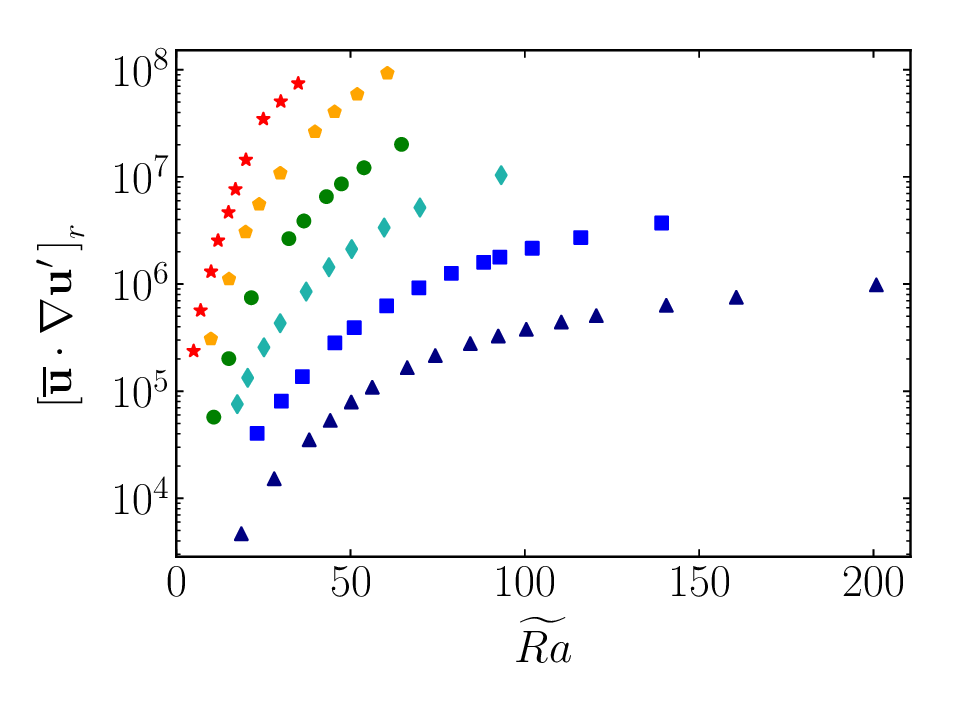}}
\subfloat[][]{\includegraphics[width=0.33\textwidth]{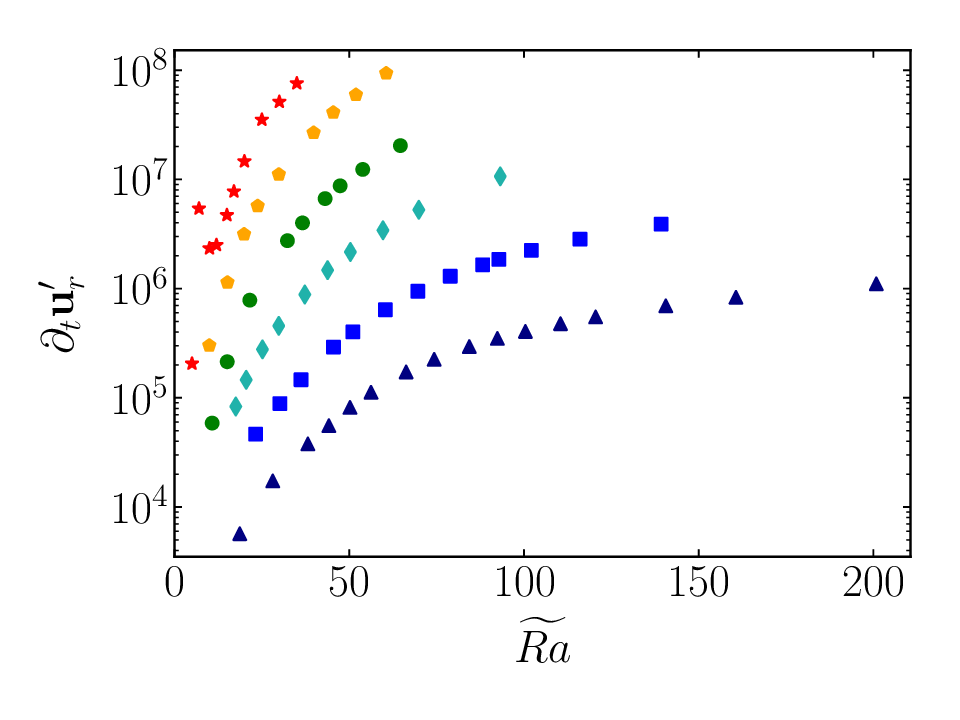}} 
\subfloat[][]{\includegraphics[width=0.33\textwidth]{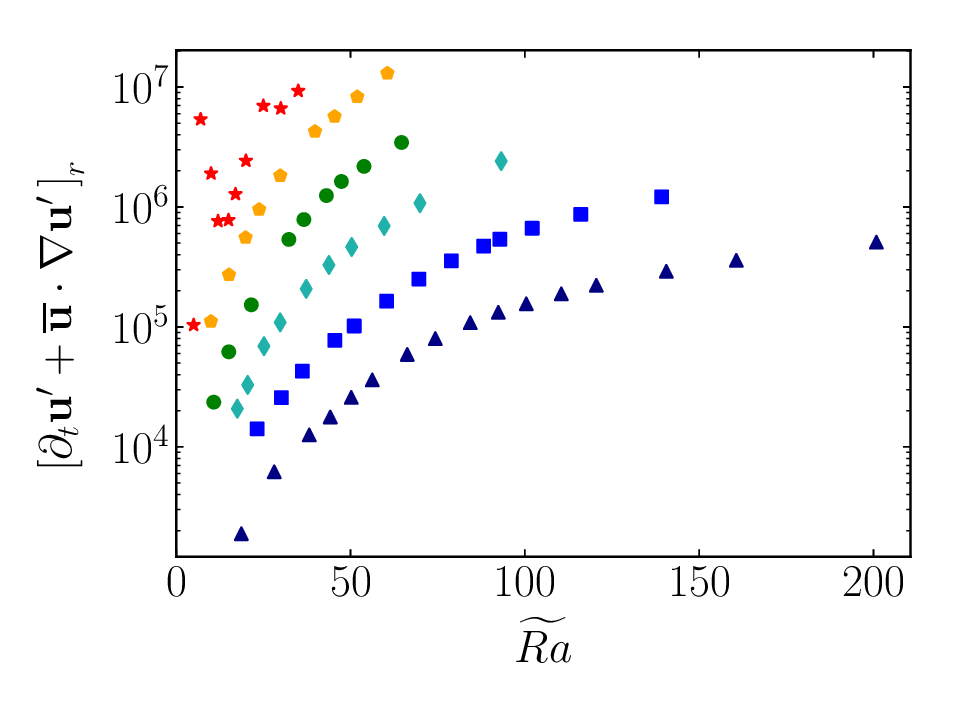}}
\hspace{0mm}
\subfloat[][]{\includegraphics[width=0.33\textwidth]{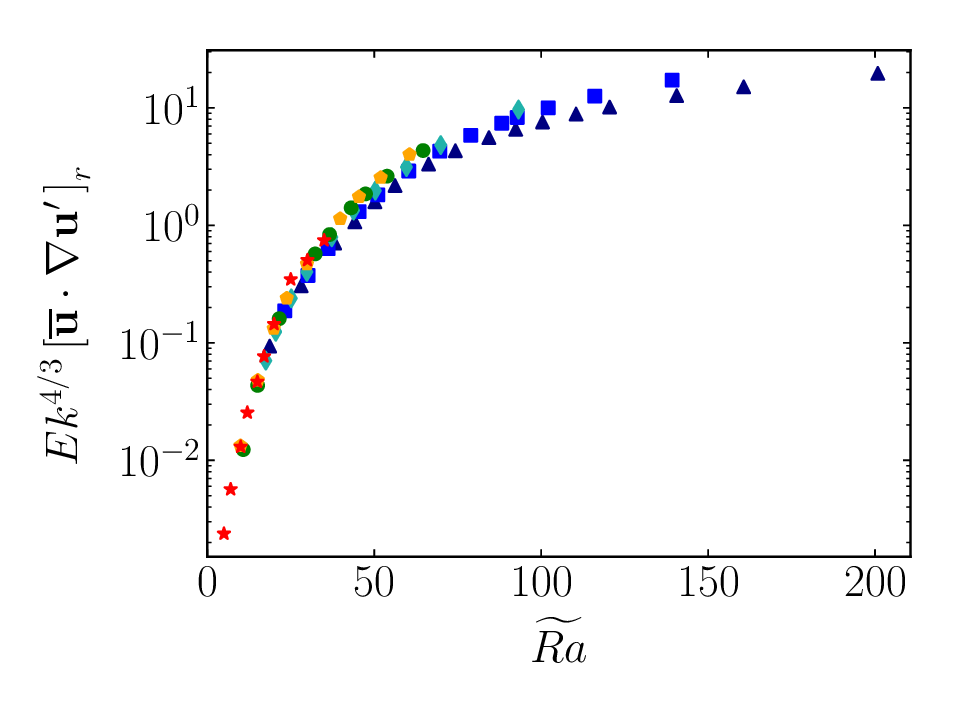}}
\subfloat[][]{\includegraphics[width=0.33\textwidth]{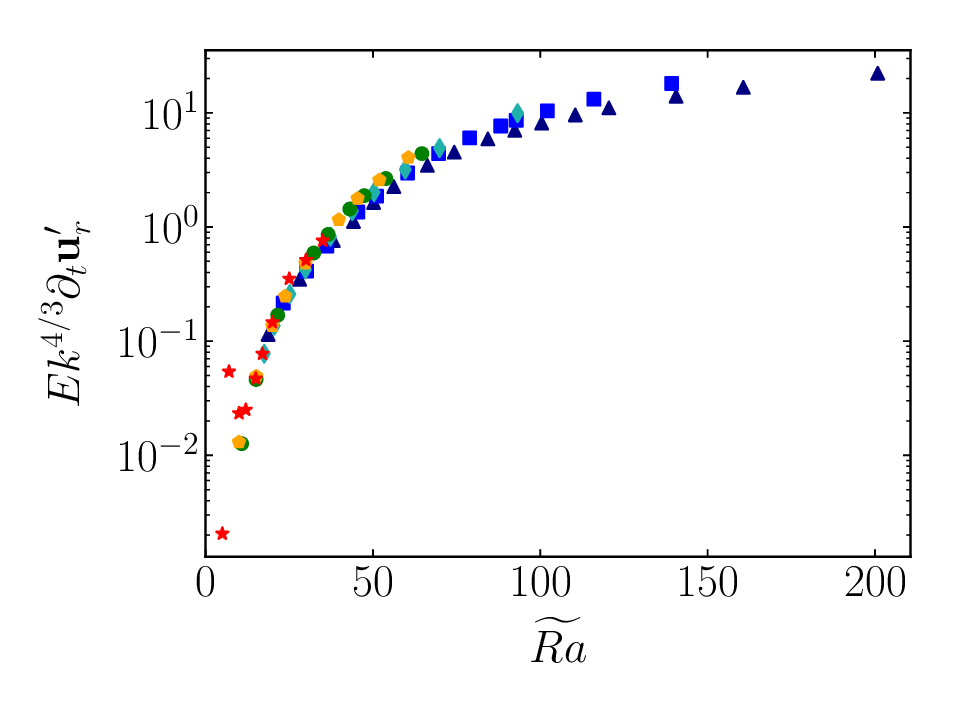}} 
\subfloat[][]{\includegraphics[width=0.33\textwidth]{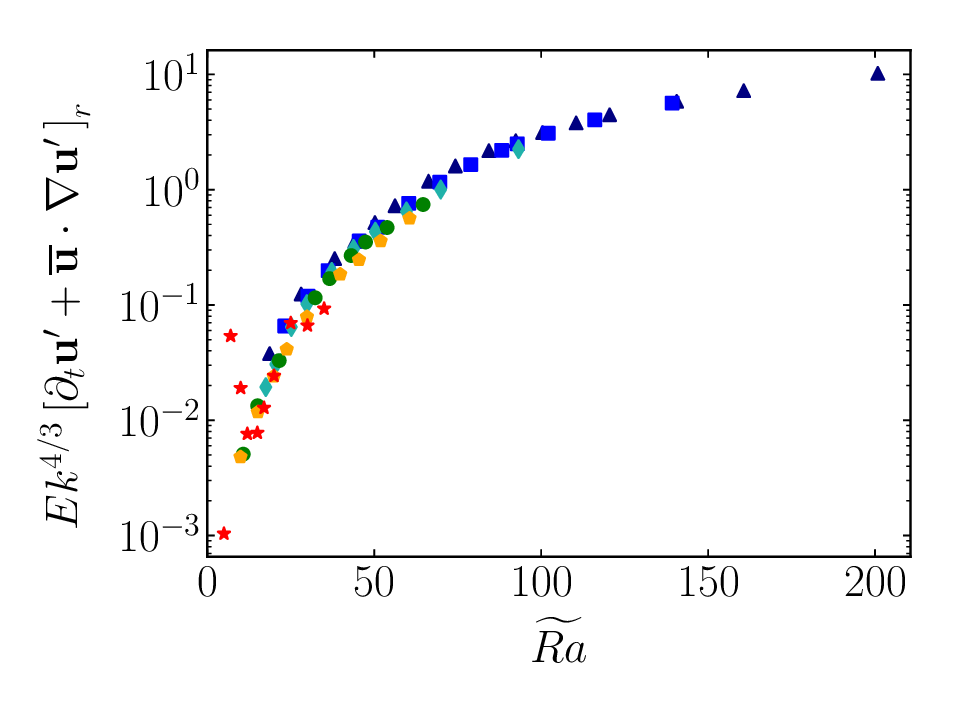}}

\caption{Time-averaged volume rms of various terms from the radial component of the fluctuating momentum equation: (a) advection by the zonal flow;  (b) time derivative (inertia); (c) the sum of the time derivative and advection by the zonal flow; (d) rescaled advection by the zonal flow; (e) rescaled time derivative; (f) rescaled sum of the time derivative and advection by the zonal flow. The symbols are the same as defined in figure \ref{F:Re_fluct}.}
\label{F:fluct_forces_advection}
\end{center}
\end{figure}

\begin{figure}
 \begin{center}
 \subfloat[][]{\includegraphics[width=0.48\textwidth]{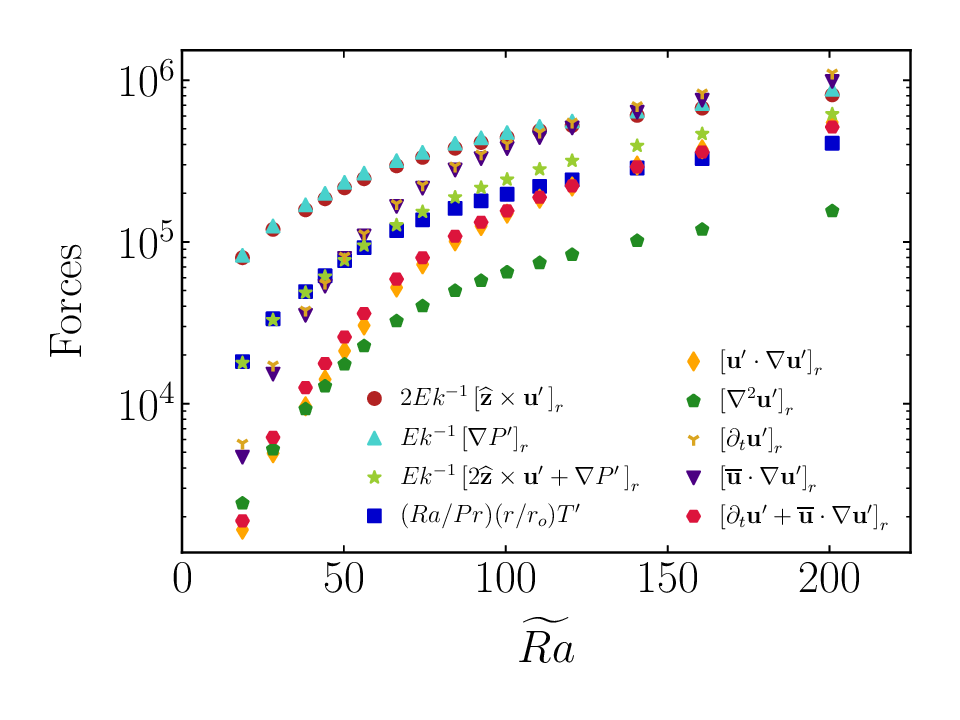}}
  \subfloat[][]{\includegraphics[width=0.48\textwidth]{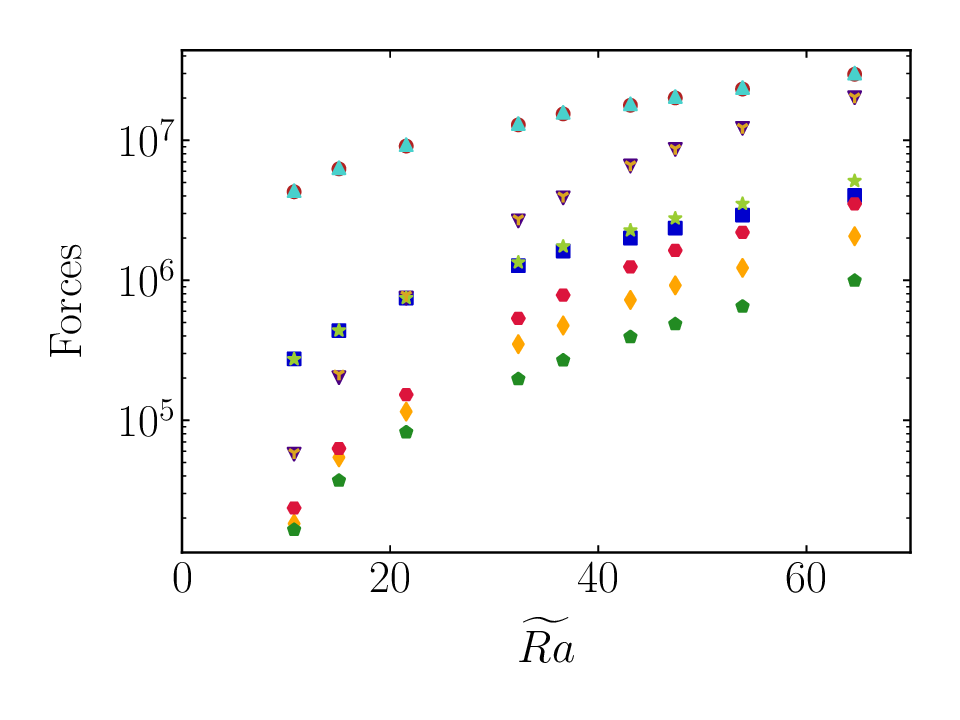}}
\caption{Volume rms of several terms from the fluctuating radial momentum equation averaged in time for thick shell simulations: (a) $Ek=3\times 10^{-4}$; (b) $Ek=10^{-5}$.}
\label{F:fluct_forces_Ra_comp}
\end{center}
\end{figure}

\begin{figure}
 \begin{center}
 \subfloat[][]{\includegraphics[width=0.48\textwidth]{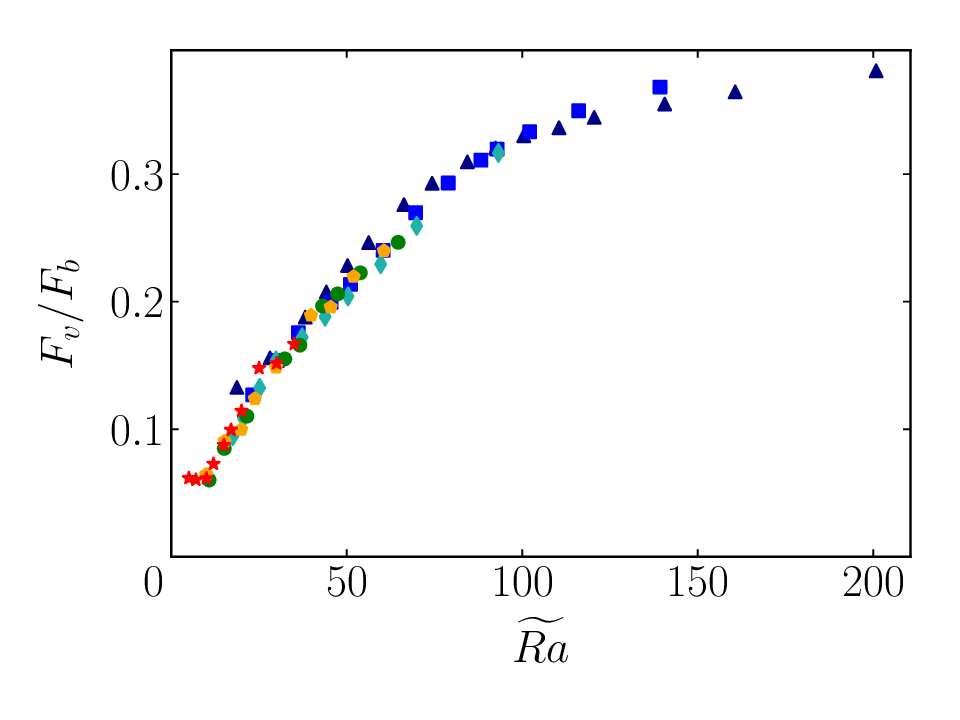}}
  \subfloat[][]{\includegraphics[width=0.48\textwidth]{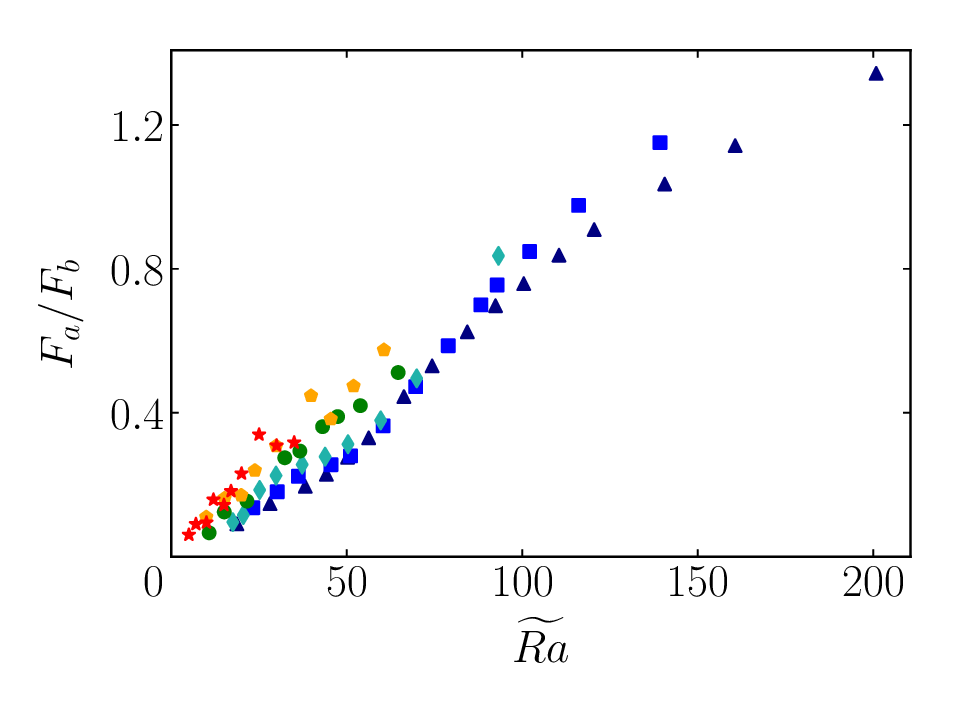}}
\caption{Ratios of forces in the fluctuating radial momentum equation where $F_v=\left[\nabla^2 \mathbf{u}'\right]_r$, $F_a=\left[\mathbf{u}'\cdot\nabla\mathbf{u}'\right]_r$, and $F_b=(Ra/Pr)(r/r_0)T'$:
(a) ratio of the viscous force to the buoyancy force; (b) ratio of the fluctuating advective term to the buoyancy force. The symbols are the same as defined in figure \ref{F:Re_fluct}.}
\label{F:force_ratio}
\end{center}
\end{figure}

In this section, we investigate the scaling behaviour of the forces present in the radial component of the fluctuating momentum equation (similar scaling was observed in the other two components). Figure \ref{F:fluct_forces_Ek_comp} shows the time-average and rms over the entire domain of the viscous force, buoyancy force, and the fluctuating advective term. For the fluctuating advective term, we remove the spherically symmetric $l=0$ mode as this mode is not dynamically relevant. From the arguments in the theory section, we expect these three terms to all scale as $Ek^{-1}$ in our non-dimensional units. We test these scalings by multiplying the data in figure \ref{F:fluct_forces_Ek_comp}(a,b,c) by $Ek$, which is shown in figure \ref{F:fluct_forces_Ek_comp}(d,e,f). We see that the predicted Ekman number scalings are consistent with the data. Similar to the heat equation terms in the previous section, the collapse for the advective term is not as good as the collapse for the other terms. This difference either indicates that the advective term is converging more slowly than the other terms or that the advective term is converging to a slightly different scaling than predicted by asymptotic theory. We also note that the scaling of the advective term is time-dependent; removing the intervals of time in which strong convective bursts are occurring in the time series produces a scaling for the fluctuating advective term that more closely follows the predicted $Ek^{-1}$ scaling. This effect suggests that the relaxation oscillations are playing some role in the scaling, which may be due to the larger Rossby numbers that occur during these times. The Coriolis force is not shown since its scaling follows from the scaling of the fluctuating velocity given in figure \ref{F:Re_fluct} -- in our units it scales as $Ek^{-4/3}$.

These scalings suggest that, for a fixed value of the reduced Rayleigh number, the ratio of the viscous force, buoyancy force, and fluctuating advection term remain approximately fixed as the Ekman number is decreased. On the other hand, the ratio of either the viscous force, the buoyancy force, or the fluctuating advective term to the Coriolis force will scale as $Ek^{1/3}$. While we have only shown the radial component of the fluctuating force balance, we note that the $\theta$ and $\phi$ components of the fluctuating force balance show similar behaviour. We also tested removing the thermal boundary layers, and did not observe a qualitative change in any of the trends shown.

Some terms from the momentum equation follow a stronger Ekman number scaling due to the presence of the zonal flow, and these terms are shown in figure \ref{F:fluct_forces_advection}. The zonal flow strongly advects the fluctuating velocity, which is unbalanced and causes large accelerations in the small-scale fluid structures. Because the zonal flow scales as $Ek^{-2/3}$, the fluctuating velocity scales as $Ek^{-1/3}$, and the length scale of the fluctuating velocity scales as $Ek^{1/3}$, we expect the advection by the zonal flow to scale as $Ek^{-4/3}$, which we also expect for the scaling of the time derivative. Figure \ref{F:fluct_forces_advection}(d,e) shows the collapse of the mean advection term and the time derivative term for this scaling. We see that both the advection by the zonal flow and the time derivative follow a $Ek^{-4/3}$ scaling, which we note is the same scaling followed by the Coriolis force. Therefore, the advection by the zonal flow and the time derivative appear at leading order asymptotically, and the sum of these two terms is somewhat smaller than either individually, as shown in figure \ref{F:fluct_forces_advection}(c,f) and in figure \ref{F:fluct_forces_Ra_comp}.


Figure \ref{F:fluct_forces_Ra_comp} shows how the different terms in the fluctuating radial momentum equation depend on the reduced Rayleigh number for two Ekman numbers. For both $Ek=3\times 10^{-4}$ and $Ek=10^{-5}$, the buoyancy force is about an order of magnitude larger than viscosity or advection at small $\Rat$, and both viscosity and advection grow more rapidly with Rayleigh number as compared to the buoyancy force. For $Ek=3\times 10^{-4}$, advection becomes larger than buoyancy near $\Rat \approx 140$ and continues growing larger than buoyancy. For $Ek=10^{-5}$, we do not reach large enough Rayleigh numbers for advection to become as large as buoyancy. Therefore, we do not observe a CIA balance in our simulations since the buoyancy force and advection scale differently with Rayleigh number for the parameter space we surveyed. 

In figure \ref{F:force_ratio} we plot the ratio of the viscous force to the buoyancy force, and the ratio of the fluctuating advective term to the buoyancy force. We find that the ratio of the viscous force to buoyancy is an increasing function of the Rayleigh number, which suggests that viscosity does not become negligible at more extreme parameters. This behaviour is in contrast to the diffusion-free scaling that is used in many previous studies, which assume that viscosity is negligible. Therefore, the trends we observe in our data do not support viscosity becoming negligible, although it is possible that this trend changes at higher values of the reduced Rayleigh number than we achieved in this study. 

\subsection{Heat flow and dissipation}
\begin{figure}
 \begin{center}
 \subfloat[][]{\includegraphics[width=0.45\textwidth]{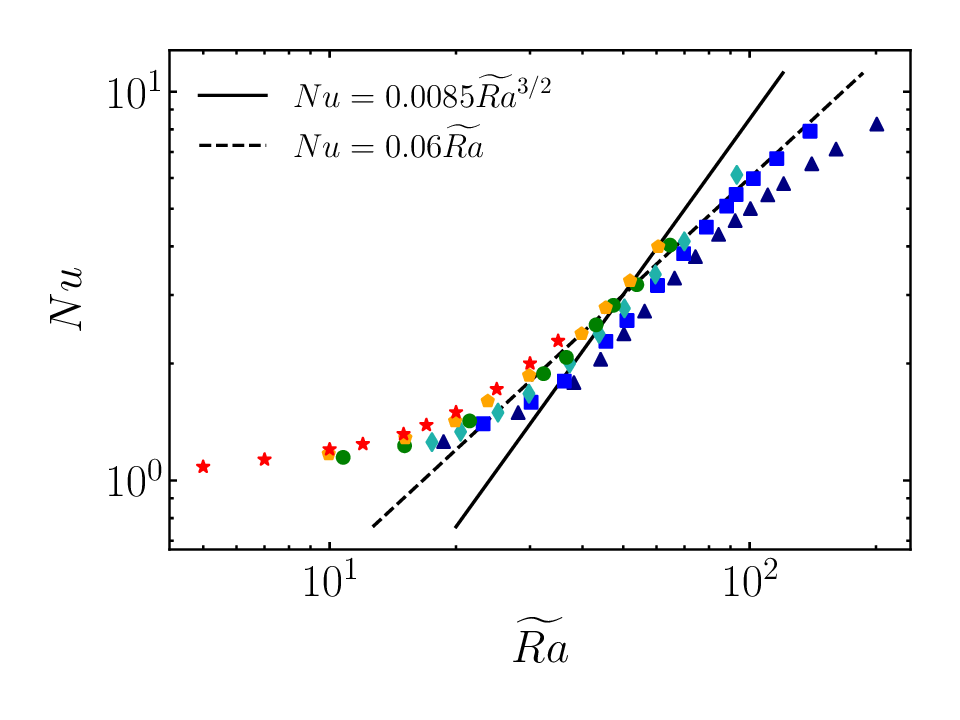}}
 \subfloat[][]{\includegraphics[width=0.45\textwidth]{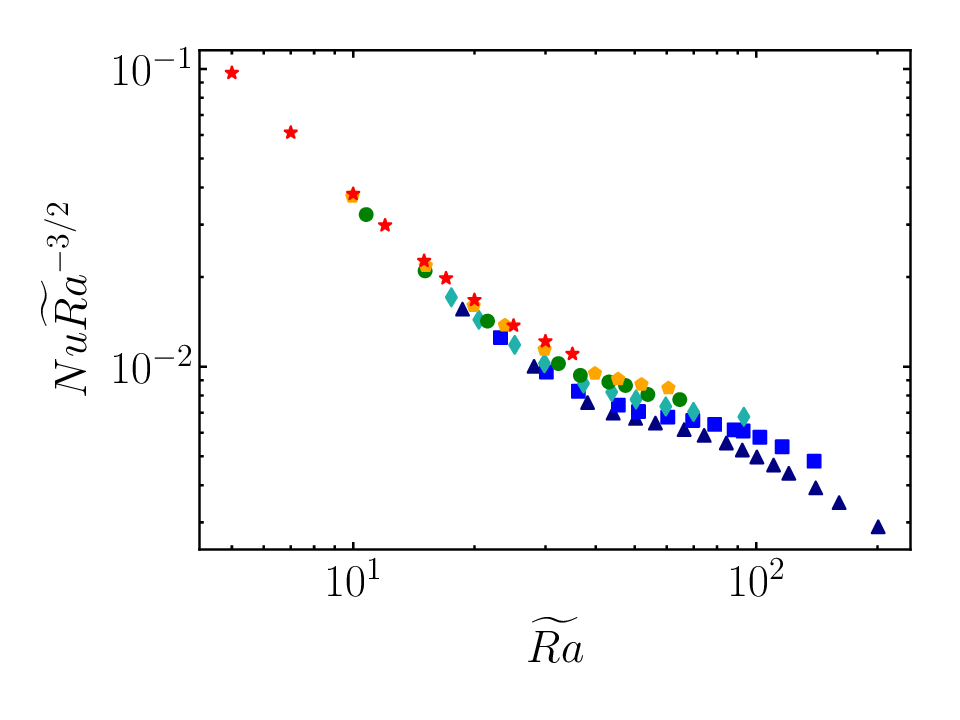}}
\caption{Time-averaged Nusselt number calculated at the outer boundary: (a) Nusselt number; (b) compensated Nusselt number. The symbols are the same as defined in figure \ref{F:Re_fluct}.}
\label{F:Nusselt}
\end{center}
\end{figure}

Figure \ref{F:Nusselt} shows the Nusselt number for all of the $\eta=0.35$ cases from this paper with two scalings shown for reference. The solid line shows the predicted scaling for diffusion-free heat transport, $Nu \sim \Rat{\vphantom{Ra}}^{3/2}$, and the dashed line shows the scaling $Nu \sim \Rat$. It appears that our data has not quite yet reached the predicted $Nu \sim \Rat{\vphantom{Ra}}^{3/2}$ scaling, although some of our $Ek=3\times 10^{-6}$ cases appear to be close to this scaling. This might suggest that the $Nu \sim \Rat{\vphantom{Ra}}^{3/2}$ scaling can be reached at lower Ekman number and higher Rayleigh number.

\begin{figure}
 \begin{center}
 \subfloat[][]{\includegraphics[width=0.45\textwidth]{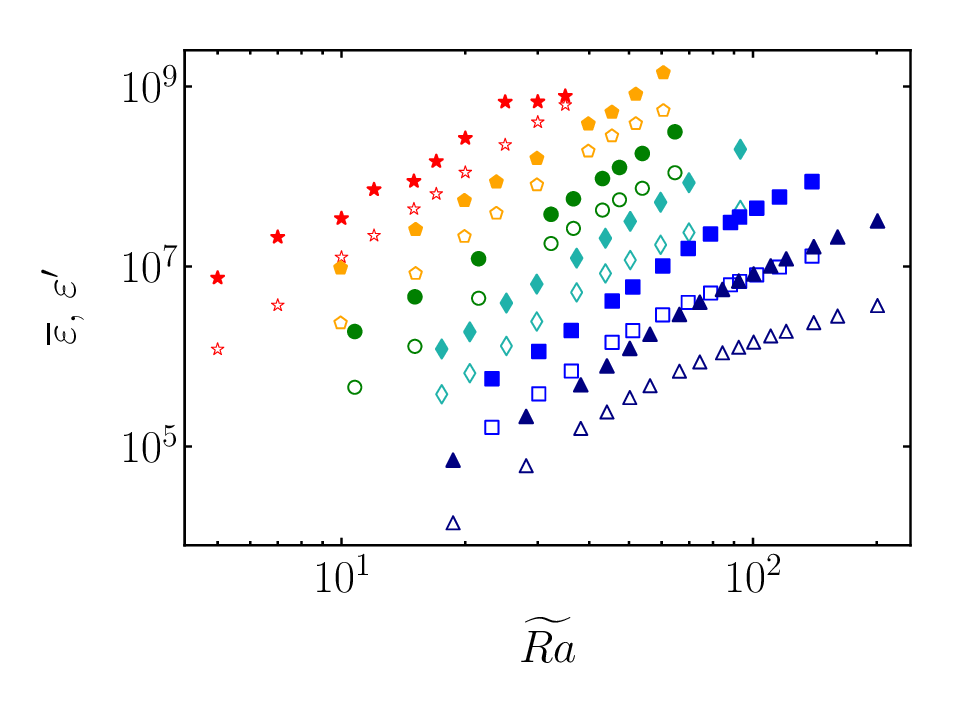}}
 \subfloat[][]{\includegraphics[width=0.45\textwidth]{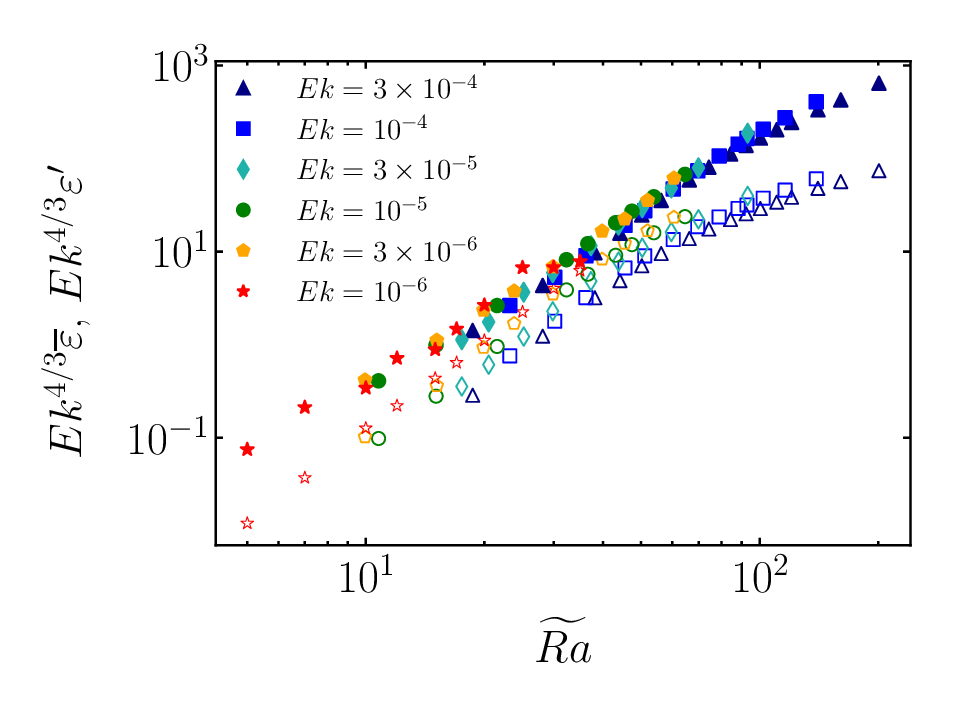}}
\caption{Viscous dissipation rates for the convective flow and the zonal flow: (a) dissipation rate; (b) rescaled dissipation rate. The filled symbols denote small-scale dissipation rate ($\varepsilon^{\prime}$) and hollow symbols denote large-scale dissipation rate ($\overline{\varepsilon}$).}
\label{F:dissipation}
\end{center}
\end{figure}

\begin{figure}
 \begin{center}
  \subfloat[][]{\includegraphics[width=0.45\textwidth]{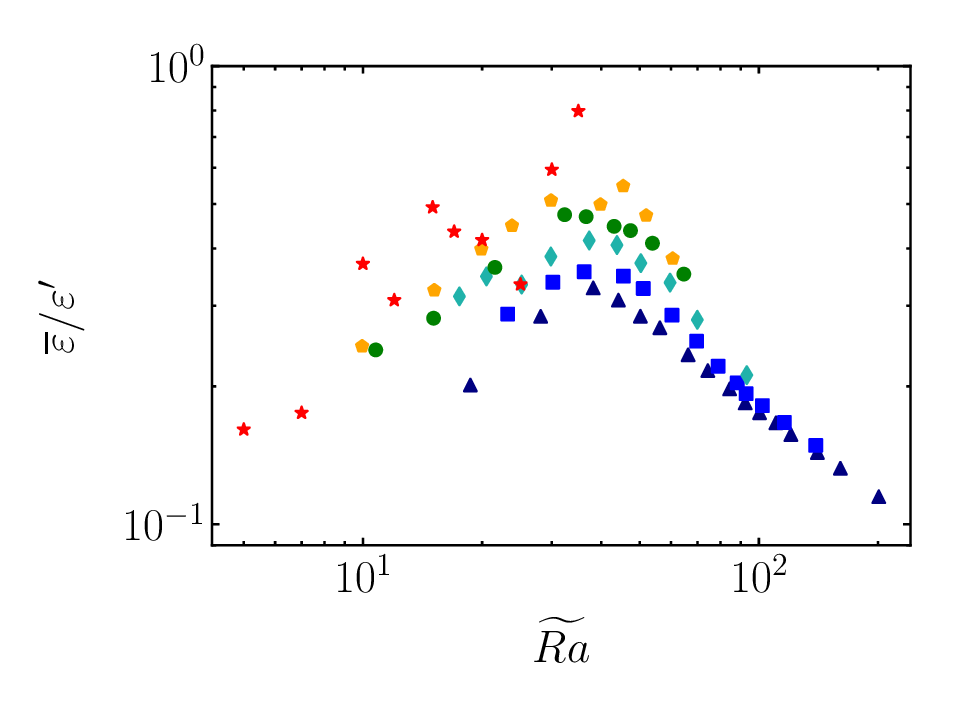}}
\caption{Ratio of the large scale viscous dissipation rate to the small-scale dissipation rate. The symbols are the same as defined in figure \ref{F:Re_fluct}.}
\label{F:dissipation_ratio}
\end{center}
\end{figure}

Figure \ref{F:dissipation}(a) shows the dissipation by both the convective flow and the mean flow. The dissipation by the convective flow is greater than the dissipation by the mean flow for all of the cases we studied. Figure \ref{F:dissipation}(b) shows that the dissipation follows the expected scaling of $Ek^{-4/3}$, and that the convective dissipation follows the scaling $Ek^{-4/3}$ slightly better than the mean flow. Figure \ref{F:dissipation_ratio} shows the ratio of the dissipation by the mean flow to the dissipation by the convective flow. We see that for all Ekman numbers, this ratio reaches a maximum when $\Rat \sim 30$, and that for higher values of the reduced Rayleigh number the ratio of mean to convective dissipation decreases. This trend might suggest that for large values of the reduced Rayleigh number, the dissipation is mainly controlled by the convection. There is also a trend where the maximum value of the ratio of the mean dissipation to the convective dissipation increases as the Ekman number is decreased. However, this trend is rather weak, with only about a factor of two change in the ratio while the change in Ekman number was two orders of magnitude.

\section{Discussion}
\label{S:discuss}


A systematic investigation of rotating convection in a spherical shell with stress-free boundary conditions was carried out. The scaling behaviour of various quantities was investigated for varying Rayleigh number and Ekman number, and two different aspect ratios were employed. An emphasis was placed on characterising the asymptotic nature of the system as the Ekman number is reduced. 
Here we have utilised the known scalings obtained from both the linear theory of rotating spherical convection \citep{cJ00,eD04} and the closely related nonlinear reduced model for rotating Rayleigh-B\'enard convection \citep[e.g.][]{mS06}. Overall we find good agreement between the asymptotic scalings and the DNS.

In general, we find that the asymptotic nature of the system can be demonstrated when plotting asymptotically rescaled quantities as a function of the reduced Rayleigh number, $\Rat$. Using this approach, we find that the convective flow speeds, as measured by the large-scale Reynolds number, scale as $Re_c = O(Ek^{-1/3})$ for both thin and thick shell simulations. This scaling can be deduced from the governing equations by requiring the convection to be geostrophically balanced to leading order, with the buoyancy force entering the next asymptotic order. As discussed by previous work, the amplitude of the zonal flow is limited solely by large-scale viscous diffusion; thus, we do not anticipate a diffusion-free scaling for this component of the velocity field at any combination of Ekman or Rayleigh number in the rotationally constrained regime. By balancing the large-scale viscous diffusion of zonal momentum with the Reynolds stresses, and noting that $Re_c = O(Ek^{-1/3})$, we deduce that the zonal flow speeds scale as $Re_z = O(Ek^{-2/3})$. The numerical data supports a trend toward this scaling as $Ek \rightarrow 0$. Moreover, we find that because of this diffusion-Reynolds-stress-balance, the correlations in the convective flow field are independent of the Ekman number. As noted in previous work \citep{uC02,tG12}, these correlations decrease rapidly with increasing Rayleigh number (approximately as $\sim \Rat{\vphantom{Ra}}^{-3/2}$). 


Kinetic energy spectra exhibit a trend with increasing Rayleigh number in which energy builds at nearly all spatial scales, with no clear preference for scales significantly larger than the linear convective instability, as indicated by a $Ek^{1/3}$ or $Ek^{2/9}$ scaling. Using the length scale calculation approach of \cite{uC06} and \cite{tG16}, we find that the dominant convective length scale decreases with increasing $\Rat$ and scales approximately as $Ek^{1/3}$. At the largest accessible Rayleigh numbers, but also the largest Ekman numbers, this length scale appears to saturate. The Taylor microscale also scales as $Ek^{1/3}$ and shows a very similar trend in comparison to the spherical-harmonic-based length scale; after initially decreasing with increasing $\Rat$ it tended to level off at the largest accessible Rayleigh numbers. Moreover, both of these length scales remain comparable in value across the entire range of parameters that were simulated, and the collapse of the asymptotically rescaled length scales suggests that this trend continues for more extreme parameter values. The largest length scales present in the fluctuating kinetic energy spectra scale weaker than the primary convective instability and may be consistent with the $Ek^{2/9}$ radial scale from linear theory \citep{eD04}, although a $Ek^{1/6}$ scaling is also consistent with out data. This $Ek^{2/9}$ length scale increases with increasing Rayleigh number, and approximately follows the scaling $\ell^{\prime}_{peak} \sim Ro_c^{1/4}$. Our results suggest that there are multiple length scales present in rotating spherical convection, and that these length scales exhibit different dependencies on the input parameters. Nevertheless, all of these length scales appear to depend on the Ekman number and are therefore viscously controlled to some degree.


Several previous investigations have utilized no-slip boundary conditions that yield different scaling behaviour of the length scales in comparison to the present study. 
\citet{tG16} computed length scales over a broad range of parameters with $\eta=0.6$. These authors used a similar definition of the spherical-harmonic-based length scale (equation \eqref{E:length}), but also included $m=0$ in their calculations. In general, they found non-monotonic behaviour in their values of $\ell_{sh}$, though they did find an increase in the convective length scale provided $Ra$ was sufficiently large and $Ek$ was sufficiently small. We have found that including the $m=0$ component in the length scale calculation causes the length scale to increase with increasing Rayleigh number over some range of $\Rat$, suggesting it is necessary to remove this component when deducing length scale behaviour of the small scale convection. 
Using constant heat flux thermal boundary conditions, \citet{rL20} also used the full spectrum version of equation \eqref{E:length} (i.e.~they included $m=0$) and found an increase in $\ell_{sh}$ at sufficiently small Ekman number.


The mechanical boundary conditions that are employed in the model might play an important role in the scaling behaviour of the convective length scales.
No-slip boundary conditions result in the formation of Ekman layers, which tend to significantly damp the large scale flows \citep{aS77,uC02}. One might expect that the strong radial shear generated by the zonal flow limits the size to which convective structures can grow, therefore yielding distinct scaling behaviour in comparison to simulations in which the zonal flow is suppressed. It is also possible that a transition to a regime in which the convective length scale increases occurs in a parameter regime outside of that used in the presented study. However, our data shows no evidence toward such a transition. Another possibility for the observed differences in the length scale could be that the $m=0$ component of the flow is influencing the length scale for some of the no-slip papers. While the zonal flow is smaller for no-slip cases, zonal flows can still be present and could influence $l_{sh}$. It is also worth noting that \cite{tG23} found that heat transport for convection in rotating spherical shells is strongly dependent on latitude. This suggests that other quantities, such as the length scale and Reynolds number, may also depend strongly on latitude. In this case, it may be illuminating to study the Rayleigh number dependence of the length scale as a function of latitude rather than as a globally averaged quantity.



The asymptotic scaling of various terms in the fluctuating momentum and heat equations was investigated. In particular, for the fluctuating momentum equation, we found that the viscous force and the buoyancy force scale as $Ek^{-1}$, and that the fluctuating-fluctuating advective term scales approximately as $Ek^{-1}$, or perhaps slightly stronger than $Ek^{-1}$ (with our large scale viscous diffusion non-dimensionalisation), whereas the Coriolis force scales as $Ek^{-4/3}$. These scalings again agree with asymptotic theory up to a small difference in the advective term. We find that viscosity does not become smaller compared to buoyancy as the reduced Rayleigh number is increased, so we do not observe a trend in which viscosity becomes negligible. We also do not observe a balance between buoyancy and advection for any of our simulations; the advective term is always growing more rapidly with Rayleigh number than the buoyancy force. Indeed, for the largest Ekman number considered for $\eta=0.35$ ($Ek=3\times 10^{-4}$), the magnitude of inertia surpasses that of the buoyancy force, though this behaviour is outside the regime of rapidly rotating convection. This finding suggests that, at least for the parameter space observed in this study, there is no CIA balance that occurs.

The force balances that were computed for our spherical simulations are similar to the results found by \cite{aG21} for rapidly rotating convection with no-slip boundaries in Cartesian coordinates. In particular, \cite{aG21} found that for large Rayleigh numbers, the viscous force was approximately as large as the buoyancy force and that the inertial term would be larger than buoyancy. Simulations of the asymptotic model for the plane layer also show that the ratio of the viscous force to the buoyancy force approaches unity in the high Rayleigh number regime \citep{sM21,tO23}. This similarity indicates that the Rayleigh number dependence of the forces may not depend too sensitively on the geometry of the system. Some dynamo simulations also show similar scalings compared to the hydrodynamic cases. \cite{mY22b} carried out numerical simulations of a dynamo in a plane layer and also found that viscosity and buoyancy follow the same Ekman number dependence with no indication of a regime where viscosity becomes small compared to buoyancy. On the other hand, \cite{rY16} carried out numerical simulations of spherical dynamo cases with no-slip boundaries and found that the ratio of viscosity to buoyancy decreased as the Ekman number was decreased at constant $Ra/Ra_c$. Some other papers studied the spectral decomposition of the forces for spherical dynamo cases and also found that viscosity was much smaller than buoyancy at small values of the Ekman number \citep[e.g.][]{tS19,tS21}. These dynamo simulations where viscosity seems to become small compared to buoyancy at small values of the Ekman number could be following a different asymptotic scaling than we observe in the hydrodynamic cases in this paper, although dynamo plane layer simulations seem to follow similar scaling laws as the hydrodynamic model \citep{mY22b}.

\newpage
\backsection[Funding]{
The authors gratefully acknowledge funding from the National Science Foundation through grants EAR-1945270 (JAN, MAC). Several computing facilities were used to support this research. A portion of the simulations were performed on resources provided by the NASA High-End Computing (HEC) Program through the NASA Advanced Supercomputing (NAS) Division at Ames Research Center.
The Stampede2 and Anvil supercomputers at the Texas Advanced Computing Center and Perdue University, respectively, were made available through allocation PHY180013 from the Advanced Cyberinfrastructure Coordination Ecosystem: Services \& Support (ACCESS) program, which is supported by National Science Foundation grants \#2138259, \#2138286, \#2138307, \#2137603, and \#2138296. Simulations were also conducted on the Summit and Alpine supercomputers at the University of Colorado, Boulder. Summit is a joint effort of the University of Colorado Boulder and Colorado State University, supported by NSF awards ACI-1532235 and ACI-1532236. 
}

\backsection[Declaration of interests]{The authors report no conflict of interest.}
\vspace{6in}

\FloatBarrier
%
%
%

\appendix

\section{Numerical simulation data}

\begin{center}
\begin{minipage}{0.95\linewidth}
  \begin{center}
  \resizebox{\textwidth}{!}{%
    \begin{tabular}{c @{\hskip 0.11in} c @{\hskip 0.11in} c @{\hskip 0.11in} c @{\hskip 0.11in} c @{\hskip 0.11in} c @{\hskip 0.11in} c}
    \hline
    
                $Ra$                        &   $Re_z$          &    $Re_c$                       &          $\ell_{sh}$               &    $\ell_{tm}$                  &    $N_r$      &    $l_{max}$      \\
    \hline
    \hline
                 \multicolumn{7}{c}{$Ek=3\times 10^{-4}$}\\
                 $9.30\times 10^{5}$   &   $30.36 \pm 1.65$           &    $20.81 \pm 1.81$         &          $0.310 \pm 0.020$      &    $0.0777 \pm 0.0035$   &    $96$        &    $191$                \\
                 $1.40\times 10^{6}$   &   $63.94 \pm 3.56$           &    $32.63 \pm 6.14$         &          $0.277 \pm 0.023$      &    $0.0706 \pm 0.0044$   &    $96$        &    $191$                \\   
                 $1.90\times 10^{6}$   &   $102.22 \pm 5.48$         &    $44.14 \pm 8.50$         &          $0.253 \pm 0.017$      &    $0.0639 \pm 0.0036$   &    $96$        &    $191$                \\
                 $2.20\times 10^{6}$   &   $126.14 \pm 6.20$         &    $53.46 \pm 9.55$         &          $0.241 \pm 0.017$      &    $0.0606 \pm 0.0034$   &    $96$        &    $191$                \\
                 $2.50\times 10^{6}$   &   $150.95 \pm 6.49$         &    $64.57 \pm 8.64$         &          $0.233 \pm 0.013$      &    $0.0585 \pm 0.0025$   &    $96$        &    $191$                \\
                 $2.80\times 10^{6}$   &   $173.46 \pm 4.98$         &    $75.08 \pm 7.34$         &          $0.228 \pm 0.011$      &    $0.0567 \pm 0.0022$   &    $96$        &    $191$                \\
                 $3.30\times 10^{6}$   &   $205.85 \pm 5.66$         &    $94.02 \pm 7.83$         &          $0.223 \pm 0.010$      &    $0.0550 \pm 0.0019$   &    $96$        &    $191$                \\
                 $3.70\times 10^{6}$   &   $228.67 \pm 5.83$         &    $108.08 \pm 8.46$       &          $0.221 \pm 0.011$      &    $0.0541 \pm 0.0019$   &    $96$        &    $191$                \\
                 $4.20\times 10^{6}$   &   $253.89 \pm 6.56$         &    $125.22 \pm 9.05$       &          $0.219 \pm 0.010$      &    $0.0531 \pm 0.0018$   &    $96$        &    $191$                \\
                 $4.60\times 10^{6}$   &   $271.37 \pm 5.77$         &    $137.07 \pm 7.96$       &          $0.217 \pm 0.009$      &    $0.0524 \pm 0.0016$   &    $96$        &    $191$                \\
                 $5.00\times 10^{6}$   &   $286.06 \pm 5.59$         &    $149.16 \pm 7.59$       &          $0.216 \pm 0.008$      &    $0.0519 \pm 0.0015$   &    $96$        &    $255$                \\
                 $5.50\times 10^{6}$   &   $305.38 \pm 5.48$         &    $163.67 \pm 8.55$       &          $0.217 \pm 0.008$      &    $0.0514 \pm 0.0014$   &    $96$        &    $255$                \\
                 $6.00\times 10^{6}$   &   $322.14 \pm 6.06$         &    $176.53 \pm 7.65$       &          $0.215 \pm 0.008$      &    $0.0507 \pm 0.0013$   &    $96$        &    $255$                \\
                 $7.00\times 10^{6}$   &   $349.95 \pm 4.95$         &    $202.83 \pm 9.24$       &          $0.214 \pm 0.008$      &    $0.0499 \pm 0.0013$   &    $96$        &    $255$                \\
                 $8.00\times 10^{6}$   &   $374.44 \pm 5.49$         &    $225.58 \pm 8.82$       &          $0.212 \pm 0.007$      &    $0.0490 \pm 0.0011$   &    $96$        &    $255$                \\
                 $1.00\times 10^{7}$   &   $409.37 \pm 7.94$         &    $271.85 \pm 10.23$     &          $0.209 \pm 0.006$      &    $0.0480 \pm 0.0011$   &    $96$        &    $383$                \\
                 \hline
                 \multicolumn{7}{c}{$Ek=10^{-4}$}\\
                 $5.00\times 10^{6}$   &   $103.78 \pm 6.10$         &    $39.80 \pm 9.54$         &          $0.214 \pm 0.016$      &    $0.0540 \pm 0.0032$   &    $96$        &    $191$                \\
                 $6.50\times 10^{6}$   &   $159.63 \pm 13.61$       &    $49.60 \pm 19.43$       &          $0.194 \pm 0.016$      &     $0.0485 \pm 0.0037$   &   $96$         &    $191$                \\
                 $7.80\times 10^{6}$   &   $212.92 \pm 18.12$       &    $60.58 \pm 26.87$       &          $0.183 \pm 0.016$      &     $0.0454 \pm 0.0040$   &   $96$         &    $191$                \\
                 $9.80\times 10^{6}$   &   $308.16 \pm 19.86$       &    $84.35 \pm 26.51$       &          $0.172 \pm 0.012$      &     $0.0423 \pm 0.0029$   &   $96$         &    $191$                \\
                 $1.10\times 10^{7}$   &   $357.39 \pm 18.14$       &    $98.92 \pm 23.40$       &          $0.167 \pm 0.010$      &     $0.0409 \pm 0.0023$   &   $96$         &    $191$                \\
                 $1.30\times 10^{7}$   &   $433.31 \pm 15.10$       &    $125.76 \pm 19.57$     &          $0.163 \pm 0.009$      &     $0.0395 \pm 0.0017$   &   $96$         &    $255$                \\
                 $1.50\times 10^{7}$   &   $502.52 \pm 12.91$       &    $152.64 \pm 15.71$     &          $0.159 \pm 0.007$      &     $0.0384 \pm 0.0012$   &   $96$         &    $255$                \\
                 $1.70\times 10^{7}$   &   $563.11 \pm 10.00$       &    $179.78 \pm 11.73$     &          $0.157 \pm 0.006$      &     $0.0376 \pm 0.0010$   &   $96$         &    $319$                \\
                 $1.90\times 10^{7}$   &   $618.83 \pm 10.85$       &    $205.16 \pm 13.94$     &          $0.156 \pm 0.005$      &     $0.0370 \pm 0.0009$   &   $96$         &    $319$                \\
                 $2.00\times 10^{7}$   &   $643.66 \pm 11.61$       &    $218.32 \pm 14.74$     &          $0.155 \pm 0.005$      &     $0.0367 \pm 0.0009$   &   $96$         &    $319$                \\
                 $2.20\times 10^{7}$   &   $693.42 \pm 11.38$       &    $240.61 \pm 11.80$     &          $0.155 \pm 0.005$      &     $0.0362 \pm 0.0008$   &   $96$         &    $383$                \\
                 $2.50\times 10^{7}$   &   $764.63 \pm 11.21$       &    $272.75 \pm 12.89$     &          $0.153 \pm 0.005$      &     $0.0355 \pm 0.0008$   &   $96$         &    $383$                \\
                 $3.00\times 10^{7}$   &   $862.91 \pm 11.30$       &    $324.65 \pm 12.96$     &          $0.152 \pm 0.004$      &     $0.0347 \pm 0.0006$   &   $96$         &    $511$                \\
                 \hline
                 \multicolumn{7}{c}{$Ek=3\times 10^{-5}$}\\
                 $1.88\times 10^{7}$  &  $157.65 \pm 17.55$         &    $42.60 \pm 18.61$       &          $0.159 \pm 0.015$      &     $0.0403 \pm 0.0037$   &   $96$         &    $191$                 \\
                 $2.20\times 10^{7}$    &  $208.17 \pm 19.61$       &    $50.79 \pm 23.78$       &          $0.150 \pm 0.015$      &     $0.0382 \pm 0.0035$   &   $96$         &    $191$                 \\
                 $2.70\times 10^{7}$    &  $296.62 \pm 31.63$       &    $63.80 \pm 41.37$       &          $0.138 \pm 0.016$      &     $0.0349 \pm 0.0041$   &   $96$         &    $191$                 \\
                 $3.20\times 10^{7}$    &  $402.76 \pm 37.15$       &    $74.98 \pm 52.70$       &          $0.128 \pm 0.016$      &     $0.0323 \pm 0.0040$   &   $96$         &    $191$                 \\
                 $4.00\times 10^{7}$    &  $584.18 \pm 44.58$       &    $98.52 \pm 61.22$       &          $0.123 \pm 0.010$      &     $0.0303 \pm 0.0030$   &   $96$         &    $255$                 \\
                 $4.70\times 10^{7}$    &  $745.19 \pm 45.97$       &    $126.10 \pm 60.13$     &          $0.120 \pm 0.009$      &     $0.0290 \pm 0.0024$   &   $96$         &    $255$                 \\
                 $5.40\times 10^{7}$    &  $883.03 \pm 46.04$       &    $153.36 \pm 54.03$     &          $0.117 \pm 0.009$      &     $0.0281 \pm 0.0018$   &   $96$         &    $255$                 \\
                 $6.40\times 10^{7}$    &  $1068.66 \pm 45.24$     &    $192.98 \pm 50.60$     &          $0.113 \pm 0.008$      &     $0.0271 \pm 0.0014$   &   $96$         &    $319$                 \\
                 $7.50\times 10^{7}$    &  $1245.90 \pm 40.44$     &    $243.58 \pm 44.68$     &          $0.111 \pm 0.005$      &     $0.0264 \pm 0.0010$   &   $96$         &    $383$                 \\
                 $1.00\times 10^{8}$    &  $1578.58 \pm 22.90$     &    $363.96 \pm 28.11$     &          $0.110 \pm 0.004$      &     $0.0255 \pm 0.0006$   &   $144$       &     $511$                 \\
    \hline                                                       
    \hline
                                           
\end{tabular}}
  \end{center}
\captionof{table}{Summary of the cases with $\eta=0.35$. $N_r$ is the number of radial grid points used in the simulation and $l_{max}$ is the maximum spherical harmonic degree used in the simulation. The standard deviation in time of given quantities is shown after the `$\pm$'.}
  \label{T:sims_thick_v0}
\end{minipage}
\end{center}
\setcounter{table}{0}

\begin{center}
\begin{minipage}{0.95\linewidth}
  \begin{center}
  \resizebox{\textwidth}{!}{%
    \begin{tabular}{c @{\hskip 0.11in} c @{\hskip 0.11in} c @{\hskip 0.11in} c @{\hskip 0.11in} c @{\hskip 0.11in} c @{\hskip 0.11in} c}
    \hline
    
                $Ra$                        &   $Re_z$          &    $Re_c$                       &          $\ell_{sh}$               &    $\ell_{tm}$                  &    $N_r$      &    $l_{max}$      \\
    \hline
    \hline
                 \multicolumn{7}{c}{$Ek=10^{-5}$}\\
                 $5.00\times 10^{7}$    &  $167.57 \pm 21.90$       &    $39.25 \pm 24.17$       &          $0.1160 \pm 0.0111$  &     $0.0311 \pm 0.0031$   &   $96$         &    $191$               \\
                 $7.00\times 10^{7}$    &  $291.25 \pm 35.19$       &    $53.62 \pm 42.30$       &          $0.1117 \pm 0.0137$  &     $0.0289 \pm 0.0036$   &   $96$         &    $191$               \\
                 $1.00\times 10^{8}$    &  $545.68 \pm 45.56$       &    $80.19 \pm 57.54$       &          $0.1036 \pm 0.0109$  &     $0.0263 \pm 0.0028$   &   $144$       &    $319$               \\
                 $1.50\times 10^{8}$    &  $1089.39 \pm 69.50$     &    $127.04 \pm 99.16$     &          $0.0984 \pm 0.0054$  &     $0.0230 \pm 0.0021$   &   $144$       &    $383$               \\
                 $1.70\times 10^{8}$    &  $1334.30 \pm 76.57$     &    $150.91 \pm 104.60$   &          $0.0977 \pm 0.0047$  &     $0.0222 \pm 0.0015$   &   $144$       &    $383$               \\
                 $2.00\times 10^{8}$    &  $1677.26 \pm 101.09$   &    $184.54 \pm 119.67$   &          $0.0861 \pm 0.0071$  &     $0.0205 \pm 0.0016$   &   $144$       &    $383$               \\ 
                 $2.20\times 10^{8}$    &  $1890.52 \pm 112.78$   &    $207.11 \pm 124.26$   &          $0.0850 \pm 0.0081$  &     $0.0199 \pm 0.0015$   &   $144$       &    $383$               \\
                 $2.50\times 10^{8}$    &  $2190.80 \pm 128.24$   &    $248.17 \pm 136.41$   &          $0.0827 \pm 0.0069$  &     $0.0193 \pm 0.0013$   &   $144$       &    $511$               \\
                 $3.00\times 10^{8}$    &  $2649.57 \pm 129.37$   &    $331.31 \pm 133.84$   &          $0.0819 \pm 0.0054$  &     $0.0190 \pm 0.0010$   &   $144$       &    $511$               \\
                 \hline
                 \multicolumn{7}{c}{$Ek=3\times 10^{-6}$}\\
                 $2.30\times 10^{8}$    &  $377.82 \pm 31.31$       &    $57.81 \pm 44.80$       &          $0.0802 \pm 0.0084$  &     $0.0216 \pm 0.0024$   &   $140$       &    $359$               \\
                 $3.50\times 10^{8}$    &  $735.92 \pm 59.95$       &    $81.72 \pm 83.93$       &          $0.0743 \pm 0.0099$  &     $0.0194 \pm 0.0029$   &   $140$       &    $399$               \\
                 $4.60\times 10^{8}$    &  $1199.93 \pm 66.06$     &    $118.52 \pm 93.93$     &          $0.0740 \pm 0.0085$  &     $0.0187 \pm 0.0022$   &   $192$       &    $511$                \\
                 $5.50\times 10^{8}$    &  $1627.09 \pm 86.01$     &    $143.85 \pm 132.06$   &          $0.0672 \pm 0.0086$  &     $0.0171 \pm 0.0024$   &   $192$       &    $511$                \\
                 $6.90\times 10^{8}$    &  $2358.89 \pm 99.82$     &    $184.17 \pm 167.07$   &          $0.0615 \pm 0.0081$  &     $0.0156 \pm 0.0020$   &   $256$       &    $639$                \\
                 $9.20\times 10^{8}$    &  $3646.15 \pm 166.26$   &    $254.77 \pm 221.77$   &          $0.0612 \pm 0.0071$  &     $0.0142 \pm 0.0013$   &   $256$       &    $639$                 \\
                 $1.05\times 10^{9}$    &  $4273.72 \pm 188.38$   &    $309.78 \pm 231.08$   &          $0.0589 \pm 0.0046$  &     $0.0138 \pm 0.0012$   &   $256$       &    $767$                 \\
                 $1.20\times 10^{9}$    &  $5040.75 \pm 217.84$   &    $366.99 \pm 244.04$   &          $0.0585 \pm 0.0050$  &     $0.0136 \pm 0.0010$   &   $256$       &    $863$                 \\
                 $1.40\times 10^{9}$    &  $5901.61 \pm 250.97$   &    $472.07 \pm 265.69$   &          $0.0580 \pm 0.0046$  &     $0.0133 \pm 0.0009$   &   $256$       &    $863$                 \\
                 \hline
                 \multicolumn{7}{c}{$Ek=10^{-6}$}\\
                 $5.00\times 10^{8}$    &  $227.03 \pm 8.32$        &    $47.30 \pm 14.01$       &          $0.0603 \pm 0.0024$ &      $0.0165 \pm 0.0008$   &   $144$       &    $431$                  \\
                 $7.00\times 10^{8}$    &  $451.07 \pm 26.64$      &    $66.29 \pm 42.08$       &          $0.0635 \pm 0.0041$ &      $0.0161 \pm 0.0014$   &   $144$       &    $431$                  \\
                 $1.00\times 10^{9}$    &  $865.04 \pm 47.76$      &    $90.02 \pm 75.50$       &          $0.0557 \pm 0.0069$ &      $0.0153 \pm 0.0020$   &   $192$       &    $575$                  \\
                 $1.20\times 10^{9}$    & $1165.20 \pm 64.33$     &    $104.45 \pm 102.12$    &          $0.0547 \pm 0.0080$ &      $0.0145 \pm 0.0023$   &   $192$       &    $575$                   \\
                 $1.50\times 10^{9}$    & $1682.51 \pm 76.85$     &    $126.77 \pm 128.89$    &          $0.0499 \pm 0.0076$ &      $0.0138 \pm 0.0021$   &   $192$       &    $575$                   \\
                 $1.70\times 10^{9}$    & $2079.70 \pm 81.97$     &    $154.19 \pm 146.21$    &          $0.0520 \pm 0.0085$ &      $0.0138 \pm 0.0017$   &   $192$       &    $639$                   \\
                 $2.00\times 10^{9}$    & $2745.75 \pm 89.92$     &    $189.02 \pm 164.60$    &          $0.0505 \pm 0.0076$ &      $0.0130 \pm 0.0015$   &   $192$       &    $639$                   \\
                 $2.50\times 10^{9}$    & $3932.11 \pm 114.99$   &    $235.48 \pm 235.13$    &          $0.0492 \pm 0.0065$ &      $0.0114 \pm 0.0015$   &   $192$       &    $639$                    \\
                 $3.00\times 10^{9}$    & $5197.22 \pm 138.03$   &    $279.56 \pm 294.64$    &          $0.0434 \pm 0.0057$ &      $0.0107 \pm 0.0013$   &   $256$       &    $767$                    \\
                 $3.50\times 10^{9}$    & $6481.03 \pm 173.10$   &    $344.18 \pm 316.83$    &          $0.0427 \pm 0.0043$ &      $0.0103 \pm 0.0012$   &   $256$       &    $767$                    \\
      \hline                                                               
      \hline

\end{tabular}}
  \end{center}
  \captionof{table}{ (Continued)}
  \end{minipage}
  \end{center}

\begin{center}
\begin{minipage}{0.95\linewidth}
 \begin{center}
  \resizebox{0.6\textwidth}{!}{%
    \begin{tabular}{c @{\hskip 0.11in} c @{\hskip 0.11in} c @{\hskip 0.11in} c @{\hskip 0.11in} c}
    \hline
    
                $Ra$                        &   $Re_z$                &    $Re_c$                &    $N_r$      &    $l_{max}$      \\
    \hline
    \hline
                 \multicolumn{5}{c}{$Ek = 10^{-3}$}\\
                 $2.40\times 10^{5}$         &   $72.75 \pm 1.60$      &    $36.92 \pm 2.83$      &    $96$       &    $287$           \\
                 $2.80\times 10^{5}$         &   $88.73 \pm 1.23$      &    $43.39 \pm 3.24$      &    $96$       &    $287$           \\
                 $3.20\times 10^{5}$         &   $103.59 \pm 1.10$     &    $50.34 \pm 3.47$      &    $96$       &    $287$           \\
                 $4.00\times 10^{5}$         &   $129.68 \pm 1.40$     &    $66.22 \pm 4.50$      &    $96$       &    $287$           \\
                 $6.00\times 10^{5}$         &   $166.64 \pm 2.93$     &    $103.71 \pm 5.50$     &    $96$       &    $287$           \\
                 \hline
                 \multicolumn{5}{c}{$3\times Ek = 10^{-4}$}\\
                 $9.30\times 10^{5}$          &   $115.96 \pm 1.76$     &    $41.72 \pm 3.01$      &    $96$       &    $323$           \\
                 $1.40\times 10^{6}$          &   $204.12 \pm 2.18$     &    $61.81 \pm 4.32$      &    $96$       &    $323$           \\
                 $1.90\times 10^{6}$          &   $295.54 \pm 2.65$     &    $86.12 \pm 5.77$      &    $96$       &    $323$           \\
                 $2.20\times 10^{6}$          &   $343.61 \pm 2.42$     &    $102.68 \pm 4.72$     &    $96$       &    $323$           \\
                 $2.50\times 10^{6}$          &   $386.59 \pm 2.52$     &    $119.17 \pm 5.10$     &    $96$       &    $323$           \\
                 $3.30\times 10^{6}$          &   $479.64 \pm 3.79$     &    $164.64 \pm 5.71$     &    $96$       &    $479$           \\
                 \hline
                 \multicolumn{5}{c}{$Ek = 10^{-4}$}\\
                 $2.80\times 10^{6}$          &   $143.37 \pm 2.08$     &    $45.95 \pm 3.67$      &    $96$       &    $511$           \\
                 $5.00\times 10^{6}$          &   $336.55 \pm 3.73$     &    $79.20 \pm 6.55$      &    $96$       &    $511$           \\
                 $6.50\times 10^{6}$          &   $468.59 \pm 4.22$     &    $99.65 \pm 7.83$      &    $96$       &    $511$           \\
                 $7.80\times 10^{6}$          &   $609.74 \pm 4.90$     &    $117.00 \pm 9.35$     &    $96$       &    $511$           \\
                 $9.80\times 10^{6}$          &   $775.45 \pm 6.01$     &    $152.47 \pm 11.39$    &    $96$       &    $511$           \\
                 $1.30\times 10^{7}$          &   $789.71 \pm 13.37$    &    $238.02 \pm 17.57$    &    $96$       &    $575$           \\
                 \hline
                 \multicolumn{5}{c}{$Ek =3\times 10^{-5}$}\\
                 $1.20\times 10^{7}$          &   $264.29 \pm 2.52$     &    $69.23 \pm 5.42$      &    $144$      &    $719$           \\
                 $1.80\times 10^{7}$          &   $564.54 \pm 5.10$     &    $94.06 \pm 11.78$     &    $144$      &    $719$           \\
                 $2.70\times 10^{7}$          &   $810.67 \pm 51.83$    &    $113.01 \pm 74.85$    &    $144$      &    $719$           \\
                 $3.20\times 10^{7}$          &   $964.48 \pm 59.99$    &    $135.17 \pm 81.81$    &    $144$      &    $719$           \\
                 $4.70\times 10^{7}$          &   $1335.08 \pm 58.80$   &    $235.85 \pm 68.34$    &    $144$      &    $863$           \\
                 $6.40\times 10^{7}$          &   $1746.58 \pm 47.76$   &    $371.95 \pm 52.73$    &    $144$      &    $1023$          \\
                 
    \hline                                                       
    \hline
                                           
\end{tabular}}
\captionof{table}{Summary of the cases with $\eta=0.7$. $N_r$ is the number of radial grid points used in the simulation and $l_{max}$ is the maximum spherical harmonic degree used in the simulation. The standard deviation in time of given quantities is shown after the `$\pm$'.}
   \label{T:sims_thin_v0}
   \end{center}
  \end{minipage}
\end{center}

\FloatBarrier

\FloatBarrier

\bibliographystyle{jfm}
\bibliography{ZonalFlows.bbl}

\end{document}